\documentclass[pdflatex,sn-mathphys-num]{sn-jnl}

\usepackage{graphicx}
\usepackage{multirow}
\usepackage{amsmath,amssymb,amsfonts}
\usepackage{amsthm}
\usepackage{mathrsfs}
\usepackage[title]{appendix}
\usepackage{xcolor}
\usepackage{textcomp}
\usepackage{manyfoot}
\usepackage{booktabs}
\usepackage{algorithm}
\usepackage{algorithmicx}
\usepackage{algpseudocode}
\usepackage{listings}

\usepackage{wrapfig}
\usepackage{tabularx}
\usepackage{array}
\usepackage{mathtools}
\usepackage{bm}
\usepackage{siunitx}
\usepackage[export]{adjustbox}
\usepackage{cleveref}
\crefname{figure}{Fig.}{Figs.}
\Crefname{figure}{Fig.}{Figs.}
\crefname{equation}{Eq.}{Eqs.}
\Crefname{equation}{Eq.}{Eqs.}
\crefname{table}{Table}{Tables}
\Crefname{table}{Table}{Tables}
\crefname{section}{Sect.}{Sects.}
\Crefname{section}{Sect.}{Sects.}
\crefname{appendix}{Appendix}{Appendices}
\Crefname{appendix}{Appendix}{Appendices}

\makeatletter
\providecommand*{\theHALG@line}{\thealgorithm.\arabic{ALG@line}}
\makeatother

\newcolumntype{L}[1]{>{\raggedright\arraybackslash}p{#1}}
\newcolumntype{C}[1]{>{\centering\arraybackslash}p{#1}}
\newcolumntype{Y}{>{\raggedright\arraybackslash}X}

\newcommand{\vect}[1]{\mathbf{#1}}

\theoremstyle{thmstyletwo}
\newtheorem{remark}{Remark}

\AtBeginDocument{\setlength{\emergencystretch}{3em}}

\graphicspath{{images/}{/home/jason/Desktop/local-image-archive/journals/2026/tl-fea/}{/home/zzhou292/Desktop/STUDY/local-image-archive/journals/2026/tl-fea/}{../../../../local-image-archive/journals/2026/tl-fea/}{../../../../../local-image-archive/journals/2026/tl-fea/}}
\makeatletter
\newcommand{\maybeincludegraphics}[2][]{%
  \IfFileExists{images/#2}{%
    \includegraphics[#1]{images/#2}%
  }{%
    \IfFileExists{/home/jason/Desktop/local-image-archive/journals/2026/tl-fea/#2}{%
      \includegraphics[#1]{/home/jason/Desktop/local-image-archive/journals/2026/tl-fea/#2}%
    }{%
      \IfFileExists{/home/zzhou292/Desktop/STUDY/local-image-archive/journals/2026/tl-fea/#2}{%
        \includegraphics[#1]{/home/zzhou292/Desktop/STUDY/local-image-archive/journals/2026/tl-fea/#2}%
      }{%
        \IfFileExists{../../../../local-image-archive/journals/2026/tl-fea/#2}{%
          \includegraphics[#1]{../../../../local-image-archive/journals/2026/tl-fea/#2}%
        }{%
          \IfFileExists{../../../../../local-image-archive/journals/2026/tl-fea/#2}{%
            \includegraphics[#1]{../../../../../local-image-archive/journals/2026/tl-fea/#2}%
          }{%
            \begingroup
            \setlength{\fboxsep}{8pt}%
            \fbox{%
              \parbox[c][0.16\textheight][c]{0.82\linewidth}{%
                \centering
                \textbf{Missing figure}\\[0.5ex]
                \small\texttt{\detokenize{#2}}%
              }%
            }%
            \endgroup
          }%
        }%
      }%
    }%
  }%
}
\makeatother

\begin{document}

\title[TL-FEA Part II: GPU Implementation]{A Total Lagrangian Finite Element
Framework for Multibody Dynamics: Part II -- GPU Implementation and
Numerical Experiments}

\author*[1]{\fnm{Zhenhao} \sur{Zhou}}\email{zzhou292@wisc.edu}
\equalcont{These authors contributed equally to this work.}

\author[2]{\fnm{Ruochun} \sur{Zhang}}\email{ruochunz@aerocae.com}
\equalcont{These authors contributed equally to this work.}

\author[1]{\fnm{Ganesh} \sur{Arivoli}}\email{arivoli@wisc.edu}

\author[1]{\fnm{Dan} \sur{Negrut}}\email{negrut@wisc.edu}

\affil*[1]{\orgdiv{Department of Mechanical Engineering},
  \orgname{University of Wisconsin--Madison},
  \orgaddress{\city{Madison}, \state{WI}, \postcode{53706}, \country{USA}}}

\affil[2]{\orgname{AEROCAE Digital Technology Ltd.},
  \orgaddress{\city{Shanghai}, \postcode{201100}, \country{China}}}

\abstract{We present the numerical methods and GPU-accelerated implementation
underlying a Total Lagrangian finite element framework for
finite-deformation flexible multibody dynamics, introduced in the companion
paper of this series. The framework supports 10-node
quadratic tetrahedral (T10) elements and ANCF beam and shell elements, with
quadrature-based hyperelastic response (St.\ Venant--Kirchhoff and
Mooney--Rivlin) and an optional Kelvin--Voigt viscous stress contribution.
Time stepping employs a velocity-based implicit backward-Euler scheme,
yielding a nonlinear residual in velocity that couples inertia, internal and
external forces, and bilateral constraints. Constraints are enforced via an
augmented Lagrangian method (ALM), alternating an inner velocity solve with
a dual-ascent multiplier update. We introduce a
two-stage GPU parallelization strategy for internal force and tangent
stiffness evaluation, and provide two inner solvers: a first-order AdamW
optimizer and a second-order Newton solver that assembles and factorizes a
sparse global Hessian on the GPU using cuDSS. A fixed-sparsity matrix
strategy eliminates repeated symbolic analysis and enables efficient
numerical refactorization across Newton iterations. For collision detection,
we present a GPU-native two-thread asynchronous algorithm operating on
triangle soups, avoiding bounding-volume hierarchies entirely. Systematic
scaling benchmarks across all three supported element types and six mesh
resolutions show that the Newton solver (on an NVIDIA GeForce RTX~5090)
achieves approximately one order of magnitude reduction in real-time factor
relative to CPU baselines (Intel i7-13700KF with MPI parallelization) at
the largest resolutions tested. The frictional contact model is validated
against closed-form rigid-body predictions through quasi-static and dynamic
impact unit tests; bilateral constraint enforcement is validated through
revolute- and spherical-joint double-pendulum tests.}


\let\origprintkeywords\printkeywords
\renewcommand{\printkeywords}{}
\maketitle
\renewcommand{\printkeywords}{\origprintkeywords}

\noindent\textbf{Article Highlights}
\begin{itemize}
  \item A two-stage GPU parallelization strategy achieves an order-of-magnitude speedup over CPU baselines at large mesh sizes.
  \item A GPU-native asynchronous collision detection algorithm operates on triangle soups without bounding-volume hierarchies.
  \item Correctness is verified through unit tests against closed-form solutions and large-scale many-body contact simulations.
\end{itemize}

\origprintkeywords

\section{Introduction}
\label{sec:introduction}

Flexible multibody dynamics (FMBD) with large deformation and large rotation has broad engineering relevance, from soft robotics and deployable space structures to tire-terrain interaction and surgical simulation. Finite element discretization in the Total Lagrangian (TL) setting provides a unified treatment of geometric nonlinearity: all kinematic and stress quantities are referred to the fixed reference configuration, shape function gradients are precomputed once, and the deformation gradient $\mathbf{F}$ serves as the natural interface between kinematics and constitutive response~\cite{belytschko00,bonet2016nonlinear}. The theoretical framework underlying this two-part work, covering element kinematics, material models, frictional contact, bilateral constraints, and their first- and second-order derivatives, is presented in the companion paper~\cite{json-ganesh-danTLFEA-1-2026}, which is currently under review at Engineering with Computers. The present paper addresses the numerical methods and GPU-accelerated implementation required to make that framework practical at simulation rates approaching or exceeding real time. The resulting implementation is released as open-source software~\cite{TotalLagrangianFEA2025}.

\paragraph{GPU acceleration for flexible body simulation}
Commodity GPU hardware offers thousands of arithmetic units well matched to the embarrassingly parallel structure of element-level computations in FEA. Early GPU-accelerated FEM work focused on explicit dynamics and linear problems~\cite{georgii2005interactive}, where per-element independence makes data-parallel execution straightforward. Position-based dynamics solvers such as Nvidia FleX~\cite{macklin2014unified} achieve real-time rates on large cloth and fluid systems by sacrificing constitutive accuracy. Simulation frameworks built on spatially sparse data structures such as Taichi~\cite{hu2019taichi} extend GPU utilization to a broad class of physical simulation problems but are generally limited to explicit or penalty-based contact. General-purpose simulation frameworks such as SOFA~\cite{duriezSOFA2012} have demonstrated GPU acceleration of specific kernels but historically retain CPU-side sparse linear solvers for the global system. By contrast, implicit second-order solvers for nonlinear solid mechanics on the GPU require assembling and factorizing a sparse global Hessian at each Newton iteration, a task for which scalable GPU-native sparse direct solvers have only recently become available. The release of cuDSS~\cite{nvidia_cudss} provides a GPU-resident direct sparse solver that makes this pathway viable at scale.
Therefore, this work bridges the gap by providing a fully GPU-resident, implicit, second-order FEA solver for large-deformation flexible multibody systems with constraints and contact.

\paragraph{Implicit time integration and constraint enforcement}
Implicit backward-Euler and generalized-$\alpha$ integrators are standard for stiff flexible body systems~\cite{GerRix94,brulsCardonaArnold2012}, but extending implicit integration to constrained systems introduces the need to solve a coupled saddle-point system at each step. Baumgarte stabilization~\cite{BaumgarteStabilization72} is widely used for constraint drift suppression but requires careful tuning of stabilization parameters. Augmented Lagrangian methods (ALM)~\cite{bertsekas1995npb,NoWr99} avoid this sensitivity by treating the penalty parameter as an outer-loop tuning knob decoupled from the inner velocity solve; the two-level structure is particularly attractive when the inner solver is itself iterative. In the flexible multibody dynamics literature, most constrained simulations rely on generalized coordinate partitioning or index-reduction, and GPU-accelerated total-lagrangian FEA implicit solvers with ALM constraint enforcement have not been reported to our knowledge.

\paragraph{Collision detection at scale}
Robust collision detection for deformable bodies typically relies on bounding-volume hierarchies (BVH), which provide $\mathcal{O}(n \log n)$ broad-phase performance on CPUs~\cite{larsson2006dynamic}. GPU-native BVH construction and traversal have been studied~\cite{lauterbach2009fastbvh}, but the tree restructuring required every time step for deformable bodies limits throughput. Spatial-hashing approaches~\cite{teschner2005collisionSurvey} achieve better GPU utilization by replacing hierarchical traversal with flat hash-table lookups, at the cost of accuracy in highly non-uniform configurations. The present work avoids BVH entirely in favor of a bin-based spatial partitioning that maps naturally to GPU thread blocks and eliminates per-step tree updates.

\paragraph{Baseline frameworks}
Two established open-source frameworks serve as CPU baselines in this work. The FEniCS Project~\cite{Alnaes2015Fenics15} is a widely used automated finite element platform that solves variational problems expressed in the Unified Form Language (UFL). It has been cross-validated against Abaqus on hyperelastic large-deformation problems with sub-percent agreement~\cite{mazier2022inverse}, and has been adopted as a CPU reference by other GPU-accelerated solvers~\cite{xue2023jaxfem}. We use it as the CPU baseline for the T10 tetrahedral element benchmarks in Section~\ref{sec:performance_benchmark}. However, FEniCS does not support ANCF elements or bilateral kinematic constraints. Project Chrono~\cite{chronoOverview2016} is an open-source, multi-physics multibody dynamics engine whose FEA module provides mature ANCF beam and shell elements~\cite{Nachbagauer2012,yamashita2015continuum} and CPU-side direct sparse solvers. Taylor~\cite{mikeANCF-comparison2023,mikeANCF-implementationAspects2023} showed that careful implementation can accelerate ANCF internal-force and Jacobian evaluations by one-to-two orders of magnitude on the CPU. We therefore use Project Chrono as the CPU baseline for the ANCF beam and shell benchmarks.

\paragraph{Scope and organization}
This paper implements and benchmarks the formulation of~\cite{json-ganesh-danTLFEA-1-2026} on GPUs. Section~\ref{sec:alm_formulation} restates the velocity-level ALM residual needed to define the numerical methods. Section~\ref{sec:collision} presents the GPU collision detection algorithm. Section~\ref{sec:solvers} describes the AdamW and Newton inner solvers and their GPU implementation. Sections~\ref{sec:performance_benchmark}--\ref{sec:large_scale} report scaling benchmarks, unit tests, and large-scale demonstrations. The contributions of this paper are summarized in Section~\ref{sec:contributions}.

\section{Augmented Lagrangian Formulation}
\label{sec:alm_formulation}

This section provides a concise summary of the velocity-level augmented Lagrangian formulation derived in Part~I~\cite{json-ganesh-danTLFEA-1-2026}; we restate only the quantities needed to define the numerical methods presented in Sections~\ref{sec:solvers}--\ref{sec:experiments}.

Let $\mathbf{q}\in\mathbb{R}^{n_q}$ collect all nodal unknowns across the deformable bodies in the system (positions, and, for ANCF elements, position gradients~\cite{shabanaYakoub2001Part1}). A backward-Euler step of size~$h$ relates the generalized velocity $\mathbf{v}$ to the updated configuration through the step map
\begin{equation}\label{eq:stepmap}
  \mathbf{q}_{n+1} = \mathbf{q}_n + h\,\mathbf{v}_{n+1}.
\end{equation}
With this update, the semi-discrete equations of motion reduce to a nonlinear residual in the velocity unknown~$\mathbf{v}\equiv\mathbf{v}_{n+1}$ (see Part~I, Eq.~(86)):
\begin{equation}\label{eq:residual}
  \begin{aligned}
    \mathbf{g}(\mathbf{v},\boldsymbol{\lambda})
    &= \frac{1}{h}\mathbf{M}(\mathbf{v}-\mathbf{v}_n) \\
    &\quad + \mathbf{f}_{\mathrm{int}}(\mathbf{q}_n+h\mathbf{v},\,\mathbf{v})
      - \mathbf{f}_{\mathrm{ext}}
      - \mathbf{f}_{\mathrm{ff}} \\
    &\quad + h\,\mathbf{C}_q(\mathbf{q}_n+h\mathbf{v})^{\!\top}
      \bigl[\boldsymbol{\lambda}+\rho\,\mathbf{c}(\mathbf{q}_n+h\mathbf{v})\bigr],
  \end{aligned}
\end{equation}
where $\mathbf{M}$ is the (constant, preassembled) consistent mass matrix, $\mathbf{f}_{\mathrm{int}}$ is the internal force vector arising from hyperelastic and viscous contributions, $\mathbf{f}_{\mathrm{ext}}$ collects applied loads specified directly as external nodal forces (including point loads and contact/friction forces treated as external), $\mathbf{f}_{\mathrm{ff}}$ accounts for mass-distributed force fields (body forces) specified per unit mass such as gravity, $\mathbf{c}(\mathbf{q})=\mathbf{0}$ denotes bilateral kinematic constraints with Jacobian $\mathbf{C}_q=\partial\mathbf{c}/\partial\mathbf{q}\in\mathbb{R}^{m\times n_q}$, $\boldsymbol{\lambda}\in\mathbb{R}^m$ are Lagrange multipliers, and $\rho>0$ is the augmented Lagrangian penalty parameter. Although $\mathbf{f}_{\mathrm{ff}}$ can be absorbed into $\mathbf{f}_{\mathrm{ext}}$, it is kept separate to emphasize its origin in the force-field virtual work; see Part~I, Sections~4.2 and~4.4 for the derivations of $\mathbf{f}_{\mathrm{ff}}$ and $\mathbf{f}_{\mathrm{ext}}$, respectively. We evaluate $\mathbf{C}_q^{\top}(\cdot)$ via a sparse representation of~$\mathbf{C}_q^{\top}$, and enforce constraints with an augmented-Lagrangian outer loop that alternates an inner solve for $\mathbf{v}$ with multiplier updates until $\|\mathbf{c}\|_2$ meets a prescribed tolerance.

As shown in Part~I, Eq.~(89), the residual~\eqref{eq:residual} is the gradient of a scalar augmented cost
\begin{equation}\label{eq:cost}
  \begin{aligned}
    \Phi_\rho(\mathbf{v},\boldsymbol{\lambda})
    &= \frac{1}{2h}(\mathbf{v}-\mathbf{v}_n)^{\!\top}\mathbf{M}(\mathbf{v}-\mathbf{v}_n) \\
    &\quad + \frac{1}{h}\Pi_{\mathrm{int}}(\mathbf{q}_n+h\mathbf{v},\,\mathbf{v}) \\
    &\quad - \mathbf{f}_{\mathrm{ext}}^{\top}\mathbf{v}
      - \mathbf{f}_{\mathrm{ff}}^{\top}\mathbf{v}
      + \boldsymbol{\lambda}^{\top}\mathbf{c}(\mathbf{q}_n+h\mathbf{v}) \\
    &\quad + \frac{\rho}{2}\bigl\|\mathbf{c}(\mathbf{q}_n+h\mathbf{v})\bigr\|_2^2,
  \end{aligned}
\end{equation}
where $\Pi_{\mathrm{int}}$ is the incremental internal-force potential satisfying $\nabla_{\mathbf{v}}\!\bigl[\tfrac{1}{h}\Pi_{\mathrm{int}}\bigr]=\mathbf{f}_{\mathrm{int}}$. Using $\nabla_{\mathbf{v}}\mathbf{c}(\mathbf{q}_n+h\mathbf{v})= h\,\mathbf{C}_q(\mathbf{q}_n+h\mathbf{v})$ and the chain rule, one verifies that $\nabla_{\mathbf{v}}\Phi_\rho(\mathbf{v},\boldsymbol{\lambda})= \mathbf{g}(\mathbf{v},\boldsymbol{\lambda})$. Consequently, for fixed multipliers~$\boldsymbol{\lambda}$, finding~$\mathbf{v}$ such that $\mathbf{g}(\mathbf{v},\boldsymbol{\lambda})=\mathbf{0}$ is equivalent to enforcing stationarity of~$\Phi_\rho$ with respect to~$\mathbf{v}$, while the outer ALM loop updates $\boldsymbol{\lambda}$ to reduce the constraint violation $\|\mathbf{c}\|_2$.

The multiplier update takes the standard form (see Part~I, Eq.~(90)):
\begin{equation}\label{eq:lambda_update}
  \boldsymbol{\lambda}^{(k+1)}
  = \boldsymbol{\lambda}^{(k)}
    + \rho\,\mathbf{c}\!\bigl(\mathbf{q}_n + h\,\mathbf{v}^{(k)},\,t_{n+1}\bigr),
\end{equation}
iterated until $\|\mathbf{c}\|_2$ meets a prescribed tolerance. This two-level structure, i.e., inner minimization plus outer multiplier correction, is detailed in Part~I, Section~7.

The formulation admits both first- and second-order inner solvers. A first-order method requires only the gradient $\mathbf{g}=\nabla_{\mathbf{v}}\Phi_\rho$; a second-order (Newton-type) method additionally requires the Hessian $\mathbf{H}=\nabla^2_{\mathbf{v}}\Phi_\rho$, whose dominant contribution is the tangent stiffness $\partial\mathbf{f}_{\mathrm{int}}/\partial\mathbf{q}$ derived for the St~Venant--Kirchhoff and Mooney--Rivlin models in Part~I, Appendix~C. In the following sections we describe both solver strategies, their GPU implementation, and the precomputation and sparse-assembly infrastructure that supports them.

\section{Collision Detection}
\label{sec:collision}

Triangle-mesh collision detection on GPUs is most often implemented using hierarchy-based culling, where bounding-volume hierarchies (BVHs) reduce the number of expensive primitive-level tests and expose substantial data parallelism in both hierarchy construction and query processing~\cite{lauterbach2009fastbvh,karras2012maxparallel,lauterbach2010gproximity,tang2011collisionstreams}. A recurring practical limitation is that proximity queries over BVHs remain highly data-dependent: traversal work and memory access patterns vary significantly across contact configurations, which can induce workload imbalance and control-flow divergence on SIMT hardware~\cite{chitalu2018bspbvh}.

For narrow-phase contact between \emph{convex} shapes, GJK and MPR-style methods are widely used due to their efficiency and generality in support-mapping form~\cite{gilbert1988,gsnethenGPG7}. However, extending these approaches to large triangle meshes typically requires convex decomposition or an equivalent strategy that converts a nonconvex mesh query into many convex queries, increasing the number of pair tests and amplifying the impact of data-dependent iteration counts. In a GPU setting, such iterative, branch-heavy control flow can exacerbate divergence and reduce throughput, particularly when the underlying geometry contains many triangles and the contact set evolves rapidly~\cite{chitalu2018bspbvh,pan2012gpuccd}.

A common GPU-friendly approximation is spherical decomposition (e.g., multi-sphere models or sphere-trees), which trades geometric fidelity for very inexpensive and easily batched distance checks~\cite{hubbard1996spheres,bradshaw2004spheretree}. While this representation can be effective for real-time collision queries, achieving high accuracy near sharp features or thin structures may require a large number of spheres, increasing memory and broad-phase cost; conversely, using fewer spheres improves performance but introduces approximation error in contact location and normal estimation~\cite{hubbard1996spheres,bradshaw2004spheretree}.

We present a two-thread asynchronous, GPU-oriented collision detection algorithm for the Total Lagrangian finite element framework. Two design choices distinguish this algorithm: (i) contact information is derived from triangle soups decomposed from the original meshes rather than from the meshes directly, increasing per-triangle data independence; and (ii) detection and physics integration are executed concurrently by two asynchronous threads, maximizing GPU utilization through enforced execution overlap.

\subsection{A Two-Thread Asynchronous Collision Detection Algorithm}

The implementation builds on the existing infrastructure of Chrono DEM-Engine~\cite{ruochun2024dem}, which uses two distinct and parallel computational threads to update the active contacts set (done by the ``kinematics thread'', abbreviated as \emph{kT}), and the integration of the equations of motion (done by the ``dynamics thread'', abbreviated as \emph{dT}), respectively. These threads are distinct from GPU threads; they are CPU-managed threads that each control a GPU stream and launch kernels. The dynamics thread processes each contact in the Active-Contact Set (ACS) at each time step to reassess the contact penetration $\delta_n$ and the ancillary information. The dynamics thread receives an ACS update when the kinematics thread finishes producing it, or if so desired, it can wait for the ACS update when the dynamics thread advances the system state too far ahead of the time stamp of the last ACS update from the kinematics thread. Through this collaboration pattern, the two threads work concurrently and the cost of contact detection is nearly hidden in the background of computation done by the dynamics thread, which continuously advances the state of the system. To avoid missing mutual contacts that might arise between successive ACS updates, we artificially enlarge all contact geometries in the system, as detailed in Section~\ref{sec:broad_phase}. This extra margin allows for the preemptive detection of potential contact pairs that might emerge in the near future.

By adding this artificial margin to all contact geometries, the kinematics thread can report false positives within the provided list of contacts, i.e., a contact between two geometries might be in the ACS, yet the two geometries are not in contact. Such false positives will be identified and ruled out by the dynamics thread when carrying out the force calculation. The thickness of contact geometry $i$'s added margin $d_i$, is determined by the simulation entities' velocity magnitude $v_i$ (which is bounded and known by the solver), the time step size $h$, which is typically small, and $n_{\text{max}}$, the maximum number of time steps the dynamics thread is allowed to advance without receiving an ACS update from the kinematics thread. Namely, $d_i = ( a  v_i + b )  h  n_{\text{max}}$, where $a$ and $b$ are configurable parameters that scale the velocity for contact safety, and take default values of $1$ and $0.5$ respectively. This ensures that no potential contact is missed within the maximum lag between ACS updates.

\begin{figure}[htb]
	\centering
	\maybeincludegraphics[width=.9\linewidth]{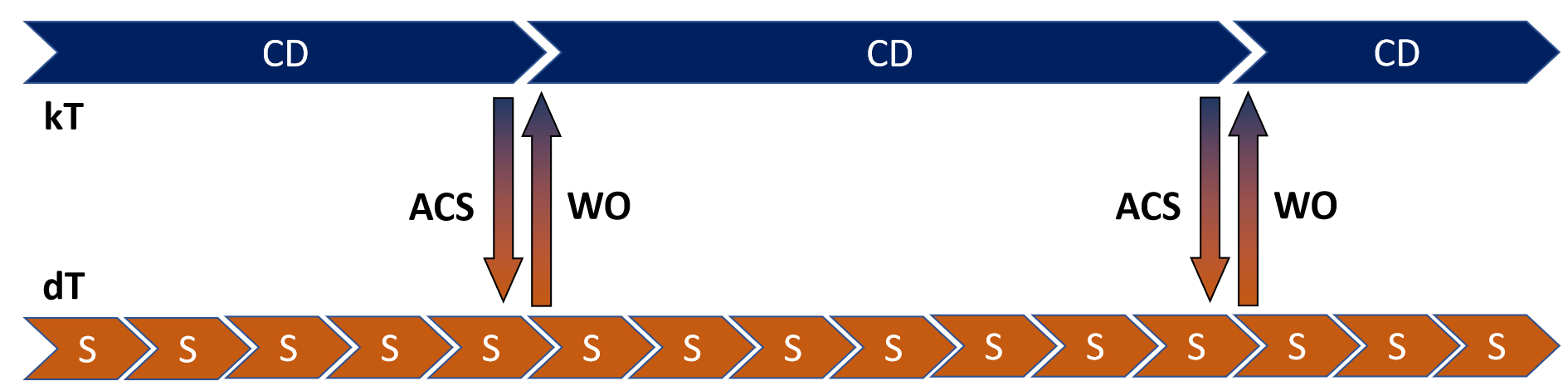}
	\caption{Two-thread asynchronous contact detection collaboration pattern, where the dynamics thread advances the physics continuously while the kinematics thread occasionally waits for updated state information to commence an ACS update} \label{fig:kTdT}
\end{figure}

The synchronization pattern between the kinematics and dynamics threads is illustrated in Fig.~\ref{fig:kTdT}. There, ``\textbf{S}'' represents a time step that the dynamics thread executes, where the contact forces are calculated (see Section~\ref{sec:cnt_force_model}), and the system state is advanced in time. A contact detection step that the kinematics thread executes is marked with ``\textbf{CD}''. Periodically, the kinematics thread finishes a contact detection step and sends the signal to the dynamics thread, allowing the dynamics thread to receive the contact array, ``\textbf{ACS}'', from the kinematics thread. Then the dynamics thread will send a work order ``\textbf{WO}'' with the current simulation system state, for the kinematics thread to pick up and continue the next contact detection step. Before the next ``\textbf{ACS}'' update is received, the dynamics thread will use this ``\textbf{ACS}'' to execute the time steps.

\subsection{Broad-Phase Contact Detection}\label{sec:broad_phase}

The broad-phase contact detection algorithm running on the kinematics thread is an extension of that presented in~\cite{hammadTobyDan2012}. It is optimized for GPU execution, operates entirely on decomposed triangle soups without requiring mesh connectivity information, and accommodates simulation entities of vastly differing sizes. Accordingly, the supplied ACS is triangle-pair-based, matching the format expected by the dynamics thread.

\begin{figure}[htb]
	\centering
	\maybeincludegraphics[width=.9\linewidth]{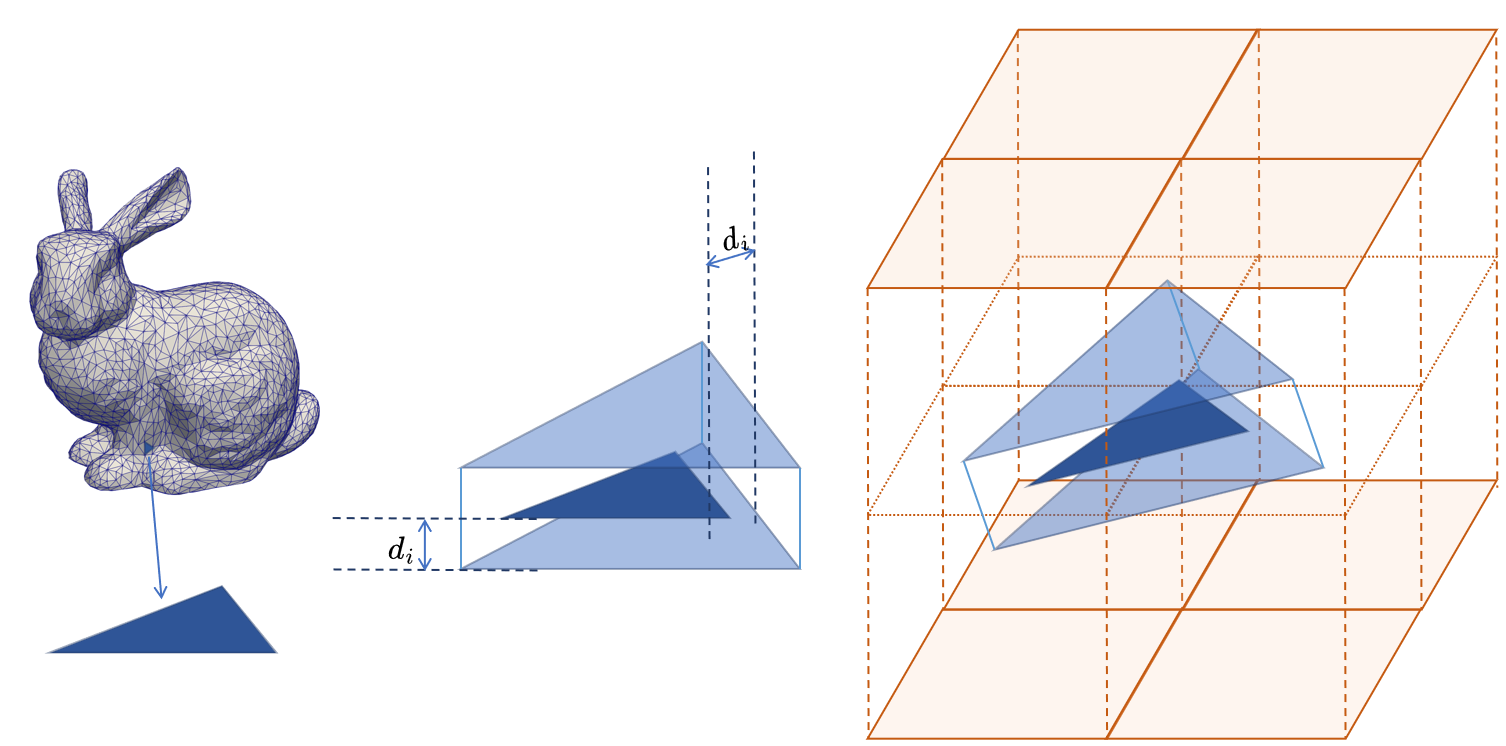}
	\caption{Left: The process of adding margin is done for every triangle of a mesh. Middle: The margin with size $d_i$ is added to the triangle, thus creating a prism as the proxy for contact detection. Right: An example of a prism in contact with eight bins} \label{fig:prism}
\end{figure}

The contact detection process in each step involves a series of tasks, executed sequentially:
\begin{enumerate}

\item At the start of each contact detection step, for each triangle facet of the meshes, the solver adds a margin to it, thus creating the \emph{future-proof} contact proxy for it. Recall that the solver determines the size of the margin $d_i$, then adds it to the triangle along all directions. This effectively creates ``prisms'' as the contact proxy, as shown in Fig.~\ref{fig:prism}. Prism contacts warrant potential underlying triangle contacts, so the broad-phase contact detection, at the implementation level, always works with prisms. This should be distinguished from the force derivation process in Section~\ref{sec:cnt_force}, which still uses the original triangle, because there, the solver works with the present time step only. In this section only, we use ``triangle'' and ``prism'' interchangeably because they are the same underlying data structure in the broad-phase contact detection.

\item Initially, all triangle facets are evaluated for potential contact using ``bins''. These bins are formed by uniformly segmenting the simulation domain into axis-aligned cubic grids. If a triangle intersects with a bin, this bin--triangle pair is recorded for subsequent processing. This is also illustrated in Fig.~\ref{fig:prism}. It is important to note that due to the allowance for variable sizes of triangles, the maximum number of bins intersecting with a triangle cannot be predetermined. This necessitates two sequential CUDA kernel executions: one to determine the count of intersecting bins per triangle for memory allocation, and another to store the bin--triangle pairs.

\item Next, we sort the bin--triangle pairs based on the bin IDs. This step groups together the triangles located within the same bin, effectively clustering entities that are adjacent within the simulation.

\item The final step involves checking triangles within the same bin for potential contacts. This is accomplished through launching CUDA blocks, each processing a bin, to derive all potential contact pairs. Similar to the previous step, two CUDA kernel calls are required: the first to ascertain the number of potential contacts per bin for array allocation, and the second to populate the ACS array with these contacts. The contacts are detected by the separating axis theorem (SAT)-based tests between prisms' base triangles. Given $n_\text{g}$ geometries in a bin, $(n_\text{g}-1)n_\text{g}/2$ checks are necessary to identify all potential contacts. This imposes a limit on $n_\text{g}$, influencing the bin size. We dynamically adjust bin sizes based on execution history for optimal performance. Duplicate contacts may be present, so they are then identified and discarded using a CUDA CUB-based device-level search.

\item Note that the triangles' corresponding patch IDs are also consolidated into arrays as part of the contact information. The patch ID is an artificial index assigned to triangles, such that the triangles sharing the same patch ID are considered to be ``in the same contact'', i.e., in the same connected ``island'' of mesh surface triangles. We also call the union of the connected triangles that share a patch ID a ``patch''. To generate patch IDs, a GPU-friendly ``index flooding'' (label propagation) over the triangle adjacency graph is carried out: Each active triangle is given an initial label equal to its own index, and over several iterations, each triangle updates its label by taking the minimum label among itself and its neighboring triangles (using the precomputed neighbor lists). This repeatedly propagates the smallest index through connected components, so triangles connected by shared edges converge to the same label while disconnected components keep distinct labels. Finally, that converged ``island ID'' is combined with higher-level grouping keys (such as contact type identifiers), and the contacts are sorted so a prefix-scan can assign a contiguous patch ID.

\end{enumerate}

\begin{figure}[htb]
	\centering
	\maybeincludegraphics[width=.9\linewidth]{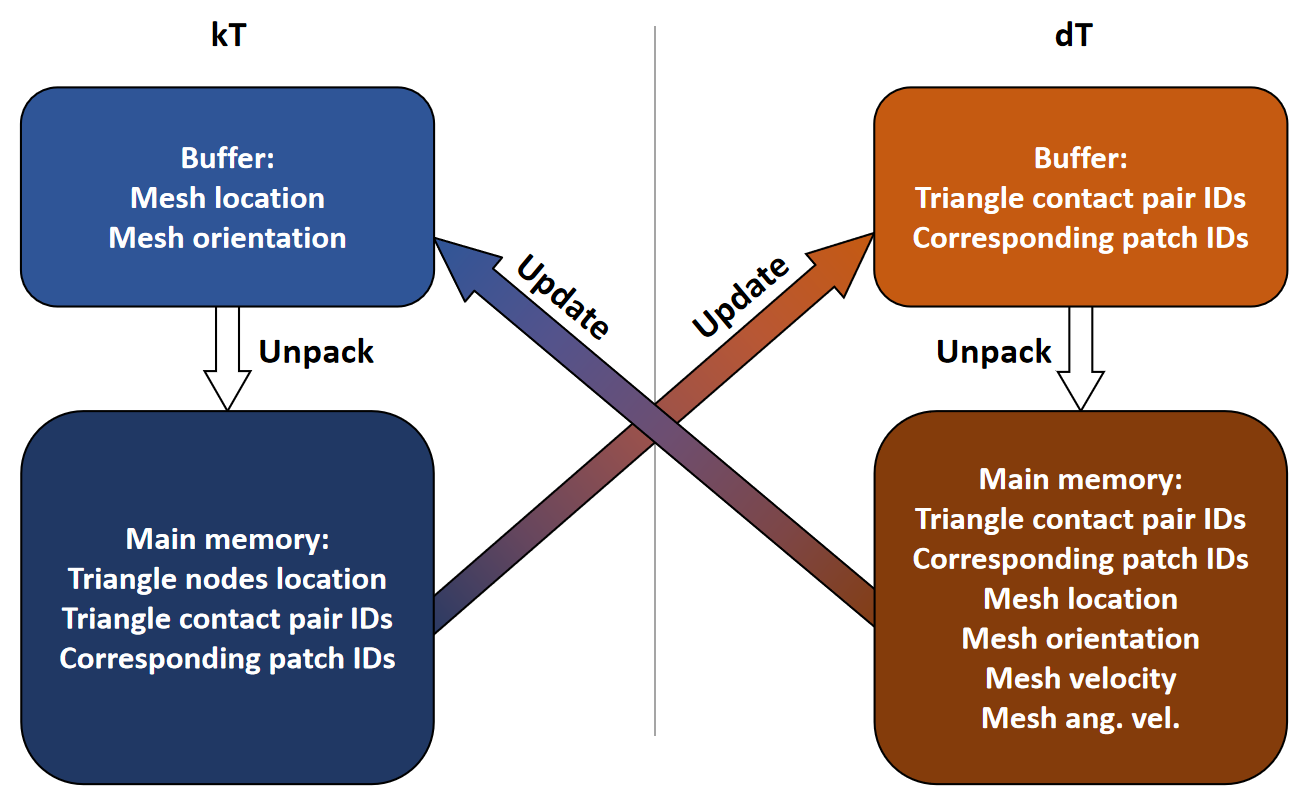}
	\caption{The collaboration pattern of the kinematics and dynamics thread, where neither of them directly modifies the working memory of the other} \label{fig:async_update}
\end{figure}

The broad-phase contact information is transferred to the dynamics thread at the end of each kinematics thread step. Since the two threads are working asynchronously, the contact information is transferred to a buffer memory. Then the dynamics thread will be notified and copy the contact information to its working memory. The dynamics thread carries out a similar routine when updating the kinematics thread with new element positions. Neither of them directly modifies the working memory of the other to avoid race conditions. This is illustrated in Fig.~\ref{fig:async_update}.

\subsection{Narrow-Phase Detection Verification and Force Derivation}\label{sec:cnt_force}

\begin{figure}[htb]
	\centering
	\maybeincludegraphics[width=.9\linewidth]{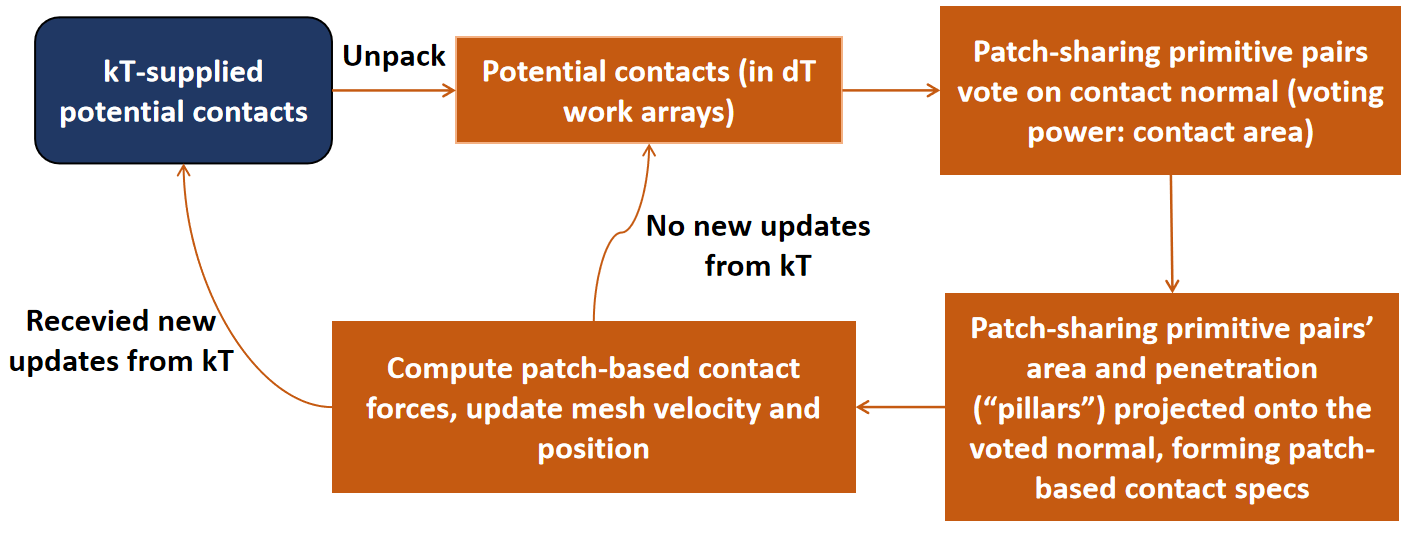}
	\caption{The workflow of the dynamics thread in a time step } \label{fig:force_workflow}
\end{figure}

From the kinematics thread-supplied potential ACS array, the dynamics thread detects the physical contacts and derives the contact force. The overall workflow is summarized in Fig.~\ref{fig:force_workflow}.

Sequentially, the following steps are executed in one dynamics step:
\begin{enumerate}
\item The solver computes the contribution of each triangle contact pair, specifically the penetration depth, contact point, area, and normal direction. We call these contacts the primitive-based contacts, because triangles are the ``primitives'' to form patches. This is to be differentiated from the patch-based contacts we introduce later. The computation is done via a projection-based method and is illustrated in Fig.~\ref{fig:projection}. For a potentially in-contact triangle pair, one triangle is projected onto the other's plane, then the projection is clipped against the projectee, forming a polygon shown in purple. The polygon's area is then the contact area, the projection direction becomes the contact normal, and the maximum projection distance becomes the penetration depth. Note that the result is in general different when swapping the projector and the projectee, so we select the one that gives the smaller penetration depth. Finally, the contact point is selected to be the estimated center of the projection ``pillar'', i.e., the volume swept by the projector triangle until it reaches the projectee's plane. This way, the primitive contacts are resolved using only the triangles in question, and never require adjacent triangles or mesh topology information, greatly reducing the per-thread data bandwidth.

\begin{figure}[htb]
	\centering
	\maybeincludegraphics[width=.9\linewidth]{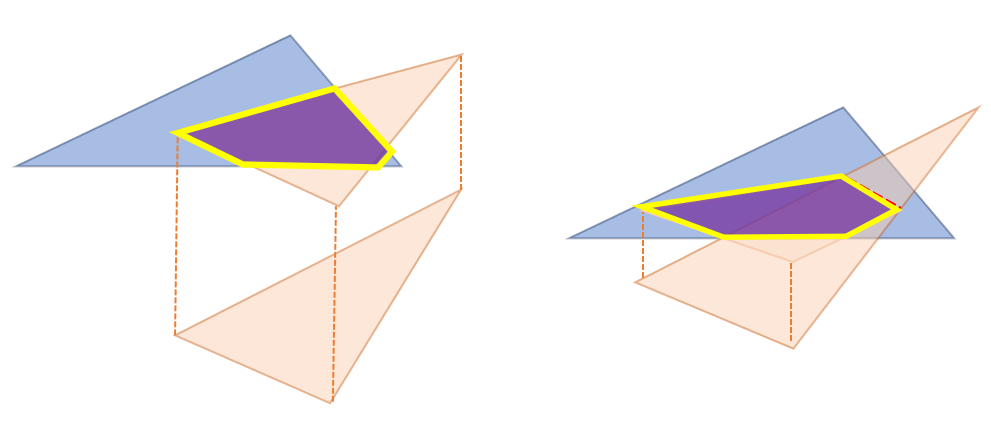}
	\caption{The primitive contact's penetration depth, contact point, area, and normal direction are derived by projecting a triangle onto the other's plane, then clipping against the latter. This represents the ``contribution'' of the primitive pair in its belonging patch-based contact pair. Note this is done when one triangle is completely submerged in the other's owner mesh (shown on the left), as well as when the two triangles are in physical contact (shown on the right)} \label{fig:projection}
\end{figure}

\item The solver determines the patch-based contact information through a reduction process, i.e., by merging the contributions of the triangles that belong to the same patch. This consolidation step is necessary to ensure stable and traceable contacts for the Mindlin friction model (introduced in Section~\ref{sec:cnt_force_model}), which can only be enforced consistently at the patch level.

The first step of this reduction process is to determine the patch-based contact normal using a voting scheme, where each primitive contact contributes with a weight proportional to its contact area. This concept is illustrated in the left part of Fig.~\ref{fig:merge_cnt}. Specifically, the patch-based contact normal $\vect{n}_{\text{patch}}$ is computed from its $c$ associated primitive contacts as
\begin{equation}
\vect{n}_{\text{patch}} =
\frac{\sum_{k=1}^{c} A^k_{\text{prim}} \vect{n}^k_{\text{prim}}}
{\left|\sum_{k=1}^{c} A^k_{\text{prim}} \vect{n}^k_{\text{prim}}\right|},
\end{equation}
where $A^k_{\text{prim}}$ denotes the area associated with a primitive contact. In this way, the resulting contact normal reflects the overall orientation of the contacting surface region, providing a physically consistent representation of the patch interaction.

\item The next step of the reduction process is to project the penetration depth, contact point, area, and normal direction of each primitive pair onto the voted normal direction, and then combine these projected quantities to obtain their patch-level counterparts. This procedure is also illustrated in Fig.~\ref{fig:merge_cnt}. Conceptually, each primitive contact contributes to the patch-based contact through a pillar-like projection of its influence. Specifically, the patch-based contact area is obtained as the sum of the projected primitive contact areas, the penetration depth is taken as the maximum projected penetration among the primitives, and the contact point and normal direction are computed as weighted averages of their projected primitive counterparts, with the weight proportional to the volume of each pillar.

This algorithm has the advantage that all reduction operations can be performed using efficient CUDA device-level primitives, without requiring atomic operations. Together with the absence of neighbor searches or on-the-fly mesh topology queries, this design leverages the highly parallel nature of GPU hardware.

We emphasize that the result of the reduction constitutes the patch-based contact information, i.e., the effective contact used by the force model described in Section~\ref{sec:cnt_force_model}. At the same time, the primitive-level contact data are retained and remain available for user queries. Although not required in the present study, such fine-grained information can be useful in applications involving tearing, wear, and material fatigue.

\begin{figure}[htb]
	\centering
	\maybeincludegraphics[width=.9\linewidth]{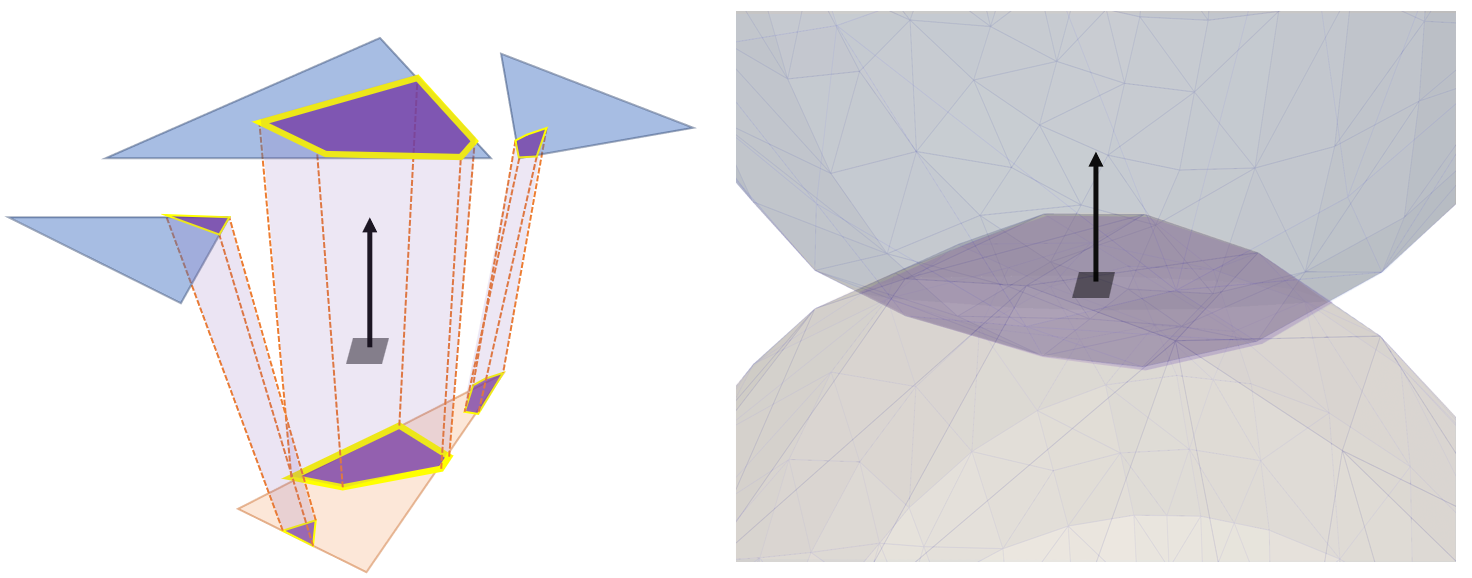}
	\caption{Left: For a patch (i.e., connected island of triangles that are involved in a contact)-based contact, the solver pools the contact normals of all involved primitive contacts, and decides an overall normal direction (shown with black arrow) using a voting process. Then each contact ``pillar'' (the volume swept by the projection of triangles, shown with purple) is projected onto this overall direction to calculate its contribution to this patch-based contact. Right: The summed contribution of all primitive contact pillars approximately recovers the physical contact volume, which is marked with purple} \label{fig:merge_cnt}
\end{figure}

\end{enumerate}

This algorithm is shown to produce accurate and stable contact physics in the unit tests documented in Section~\ref{sec:experiments}.

\subsection{Contact Force Models}\label{sec:cnt_force_model}

Contact interactions are described using a Hertz--Mindlin-type nonlinear spring--dashpot formulation with Coulomb friction and rolling resistance. For two contacting bodies $i$ and $j$, the normal and tangential forces are expressed as
\begin{subequations}
\begin{align}
\vect{F}_n &= \sqrt{A/\pi}
\left(k_n \vect{d}_n - \gamma_n \bar{m}\vect{v}_n \right), \\
\vect{F}_t &= \sqrt{A/\pi}
\left(-k_t \vect{d}_t - \gamma_t \bar{m}\vect{v}_t \right),
\quad \|\vect{F}_t\| \le \mu \|\vect{F}_n\|, \\
\bar{m} &= m_i m_j/(m_i + m_j),
\end{align}
\end{subequations}
where $\vect{d}_n$ and $\vect{d}_t$ are the normal and tangential displacement, $A$ is the contact area, $\bar{m}$ is the effective mass, and $\vect{v}_n$, $\vect{v}_t$ are the normal and tangential components of the relative contact velocity. The stiffness and damping parameters $k_n, k_t, \gamma_n, \gamma_t$, are derived from material properties (Young's modulus, Poisson's ratio, and restitution coefficient) following~\cite{jonJCND2015}. Tangential displacement evolves incrementally during contact and is limited to satisfy the Coulomb friction criterion controlled by the coefficient of friction $\mu$.

Rolling resistance is incorporated via an additional torque $\vect{\tau}$ proportional to the normal force magnitude, accounting for contact-level energy dissipation during rolling~\cite{johnson1987contact}.

Contact forces are evaluated at the patch-based contact level (as opposed to primitive-level). Recall that a mesh could have multiple disconnected contact ``islands'' thus multiple patch-based contacts, so all $n_c$ contacts associated with a specific mesh are accumulated to the mesh's center of mass. The translational and rotational dynamics of body $i$ follow
\begin{subequations}
\begin{align}
m_i \dot{\vect{v}}_i &= m_i \vect{g} + \sum_{k=1}^{n_c} \vect{F}^k, \\
I_i \dot{\vect{\omega}}_i &= \sum_{k=1}^{n_c}
\left( \vect{r}^k \times \vect{F}^k + \vect{\tau}_r^k \right),
\end{align}
\end{subequations}
where $\vect{F}^k = \vect{F}_n^k + \vect{F}_t^k$, and $\vect{r}$ is the vector from the mesh's center of mass to the contact point.

\section{Numerical Solvers}
\label{sec:solvers}

This section presents the numerical solvers used to solve the velocity-level augmented Lagrangian subproblems arising in our Total Lagrangian finite element framework. We describe both first-order and second-order methods, covering their algorithmic details and GPU implementation considerations.

\subsection{Precomputation of Constant Terms}
\label{sec:precompute_constants}

Before time integration, we precompute and cache (in device memory) all
quantities that depend only on the reference configuration, the element
interpolation, and the chosen quadrature rule. The following terms are
precomputed once and reused throughout the simulation:

\begin{itemize}
  \item $s_i(\mathbf{u}_q)$ \textbf{(shape-function values at quadrature points).}
  Scalar shape functions evaluated at quadrature nodes $\mathbf{u}_q$ in the
  parent domain. These values enter inertia terms through products
  $s_i(\mathbf{u}_q)s_j(\mathbf{u}_q)$ and force-field terms through
  $s_i(\mathbf{u}_q)$, avoiding repeated evaluation of basis polynomials (or
  barycentric forms) at every time step.

  \item $\mathbf{H}(\mathbf{u}_q)=\partial s/\partial \mathbf{u}$ and
  $\mathbf{h}_i^\top(\mathbf{u}_q)=\partial s_i/\partial \mathbf{u}$
  \textbf{(reference shape-function gradients).}
  Reference gradients of the interpolation evaluated at $\mathbf{u}_q$.
  In the Total Lagrangian setting, these operators are purely geometric and
  constant. They are reused to assemble both the deformation gradient and its
  rate via $F(\mathbf{u}_q,t)=N(t)H(\mathbf{u}_q)$ and
  $\dot{F}(\mathbf{u}_q,t)=\dot{N}(t)H(\mathbf{u}_q)$; consequently, the same
  precomputed gradients are shared by hyperelasticity and the finite-strain
  Kelvin--Voigt damping model.

  \item $\{(\mathbf{u}_q,W_q)\}_{q=1}^{n_{qp}}$ (or
  $\{((\xi_q,\eta_q,\zeta_q),W_q)\}_{q=1}^{n_{qp}}$ for $T10$)
  \textbf{(quadrature nodes and weights).}
  Quadrature rules are tabulated once per element type and integration order.
  The cached nodes and weights are reused across mass, internal-force, and
  damping integrations, eliminating repeated construction of quadrature tables.

  \item $J_q=\det(\partial X/\partial\boldsymbol{\xi})|_q$ (and, when needed,
  $(\partial X/\partial\boldsymbol{\xi})^{-1}|_q$)
  \textbf{(geometric Jacobian factors for isoparametric solids).}
  For isoparametric solid elements integrated on the parent domain
  $\hat{\Omega}$ (e.g., $T10$), the reference mapping
  $X(\boldsymbol{\xi})=\sum_a N_a(\boldsymbol{\xi})X_a$ depends only on reference
  nodal positions $X_a$. Therefore the Jacobian determinant $J_q$ (and optionally
  the inverse Jacobian) is constant and can be precomputed at each quadrature
  point.

  \item $m_{ij}=\int_{V_r}\rho_r(\mathbf{u})\,s_i(\mathbf{u})\,s_j(\mathbf{u})\,dV_r$
  and $M_e=[m_{ij}I_3]$
  \textbf{(constant inertia coefficients and element mass matrix).}
  In the Total Lagrangian formulation, $\rho_r$ and $s_i$ live in the reference
  configuration, hence $m_{ij}$ is constant and is computed once. The resulting
  consistent element mass matrix $M_e$ (and the assembled global mass matrix) is
  constant and reused throughout the simulation.

\end{itemize}

\subsection{The Sparse Assembly}
\label{sec:mass_matrix}
\label{sec:sparsity}

All sparse operators are stored using the compressed sparse row (CSR) format,
which represents a matrix $\mathbf{A}\in\mathbb{R}^{m\times n}$ by three arrays:
(i) \texttt{offsets} (row pointers) of length $m{+}1$, (ii) \texttt{columns}
(column indices) of length $\mathrm{nnz}$, and (iii) \texttt{values} of length
$\mathrm{nnz}$, where the nonzeros in row $r$ occupy the contiguous range
\texttt{offsets[$r$] : offsets[$r{+}1$]} in \texttt{columns} and \texttt{values}
\cite{Saad2003,davis2006dm}. This layout is standard in large-scale
scientific simulation software because it provides compact storage and supports
efficient sparse matrix--vector products and sparse factorizations
\cite{petsc-efficient,heroux2005trilinos,anderson2021mfem}.

In our implementation, the \emph{structure} (row offsets and column indices) of
all global sparse operators is treated as time-invariant as long as the mesh
topology and the set of active constraints remain unchanged. For the consistent
mass matrix, we construct and cache a coefficient-level sparsity pattern from
element connectivity; subsequent assemblies update only the numerical values
while preserving the same \texttt{offsets} and \texttt{columns} arrays. The
constraint Jacobian $\mathbf{C}_q$ is stored in CSR form with a fixed
pattern: for clamped Dirichlet constraints, each constraint row couples to a
single generalized degree of freedom (identity-like rows), whereas for general
constraints the nonzero structure is fully determined by the user-provided
CSR description of $\mathbf{C}_q$ and therefore remains constant throughout the
simulation. Finally, the Newton system matrix $\mathbf{H}$ is stored in a
solver-owned CSR structure whose pattern is constructed once by combining
element-induced couplings (obtained by lifting the coefficient-level adjacency
implied by element connectivity into generalized-coordinate space) with any
additional couplings implied by the nonzero structure of $\mathbf{C}_q$ (to
support the $\mathbf{C}_q^\top \mathbf{C}_q$ penalty contribution without
structural fill-in). With this fixed-pattern design, each Newton iteration updates
only CSR \emph{values} (mass, tangent, and constraint contributions), while the
symbolic analysis in the sparse direct solver can be performed once and reused
across iterations; numerical factorization then proceeds as repeated
refactorization on the same CSR pattern.

To form the coefficient-level sparsity, we enumerate all element-local index pairs
$(i,j)$ that can contribute to the same global row/column, map them to global
indices, and encode each pair as a 64-bit key. These keys are sorted on the
device and duplicates are removed, yielding the unique set of column indices per
row. A row-count pass followed by an exclusive scan produces the CSR row
offsets. This procedure is used both to allocate the time-invariant mass-matrix
structure and to provide a compact adjacency graph that is subsequently lifted to
the generalized-coordinate space when constructing the Newton-matrix sparsity
pattern.

\subsection{The Parallelization of Internal Force Evaluation}
\label{sec:parallel_fint}

In the Total Lagrangian formulation the evaluation of the global internal force vector $\mathbf{f}^{\mathrm{int}}$ consists of two conceptually distinct phases: the pointwise
evaluation of the first Piola--Kirchhoff stress tensor $\mathbf{P}$ at every quadrature point, and the subsequent assembly of element-level nodal contributions into the global force vector. Because these two phases exhibit fundamentally different data-access patterns, i.e., the former is entirely write-independent while the latter requires updates to
shared global degrees of freedom, they are parallelized on the GPU using distinct strategies. We describe the two stages next.

\paragraph{Stage 1: Parallelization of Stress Evaluation}
In Stage 1 the first Piola--Kirchhoff stress tensor $\mathbf{P}^{(e,q)}$ is evaluated independently at every pair $(e,q)$. Because each result is stored in a dedicated buffer location indexed uniquely by $(e,q)$, no write conflicts can occur and the computation is embarrassingly parallel. A GPU thread is assigned to each (element, quadrature point) pair, with a total of $n_{\mathrm{el}} \times n_{\mathrm{qp}}$ threads launched. The number of quadrature points per element $n_{\mathrm{qp}}$ equals 5 for the quadratic tetrahedral (T10) elements (5-point Keast rule), 12 for the ANCF beam elements ($3 \times 2 \times 2$ Gauss--Legendre product rule), and 48 for the ANCF shell elements ($4 \times 4 \times 3$ Gauss--Legendre product rule).

Each thread gathers the current generalized nodal coordinates $\{\mathbf{x}_{a}\}$ for element $e$ and assembles the deformation gradient at quadrature point $q$:
\begin{equation}
  \mathbf{F}^{(e,q)}
  = \sum_{a=1}^{n_{\mathrm{en}}} \mathbf{x}_{a} \otimes \nabla_{\!\mathbf{X}} N_{a}^{(e,q)}.
  \label{eq:F_assembly}
\end{equation}
The constitutive model is then evaluated to obtain the elastic stress:
\begin{equation}
  \mathbf{P}_{\mathrm{el}}^{(e,q)} = \mathcal{P}\!\left(\mathbf{F}^{(e,q)}\right),
\end{equation}
where $\mathcal{P}(\cdot)$ represents the chosen hyperelastic relation (Saint Venant--Kirchhoff or compressible Mooney--Rivlin in the present framework). When Kelvin--Voigt viscoelastic damping is active, a viscous stress contribution $\mathbf{P}_{\mathrm{vis}}^{(e,q)}$, computed from the rate of deformation, is added to yield the total stress
\begin{equation}
  \mathbf{P}^{(e,q)} = \mathbf{P}_{\mathrm{el}}^{(e,q)} + \mathbf{P}_{\mathrm{vis}}^{(e,q)}.
\end{equation}
The result is stored in the dedicated buffer entry for $(e,q)$ without requiring any inter-thread communication. The cached stress tensors are further reused in the Hessian evaluation described in Section~\ref{sec:parallel_hessian}, avoiding redundant constitutive calls.

\paragraph{Stage 2: Parallelization of Internal Force Assembly}
Given the stress buffer $\{\mathbf{P}^{(e,q)}\}$ produced by Stage 1, the contribution of element $e$ to the internal force at its local shape-function index $a$ is
\begin{equation}
  \mathbf{f}_{a}^{\mathrm{int},(e)}
  = \sum_{q=1}^{n_{\mathrm{qp}}} \mathbf{P}^{(e,q)}\,
    \nabla_{\!\mathbf{X}} N_{a}^{(e,q)}\,
    J_{0}^{(e,q)}\, w_{q},
  \label{eq:fint_local}
\end{equation}
where $J_0^{(e,q)} = \det(\partial\mathbf{X}/\partial\boldsymbol{\xi})|_q$ is the determinant of the reference-configuration Jacobian at quadrature point $q$ of element $e$ (precomputed once and fixed in the TL frame), and $w_{q}$ is the quadrature weight associated with point $q$. The global internal force is assembled by summing contributions from all elements sharing a given global degree of freedom $I$:
\begin{equation}
  \mathbf{f}_{I}^{\mathrm{int}}
  = \sum_{e \in \mathcal{S}(I)} \mathbf{f}_{a(e,I)}^{\mathrm{int},(e)},
  \label{eq:fint_global}
\end{equation}
where $\mathcal{S}(I)$ is the set of elements that share degree of freedom $I$ and $a(e,I)$ is the local shape-function index of $I$ within element $e$.

This assembly introduces write conflicts: threads associated with different elements may concurrently attempt to update the same entry in $\mathbf{f}^{\mathrm{int}}$, due to multiple elements sharing the same node in the mesh topology. The thread decomposition in Stage 2 therefore differs from Stage 1. One thread is assigned to each (element, local shape-function) pair, with $n_{\mathrm{el}} \times n_{\mathrm{en}}$ total threads. Each thread independently evaluates the local accumulation in Eq.~\eqref{eq:fint_local} by looping over all $n_{\mathrm{qp}}$ quadrature points of its element, and then scatters the result to the corresponding global DOF entries via GPU atomic addition:
\begin{equation}
  \mathbf{f}^{\mathrm{int}}\!\left[\mathcal{G}(e,a)\right]
  \mathrel{+}= \mathbf{f}_{a}^{\mathrm{int},(e)},
  \label{eq:atomic_scatter}
\end{equation}
where $\mathcal{G}(e,a)$ denotes the global DOF index triplet corresponding to local shape-function $a$ of element $e$. Atomic addition guarantees correctness in the presence of concurrent writes from different elements; the contention per node remains low because each global node is shared by only a small number of elements.

The number of shape functions per element $n_{\mathrm{en}}$ equals 10 for the T10 tetrahedral elements, 8 for the ANCF beam elements (four generalized coordinates per physical node times two nodes), and 16 for the ANCF shell elements (four generalized coordinates per physical node times four nodes).

\begin{figure}[t]
  \centering
  \maybeincludegraphics[width=\linewidth]{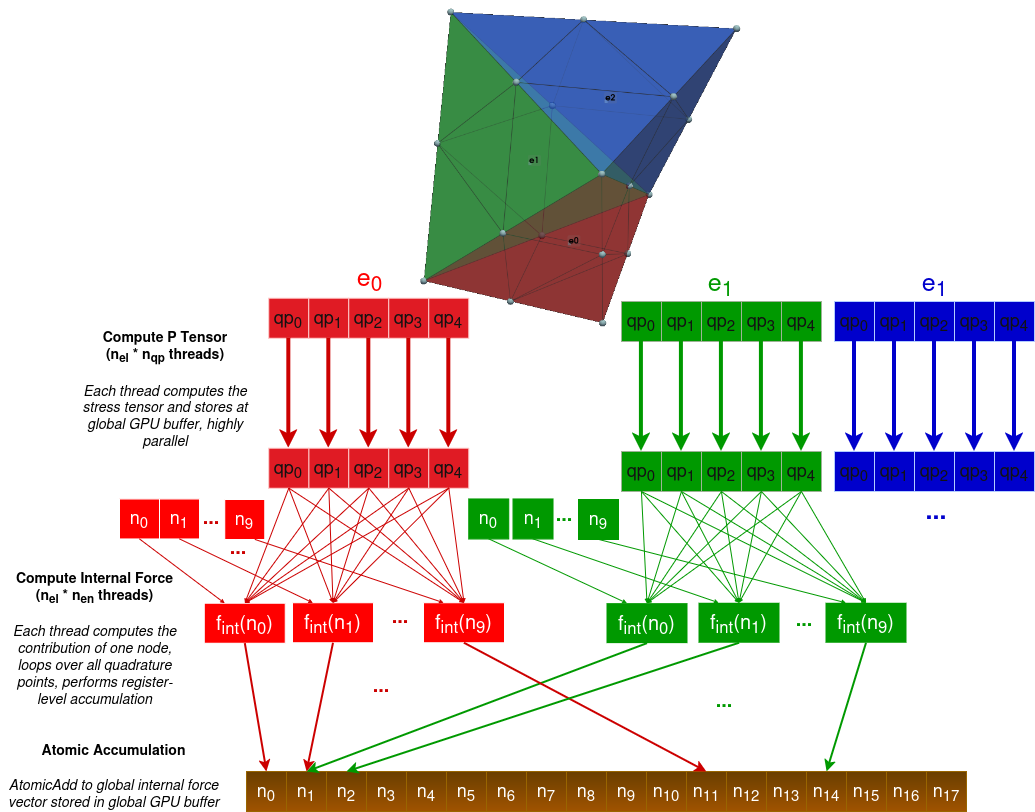}
  \caption{A visualization of the parallelization strategy used for the evaluation of the First Piola--Kirchhoff stress and the internal force vector. For the computation of the First Piola--Kirchhoff stress tensor $\mathbf{P}$, each thread is responsible for computing $\mathbf{P}^{(e,q)}$ for a specific element $e$ and quadrature point $q$, this stage is highly parallel without write conflicts. The resulting stress tensors are stored in a dedicated, continuous GPU  buffer indexed by $(e,q)$, which is then reused in the assembly of the internal force vector and the Hessian matrix. In the internal force assembly stage, each thread is responsible for computing the contribution of a specific element $e$ and local shape-function index $a$ to the internal force. The thread loops over all quadrature points of its element to compute the local contribution, and then scatters the result to the corresponding global DOF entries using atomic addition to ensure correctness in the presence of concurrent writes from different elements}
  \label{fig:compute_fint}
\end{figure}

Fig.~\ref{fig:compute_fint} provides a visualization of the two-stage GPU parallelization strategy for computing $\mathbf{P}^{(e,q)}$ and assembling $\mathbf{f}^{\mathrm{int}}$. The visualization uses a 3-element T10 mesh as an example, for ANCF beam and ANCF shell elements, the processes are similar but with different numbers of shape functions per element and different numbers of quadrature points.

\begin{sloppypar}
Two alternative assembly strategies were considered but not adopted in this work. In the first, atomic operations are avoided entirely by allocating a larger global GPU memory buffer to store the nodal contributions from each element independently; a subsequent reduce-by-key kernel then accumulates contributions across elements. For this approach to be effective, the reduction must be implemented efficiently—for instance, by leveraging a GPU library such as CUB~\cite{cubNVIDIA}. The trade-off, however, is an increased global memory footprint and the overhead associated with the reduce-by-key operation. In the second strategy, a greater degree of parallelism is achieved by launching threads on a per-node-per-quadrature-point basis, rather than looping over all quadrature points for a given nodal unknown within a single thread. Atomic additions are then used to scatter the per-node-per-quadrature-point contributions into the global internal force vector. While this approach increases the available parallelism, it also increases the number of atomic operations, which may offset the performance gains. The adopted method was selected to balance global memory footprint, the number of atomic operations, and the overall degree of parallelism, targeting NVIDIA RTX-class GPUs.
\end{sloppypar}

\subsection{The Parallelization of Gradient and Hessian Evaluation}
\label{sec:parallel_grad_hessian}
\begin{sloppypar}
Every iteration of the time integrator requires two quantities derived from the discrete augmented Lagrangian: the gradient $\nabla_{\mathbf{v}} \mathcal{L}$, and the system Hessian $\mathbf{H} = \nabla^2_{\mathbf{v}} \mathcal{L}$, which is additionally required by second-order Newton-type methods. As discussed in Section~\ref{sec:solvers_first_order}, first-order methods such as the AdamW solver advance the velocity state using the gradient alone, and the substantial cost of Hessian assembly is avoided entirely. Second-order methods, most notably the Newton solver presented in Section~\ref{sec:solvers_newton}, construct the full Hessian at every Newton iteration and solve the resulting sparse linear system for the velocity correction. The two computations are therefore described separately below.
\end{sloppypar}

\subsubsection{Gradient Evaluation}
\label{sec:parallel_grad}
\label{sec:gradient_eval}

The gradient of the discrete augmented Lagrangian with respect to the generalized velocity iterate $\mathbf{v}^{(k)}$ is a vector of length $n_{\mathrm{dof}} = 3 n_{\mathrm{coef}}$, where $n_{\mathrm{coef}}$ is the total number of coefficient indices (nodes for T10 elements; physical nodes times the number of gradient DOFs per node for ANCF elements). Its entry at global DOF index $(I, d)$, i.e., node $I$, spatial component $d$, takes the form

\begin{equation}
  g_{I,d}^{(k)}
  = \frac{1}{h} \sum_{J} M_{IJ}
    \bigl(v_{J,d}^{(k)} - v_{J,d}^{n}\bigr)
  + f_{I,d}^{\mathrm{int}}
  - f_{I,d}^{\mathrm{ext}}
  + h \sum_{c} [\mathbf{C}_q^{\top}]_{(I,d),\,c}
    \bigl(\lambda_{c} + \rho\, c_{c}\bigr),
  \label{eq:grad_L}
\end{equation}

where $h$ is the time-step size, $M_{IJ}$ is the consistent mass matrix entry between coefficient indices $I$ and $J$, $v^{n}$ is the converged velocity from the previous time step, $\mathbf{f}^{\mathrm{int}}$ is the internal force vector assembled in Section~\ref{sec:parallel_fint}, $\mathbf{f}^{\mathrm{ext}}$ is the applied external force, $\mathbf{C}_q$ is the constraint Jacobian, and $\lambda_{c}$ and $\rho$ are the dual variable and penalty parameter of the augmented Lagrangian, respectively.

A GPU thread is assigned to each scalar DOF entry of the gradient vector, with $n_{\mathrm{dof}}$ threads launched in total. Because each thread writes to a single, privately owned scalar location in the gradient vector, no write conflicts arise and no atomic operations are required.

The first term in Eq.~\eqref{eq:grad_L} is a sparse matrix-vector product involving the consistent mass matrix $\mathbf{M}$, which is preassembled once and stored in CSR format (Section~\ref{sec:mass_matrix}). Each thread traverses the CSR row corresponding to coefficient index $I$ and accumulates
\begin{equation}
  \sum_{J \in \mathcal{N}(I)} M_{IJ}\,\frac{v_{J,d}^{(k)} - v_{J,d}^{n}}{h},
\end{equation}
where $\mathcal{N}(I)$ is the set of coefficient neighbors of $I$ in the mesh
adjacency graph.  No element-level loop is needed; the mass matrix is accessed
directly in its globally assembled CSR form.

The contributions $f_{I,d}^{\mathrm{int}}$ and $f_{I,d}^{\mathrm{ext}}$ are read directly from device-resident global arrays. The internal force has already been assembled in full by the two-stage process of Section~\ref{sec:parallel_fint}, so no additional computation is required here.

When equality constraints are present, each thread additionally reads a row of the transposed constraint Jacobian $\mathbf{C}_q^{\top}$, stored as a separate CSR array on the device. Traversing the entries of that row, the thread accumulates
\begin{equation}
  h \sum_{c \in \mathcal{C}(I,d)} J_{c,(I,d)}^{\top}
    \bigl(\lambda_{c} + \rho\, c_{c}\bigr),
\end{equation}
where $\mathcal{C}(I,d)$ is the set of constraint indices whose Jacobian row has a nonzero entry at DOF $(I,d)$, and $c_{c}$ is the constraint residual.

\subsubsection{Hessian Assembly}
\label{sec:parallel_hessian}
\label{sec:hessian_assembly}

The system Hessian required by the Newton solver takes the form
\begin{equation}
  \mathbf{H}
  = \frac{1}{h}\,\mathbf{M}
  + h\,\mathbf{K}_{t}
  + h^{2}\rho\,\mathbf{C}_q^{\top}\mathbf{C}_q,
  \label{eq:hessian}
\end{equation}
\begin{sloppypar}
where $\mathbf{K}_{t}$ is the tangent stiffness matrix. The matrix $\mathbf{H}$ is symmetric positive definite and is stored in CSR format at the DOF level.
\end{sloppypar}

\begin{remark}[Positive definiteness of $\mathbf{H}$]
\begin{sloppypar}
Eq.~\eqref{eq:hessian} is a Gauss--Newton approximation of the full augmented-Lagrangian Hessian: the constraint-curvature term $h^{2}\sum_{c}\eta_{c}\,\nabla^{2}_{\mathbf{q}}c_{c}$ is omitted (see companion paper~\cite{json-ganesh-danTLFEA-1-2026}, Appendix~B). This term is indefinite for DP1-containing constraints, so its omission is what preserves the symmetric positive definite structure required by the Cholesky factorization in cuDSS. When nonlinear constraints are present, the solver attains superlinear rather than quadratic convergence near the solution.

With the curvature term absent, $\mathbf{H}$ is positive definite whenever $\mathbf{M} \succ 0$ (which holds by construction), $h > 0$, $\rho > 0$, and $\mathbf{M} + h^{2}\mathbf{K}_{t} \succ 0$. When $\mathbf{K}_{t} \succeq 0$ — as in the elastic regime of the SVK model — the last condition is satisfied for any $h$. When $\mathbf{K}_{t}$ is indefinite (e.g., under large compressive strains), a sufficient scalar bound is $h < \sqrt{\lambda_{\min}(\mathbf{M})\,/\,|\lambda_{\min}(\mathbf{K}_{t})|}$; the same step-size reductions used to improve robustness and accuracy also strengthen definiteness.
\end{sloppypar}
\end{remark}

\begin{sloppypar}
Before the first Newton iteration, a one-time symbolic analysis determines the nonzero structure of $\mathbf{H}$. The analysis starts from the coefficient-level adjacency graph, which is derived directly from element connectivity via the mass matrix sparsity pattern (see Section~\ref{sec:mass_matrix}). Each adjacent pair of coefficient indices $(I,J)$ contributes a $3 \times 3$ block of nonzeros to $\mathbf{H}$. When linear CSR constraints are present, the outer product $\mathbf{C}_q^{\top}\mathbf{C}_q$ may couple coefficient indices that are otherwise non-adjacent in the mesh, and these additional fill-in entries are incorporated into the pattern. The DOF-level CSR arrays (row offsets and column indices) are assembled on the host using a prefix-sum scan and uploaded to device memory, where they remain fixed for the entire simulation since the mesh topology does not change.
\end{sloppypar}

\begin{sloppypar}
The mass contribution $\mathbf{M}/h$ is assembled by a kernel that launches one thread per coefficient index. Each thread traverses its CSR mass row and, for every neighbor coefficient $J$, adds $M_{IJ}/h$ to the diagonal elements $H(3I+d,\,3J+d)$ for $d = 0,1,2$ (only the diagonal components of each $3\times 3$ block are nonzero due to the isotropic mass scaling). The CSR column position is located via binary search, and the value is deposited with \texttt{atomicAdd}.
\end{sloppypar}

The tangent stiffness contribution $h\mathbf{K}_{t}$ is assembled by a kernel that applies the same thread decomposition used in Stage 1 of Section~\ref{sec:parallel_fint} — one thread per (element, quadrature point) pair, with $n_{\mathrm{el}} \times n_{\mathrm{qp}}$ threads in total.

Each thread independently performs the following computation. It gathers the current nodal positions $\{\mathbf{x}_{a}\}$ and the precomputed reference gradients $\nabla_{\!\mathbf{X}} N_{a}^{(e,q)}$, then reassembles the deformation gradient
$\mathbf{F}^{(e,q)}$ (as in Eq.~\eqref{eq:F_assembly}). From $\mathbf{F}^{(e,q)}$ it evaluates the 4th-order material tangent tensor $\mathcal{A}^{(e,q)}$, whose components depend on the chosen constitutive model (Saint Venant--Kirchhoff or compressible Mooney--Rivlin). For every pair of local shape-function indices $(a,b)$, the $3\times 3$ tangent block is then
\begin{equation}
  [\mathbf{K}_{ab}]_{de}^{(e,q)}
  = \mathcal{A}_{dJeL}^{(e,q)}\,
    (\nabla_{\!\mathbf{X}} N_{a})_{J}\,
    (\nabla_{\!\mathbf{X}} N_{b})_{L}\,
    J_{0}^{(e,q)}\, w_{q},
  \label{eq:tangent_block}
\end{equation}
where repeated indices $J, L$ are summed over $\{1,2,3\}$. The blocks are accumulated into a thread-local element stiffness matrix $\mathbf{K}_{\mathrm{elem}}$, whose size is $30\times 30$ for T10 elements, $24\times 24$ for ANCF beam elements, and $48\times 48$ for ANCF shell elements.

\begin{sloppypar}
Once $\mathbf{K}_{\mathrm{elem}}$ is complete, the thread scatters each entry $h [\mathbf{K}_{\mathrm{elem}}]_{r,s}$ to its corresponding position in the global CSR array. The global row and column indices are obtained from the element-to-global node mapping; the CSR column position is located by binary search over the pre-built column index array; and the contribution is accumulated with \texttt{atomicAdd}. Concurrent writes from threads processing different elements that share a global node are correctly serialized by the atomic operation, with negligible overhead given the small number of elements per node in a well-conditioned mesh.
\end{sloppypar}

\begin{sloppypar}
The constraint contribution $h^{2}\rho\,\mathbf{C}_q^{\top}\mathbf{C}_q$ is assembled by a kernel that launches one thread per constraint. Thread $c$ accesses the CSR row of the constraint Jacobian $\mathbf{C}_q$ corresponding to constraint index $c$ and loops over all pairs of DOFs $(\mathrm{dof}_{i}, \mathrm{dof}_{j})$ whose Jacobian entries $J_{c,i}$ and $J_{c,j}$ are both nonzero. The contribution $h^{2}\rho\, J_{c,i}\, J_{c,j}$ is added to $\mathbf{H}[\mathrm{dof}_{i}, \mathrm{dof}_{j}]$ via binary search and \texttt{atomicAdd}. When different constraints share common DOFs, the resulting write conflicts are resolved by the atomic operation.
\end{sloppypar}

\subsection{First-Order Optimization Method}
\label{sec:solvers_first_order}

\begin{sloppypar}
We include an exploratory study of a first-order inner solver as an alternative to the Newton method. The motivation is as follows: in the Total Lagrangian FEA setting, each inner iteration of a first-order method reduces entirely to a gradient evaluation and a per-DOF state update. As shown in Section~\ref{sec:gradient_eval}, the gradient is embarrassingly parallel: one thread per DOF, with no inter-thread communication. This makes it a natural fit for GPU execution without the Hessian assembly and sparse factorization required by Newton. The question we investigate is whether this per-iteration cheapness is sufficient to offset the slower convergence rate. We employ AdamW~\cite{LoshchilovHutter2019AdamW}, an adaptive gradient method that combines exponential moving averages of first and second moments with decoupled weight decay; its per-DOF update is likewise fully parallel with no inter-thread communication. We provide AdamW as a documented reference point for gradient-only GPU solvers that will be compared in this contribution with the second-order method introduced in the next section.
\end{sloppypar}

\begin{algorithm}[H]
\caption{Augmented Lagrangian Method for Velocity-Level Constraints}
\label{alg:alm}
\begin{algorithmic}[1]
\Require $\mathcal{L}_\rho(\boldsymbol{v};\boldsymbol{\lambda})
  = \Phi(\boldsymbol{v})
  + \boldsymbol{\lambda}^\top\mathbf{c}(\mathbf{q})
  + \tfrac{\rho}{2}\|\mathbf{c}(\mathbf{q})\|^2$,\quad
  $\mathbf{q} = \mathbf{q}_n + h\boldsymbol{v}$
\State Initialize $\boldsymbol{\lambda}^0 \leftarrow \mathbf{0}$,\;
  $\rho > 0$
\For{$k = 0, 1, \ldots, K_{\max}$}
  \State $\boldsymbol{v}^{k+1} \leftarrow
    \arg\min_{\boldsymbol{v}}\;\mathcal{L}_\rho(\boldsymbol{v};\,\boldsymbol{\lambda}^k)$
    \Comment{Primal subproblem; solved via AdamW or Newton}
  \State $\mathbf{q}^{k+1} \leftarrow \mathbf{q}_n + h\boldsymbol{v}^{k+1}$
    \Comment{Backward-Euler step map}
  \State $\boldsymbol{\lambda}^{k+1} \leftarrow
    \boldsymbol{\lambda}^k + \rho\,\mathbf{c}(\mathbf{q}^{k+1})$
    \Comment{Dual ascent; drives constraint satisfaction}
  \If{$\|\mathbf{c}(\mathbf{q}^{k+1})\| \leq \varepsilon_{\mathrm{out}}$}
    \textbf{break}
    \Comment{Constraints satisfied}
  \EndIf
\EndFor
\State \Return $\boldsymbol{v}_{n+1} \leftarrow \boldsymbol{v}^{k+1}$,\;
  $\mathbf{q}_{n+1} \leftarrow \mathbf{q}^{k+1}$
\end{algorithmic}
\end{algorithm}

\begin{algorithm}[H]
\caption{One Time Step: AdamW Solver}
\label{alg:adamw_minimize_alm}
\begin{algorithmic}[1]
\Require Precomputed on device: $\nabla_{\mathbf{X}}\mathbf{N}^{(e,q)}$,
  $J_0^{(e,q)}$; $\mathbf{M}$ and $\mathbf{C}_q$ in CSR format; device arrays for
  $\mathbf{f}_{\mathrm{ext}}$
\State $\mathbf{q} \leftarrow \mathbf{q}_n$
\For{$k = 1, \ldots, K_{\max}$}
  \Comment{ALM outer loop; drives constraint satisfaction}
  \State $\mathbf{g} \leftarrow \mathbf{0}$,\;
    $\mathbf{m} \leftarrow \mathbf{0}$,\;
    $\mathbf{s} \leftarrow \mathbf{0}$
    \Comment{Reset gradient and moment accumulators}
  \For{$l = 1, \ldots, L_{\max}$}
    \Comment{AdamW inner loop; minimizes $\mathcal{L}_\rho$ w.r.t.\ $\boldsymbol{v}$}
    \State \textbf{[GPU, $n_v$]}\;
      $\mathbf{m} \leftarrow \beta_1\mathbf{m} + (1-\beta_1)\mathbf{g}$
      \Comment{First moment}
    \State \textbf{[GPU, $n_v$]}\;
      $\mathbf{s} \leftarrow \beta_2\mathbf{s} + (1-\beta_2)\mathbf{g}{\odot}\mathbf{g}$
      \Comment{Second moment}
    \State \textbf{[GPU, $n_v$]}\;
      $\hat{\mathbf{m}} \leftarrow \mathbf{m}/(1-\beta_1^l)$,\;
      $\hat{\mathbf{s}} \leftarrow \mathbf{s}/(1-\beta_2^l)$
      \Comment{Bias correction}
    \State \textbf{[GPU, $n_v$]}\;
      $\boldsymbol{v} \leftarrow
        (1-\alpha_l\lambda_{\mathrm{wd}})\boldsymbol{v}
        - \alpha_l\,\hat{\mathbf{m}}/(\sqrt{\hat{\mathbf{s}}}+\varepsilon)$
      \Comment{Velocity update; weight decay decoupled}
    \State \textbf{[GPU, $n_v$]}\;
      $\mathbf{q} \leftarrow \mathbf{q}_n + h\boldsymbol{v}$
      \Comment{Backward-Euler step map}
    \State \textbf{[GPU, $n_{el}{\times}n_{qp}$]}\;
      Evaluate $\mathbf{F}^{(e,q)}$ and $\mathbf{P}^{(e,q)}$
      \Comment{PK1 stress per quadrature point}
    \State \textbf{[GPU, $n_{el}{\times}n_{en}$]}\;
      Accumulate $\mathbf{f}_{\mathrm{int}}$ atomically
      \Comment{Internal forces computation}
    \State \textbf{[GPU, $n_c$]}\;
      Evaluate $\mathbf{c}(\mathbf{q})$
      \Comment{Bilateral constraint residual}
    \State \textbf{[GPU, $n_v$]}\;
      $\mathbf{g} \leftarrow
        \tfrac{1}{h}\mathbf{M}(\boldsymbol{v}-\boldsymbol{v}_n)
        + \mathbf{f}_{\mathrm{int}}
        - \mathbf{f}_{\mathrm{ext}}
        + h\mathbf{C}_q^{\top}(\boldsymbol{\lambda}+\rho\,\mathbf{c})$
      \Comment{Compute gradient}
    \State Transfer $\|\mathbf{g}\|$ to host;\;
      \textbf{if} $\|\mathbf{g}\| \leq \varepsilon_{\mathrm{in}}(1+\|\boldsymbol{v}\|)$
      \textbf{or} $\|\mathbf{g}\| \leq \varepsilon_{\mathrm{rel}}\|\mathbf{g}_0\|$:
      \textbf{break}
      \Comment{Inner convergence}
  \EndFor
  \State $\texttt{inner\_converged} \leftarrow$ \textbf{true} if the inner stopping criterion was met
  \State $\boldsymbol{v}_n \leftarrow \boldsymbol{v}$;\;
    $\mathbf{q} \leftarrow \mathbf{q}_n + h\boldsymbol{v}$
    \Comment{Commit velocity; update position}
  \State \textbf{[GPU, $n_c$]}\; Re-evaluate $\mathbf{c}(\mathbf{q})$
  \State \textbf{[GPU, $n_c$]}\;
    $\boldsymbol{\lambda} \leftarrow \boldsymbol{\lambda} + \rho\,\mathbf{c}(\mathbf{q})$
    \Comment{Dual ascent step}
  \State Transfer $\|\mathbf{c}\|$ to host;\;
    \textbf{if} $\texttt{inner\_converged}$ \textbf{and} $\|\mathbf{c}\| \leq \varepsilon_{\mathrm{out}}$: \textbf{break}
    \Comment{Outer convergence}
\EndFor
\State \Return $\mathbf{q}_{n+1} \leftarrow \mathbf{q}$,\;
  $\boldsymbol{v}_{n+1} \leftarrow \boldsymbol{v}_n$
\end{algorithmic}
\end{algorithm}

The AdamW solver implements the primal minimization in each outer augmented-Lagrangian
iteration of Algorithm~\ref{alg:alm}; Algorithm~\ref{alg:adamw_minimize_alm} lists
the resulting fused time-step routine. At the start of each outer iteration $k$,
the gradient vector $\mathbf{g}$ and the first and second moment accumulators
$\mathbf{m}$ and $\mathbf{s}$ are reset to zero, and the reference nodal positions
$\mathbf{q}_n$ are kept resident on-device to support repeated evaluations of the
backward-Euler step map. Each inner iteration $l$ then proceeds as a sequence of
GPU phases: (i) the velocity update applies the bias-corrected AdamW rule with
decoupled weight decay; (ii) the updated configuration
$\mathbf{q} = \mathbf{q}_n + h\boldsymbol{v}$ is computed; (iii) the first
Piola--Kirchhoff stress $\mathbf{P}^{(e,q)}$ is evaluated at every quadrature
point (Stage~1 of Section~\ref{sec:parallel_fint}); (iv) the internal force
$\mathbf{f}^{\mathrm{int}}$ is assembled (Stage~2 of
Section~\ref{sec:parallel_fint}); (v) the bilateral constraint residual
$\mathbf{c}(\mathbf{q})$ is evaluated; and (vi) the gradient $\mathbf{g}$ is
computed. Convergence of the inner loop is assessed at a prescribed check
interval by computing $\|\mathbf{g}\|_2$ on-device and transferring the scalar to
the host to evaluate a combined absolute--relative stopping criterion; the inner
loop terminates when $\|\mathbf{g}\| \leq \varepsilon_{\mathrm{in}}(1 +
\|\boldsymbol{v}\|)$ or $\|\mathbf{g}\| \leq
\varepsilon_{\mathrm{rel}}\|\mathbf{g}_0\|$, where $\mathbf{g}_0$ denotes the
first gradient evaluation within the current outer iteration. After the inner
loop, the committed velocity $\boldsymbol{v}^n \leftarrow \boldsymbol{v}$ and
position $\mathbf{q}$ are updated, followed by the dual ascent step
$\boldsymbol{\lambda} \leftarrow \boldsymbol{\lambda} + \rho\,\mathbf{c}(\mathbf{q})$,
consistent with the multiplier update in Eq.~\eqref{eq:lambda_update}. Outer convergence
requires both that the inner loop has converged and that $\|\mathbf{c}\| \leq
\varepsilon_{\mathrm{out}}$.

A key property of this implementation is that the gradient evaluation in
Algorithm~\ref{alg:adamw_minimize_alm} is fully data-parallel. One GPU thread
is assigned to each scalar DOF entry $(I, d)$, with $n_{\mathrm{dof}} = 3
n_{\mathrm{coef}}$ threads launched in total. Each thread independently evaluates
$g_{I,d}^{(k)}$ (Eq.~\eqref{eq:grad_L}): it traverses the CSR row of the
preassembled mass matrix $\mathbf{M}$ to accumulate the inertia contribution,
reads $f_{I,d}^{\mathrm{int}}$ and $f_{I,d}^{\mathrm{ext}}$ directly from
device-resident global arrays, and traverses the corresponding row of the
transposed constraint Jacobian $\mathbf{C}_q^\top$ stored in CSR format to
accumulate the constraint contribution. Because each thread writes exclusively
to the single device-memory entry $g_{I,d}$, no write conflicts arise and no
atomic operations are required. This per-thread independence extends to the
AdamW velocity update in Algorithm~\ref{alg:adamw_minimize_alm}: the moment
estimates $\mathbf{m}$ and $\mathbf{s}$ are maintained in persistent device-side
buffers of length $n_{\mathrm{dof}}$, and each thread reads and overwrites only
the entries at its own index, computing the bias-corrected update and writing
the new velocity $v_{I,d}$ without any inter-thread communication. The entire
gradient evaluation and velocity update are therefore fully parallel across all
DOFs.

\subsection{Second-Order Optimization Method}
\label{sec:solvers_newton}

The velocity-level augmented Lagrangian subproblem can also be solved using a second-order Newton method. In each inner iteration (Algorithm~\ref{alg:newton_minimize_alm}), the gradient and Hessian of the augmented Lagrangian are computed with respect to the generalized velocity, the resulting sparse linear system is solved for the Newton increment, and the iterate is updated. Unlike the first-order AdamW solver, the Newton method requires assembling and factorizing the full sparse Hessian.

The dominant contribution to the Hessian is the tangent stiffness matrix $\mathbf{K}_t$, whose derivation for the St.\ Venant--Kirchhoff and Mooney--Rivlin material models is presented in the companion paper~\cite{json-ganesh-danTLFEA-1-2026}; the Hessian assembly procedure is described in Section~\ref{sec:parallel_hessian}.

Once the assembled Hessian $\mathbf{H}$ is available in CSR format, each Newton
iteration requires the solution of the linear system
\begin{equation}
  \mathbf{H}^{(k)}\mathbf{p}^{(k)} = -\mathbf{g}^{(k)},
  \label{eq:newton_linear_system}
\end{equation}
for the velocity correction $\mathbf{p}^{(k)}$. The solution is obtained using
cuDSS~\cite{nvidia_cudss}, which
operates entirely in device memory and is designed to exploit the data
parallelism available on modern GPU architectures. The Hessian is registered
with cuDSS as a symmetric positive definite (SPD) matrix, with only the upper
triangular part stored, allowing cuDSS to employ a Cholesky factorization
$\mathbf{H} = \mathbf{L}\mathbf{L}^\top$ rather than a general LU
decomposition, thereby reducing both storage and arithmetic cost. Iterative
refinement is disabled in the solver configuration, as the outer Newton
iterations themselves provide sufficient convergence control without the
additional overhead of post-solve refinement passes.

The solution procedure is organized into three distinct phases that differ
substantially in their computational cost and in the frequency with which they
must be executed.

\paragraph{Analysis phase}
The analysis phase is performed once, prior to the first Newton iteration of
the simulation. Given the symbolic sparsity pattern of $\mathbf{H}$---which,
as established in Section~\ref{sec:sparsity}, remains fixed throughout the
simulation---cuDSS computes a fill-reducing column reordering, performs the
symbolic factorization, and allocates the internal data structures required for
numerical factorization. Because the CSR row offsets and column indices of
$\mathbf{H}$ do not change across Newton iterations or time steps, the cost of
this analysis is incurred exactly once and is amortized over the entire
simulation run. After the analysis completes, an internal flag is reset to
indicate that no numerical factorization has yet been performed, ensuring that
the very first factorization call unconditionally executes the full
factorization phase rather than refactorization.

\paragraph{Factorization phase}
At each Newton iteration, once the CSR values of $\mathbf{H}$ have been
updated by the assembly kernels described in Section~\ref{sec:hessian_assembly},
cuDSS performs numerical factorization using the symbolic structure established
during the analysis phase. Two factorization modes are employed depending on
call history. On the first Newton iteration of the simulation, cuDSS executes a
full numerical Cholesky factorization that initializes all internal factor
buffers. On every subsequent Newton iteration, cuDSS instead executes a
\emph{refactorization}: this mode reuses the symbolic structure, including the
fill-reducing permutation and pre-allocated factor storage, established during
the analysis phase, and only updates the numerical values of the triangular
factor $\mathbf{L}$. Because the symbolic work, which involves graph traversal,
ordering algorithms, and structural memory allocation, constitutes the dominant
setup cost of a sparse direct factorization, refactorization is substantially
faster than a full factorization for the same sparsity pattern. This efficiency
gain is precisely what makes the fixed-sparsity strategy described in
Section~\ref{sec:sparsity} valuable: the sparsity pattern is constructed once
and held constant so that every Newton iteration after the first can take
advantage of the cheaper refactorization path. Both factorization modes operate
entirely on device memory, avoiding host--device data transfers and reusing the
precomputed permutation arrays from the analysis step.

\paragraph{Solve phase}
Given the factored representation $\mathbf{H} = \mathbf{L}\mathbf{L}^\top$,
cuDSS performs forward and backward triangular solves to compute the Newton
step $\mathbf{p}^{(k)}$. The right-hand side vector $\mathbf{g}^{(k)}$ resides
in device memory as produced by the gradient evaluation
(Section~\ref{sec:gradient_eval}), and the computed step is written directly to
device memory for use in the velocity update of Algorithm~\ref{alg:newton_minimize_alm}.

This three-phase design, with analysis executed once per simulation, a single
full factorization on the first call, and numerical refactorization repeated at
every subsequent Newton iteration -- aligns naturally with the fixed-pattern,
repeated-refactorization strategy described in Section~\ref{sec:sparsity}. The
dominant per-iteration cost of the Newton method is consequently the Hessian
assembly and numerical refactorization, both of which execute entirely on the
GPU without intermediate host transfers.

\paragraph{Line search (optional)}
In highly nonlinear regimes such as near contact events or at large
deformation increments, the full Newton step $\delta\boldsymbol{v}$ can
overshoot a local minimum of the augmented Lagrangian, causing the gradient norm
to increase rather than decrease. An optional Armijo backtracking line search
guards against this. After computing $\delta\boldsymbol{v}$, a step size
$\alpha \in (0,1]$ is sought such that the trial gradient satisfies the
sufficient decrease condition
\begin{equation}
  \phi(\alpha) \;\leq\; (1 - 2c_1\alpha)\,\phi_0,
  \qquad \phi(\alpha) := \tfrac{1}{2}\|\mathbf{g}^{\mathrm{tr}}(\alpha)\|^2,
  \quad \phi_0 := \tfrac{1}{2}\|\mathbf{g}\|^2,
  \label{eq:armijo}
\end{equation}
where $\mathbf{g}^{\mathrm{tr}}(\alpha)$ is the gradient re-evaluated at the
trial point $(\boldsymbol{v} + \alpha\,\delta\boldsymbol{v},\;\mathbf{q}_n +
h(\boldsymbol{v}+\alpha\,\delta\boldsymbol{v}))$. Starting from $\alpha = 1$,
the step is reduced by a factor $\beta \in (0,1)$ at each backtracking iteration
until Eq.~\eqref{eq:armijo} is satisfied or a maximum of $j_{\max}$ reductions is
reached; in the latter case the last trial point is accepted regardless. Typical
values are $c_1 = 10^{-4}$, $\beta = 0.5$, and $j_{\max} = 10$. Each
re-evaluation requires one pass through the internal force, constraint, and
gradient assembly kernels, so the per-backtrack cost is comparable to a single
gradient evaluation step. When the problem is well-conditioned and step sizes
remain near unity, the line search terminates immediately at $j = 0$ and incurs
no additional cost beyond the single trial evaluation. The complete per-step
procedure is summarized in Algorithm~\ref{alg:newton_minimize_alm}.

\begin{algorithm}[H]
\caption{One Time Step: Newton Solver}
\label{alg:newton_minimize_alm}
\begin{algorithmic}[1]
\Require Precomputed on device: $\nabla_{\mathbf{X}}\mathbf{N}^{(e,q)}$,
  $J_0^{(e,q)}$; $\mathbf{M}$, $\mathbf{C}_q$, $\mathbf{H}$ in CSR format;
  $\mathbf{f}_{\mathrm{ext}}$; cuDSS handle with preanalyzed sparsity pattern
\State $\mathbf{q} \leftarrow \mathbf{q}_n$,\; $\boldsymbol{v} \leftarrow \boldsymbol{v}_n$
\For{$k = 1, \ldots, K_{\max}$}
  \Comment{ALM outer loop; drives constraint satisfaction}
  \For{$l = 1, \ldots, L_{\max}$}
    \Comment{Newton inner loop; second-order solve of $\mathcal{L}_\rho$}
    \State \textbf{[GPU, $n_{el}{\times}n_{qp}$]}\;
      Evaluate $\mathbf{F}^{(e,q)}$ and $\mathbf{P}^{(e,q)}$
      \Comment{PK1 stress per quadrature point}
    \State \textbf{[GPU, $n_{el}{\times}n_{en}$]}\;
      Accumulate $\mathbf{f}_{\mathrm{int}}$ atomically
      \Comment{Internal forces}
    \State \textbf{[GPU, $n_c$]}\;
      Evaluate $\mathbf{c}(\mathbf{q})$
      \Comment{Bilateral constraint residual}
    \State \textbf{[GPU, $n_v$]}\;
      $\mathbf{g} \leftarrow
        \tfrac{1}{h}\mathbf{M}(\boldsymbol{v}-\boldsymbol{v}_n)
        + \mathbf{f}_{\mathrm{int}}
        - \mathbf{f}_{\mathrm{ext}}
        + h\mathbf{C}_q^{\top}(\boldsymbol{\lambda}+\rho\,\mathbf{c})$
      \Comment{Gradient}
    \State Transfer $\|\mathbf{g}\|$ to host;\;
      \textbf{if} $l=1$: $\|\mathbf{g}_0\| \leftarrow \|\mathbf{g}\|$
    \State \textbf{if} $\|\mathbf{g}\| \le
      \max(\varepsilon_{\mathrm{atol}},\,\varepsilon_{\mathrm{rtol}}\|\mathbf{g}_0\|)$:
      \textbf{break}
      \Comment{Inner convergence (abs.\ or rel.)}
    \State \textbf{[GPU]}\;
      $\mathbf{H} \leftarrow \tfrac{1}{h}\mathbf{M} + h\mathbf{K}_t + h^2\rho\,\mathbf{C}_q^{\top}\mathbf{C}_q$
      \Comment{Hessian assembly}
    \State \textbf{[cuDSS]}\;
      Numeric (re)factorize $\mathbf{H}$
      \Comment{Reuses sparsity pattern}
    \State \textbf{[cuDSS]}\;
      Solve $\mathbf{H}\,\delta\boldsymbol{v} = -\mathbf{g}$
      \Comment{Newton direction}
    \If{Armijo line search enabled}
      \State $\phi_0 \leftarrow \tfrac{1}{2}\|\mathbf{g}\|^2$;\; $\alpha \leftarrow 1$
      \For{$j = 1, \ldots, N_{\mathrm{bt}}$}
        \Comment{Backtracking; $\beta=0.5$,\; $c_1=10^{-4}$}
        \State \textbf{[GPU, $n_v$]}\;
               $\boldsymbol{v}^{\mathrm{tr}} \leftarrow \boldsymbol{v}+\alpha\,\delta\boldsymbol{v}$;\;
               $\mathbf{q}^{\mathrm{tr}} \leftarrow \mathbf{q}_n+h\boldsymbol{v}^{\mathrm{tr}}$
               \Comment{Trial step}
        \State \textbf{[GPU]}\;
               Re-eval $\mathbf{F},\mathbf{P},\mathbf{f}_{\mathrm{int}},\mathbf{c},\mathbf{g}^{\mathrm{tr}}$;\;
               $\phi \leftarrow \tfrac{1}{2}\|\mathbf{g}^{\mathrm{tr}}\|^2$
        \If{$\phi \le (1-2c_1\alpha)\,\phi_0$}
          \State $\boldsymbol{v} \leftarrow \boldsymbol{v}^{\mathrm{tr}}$;\;
                 $\mathbf{q} \leftarrow \mathbf{q}^{\mathrm{tr}}$;\; \textbf{break}
                 \Comment{Accept; exit backtracking}
        \EndIf
        \State $\alpha \leftarrow \beta\alpha$
               \Comment{Shrink step}
      \EndFor
      \State \textbf{if} no trial accepted:\;
             restore $\boldsymbol{v}$, $\mathbf{q}$;\; \textbf{break}
             \Comment{Line-search failure; exit inner loop}
    \Else
      \State \textbf{[GPU, $n_v$]}\;
             $\boldsymbol{v} \leftarrow \boldsymbol{v}+\delta\boldsymbol{v}$;\;
             $\mathbf{q} \leftarrow \mathbf{q}_n+h\boldsymbol{v}$
             \Comment{Full Newton step}
    \EndIf
  \EndFor
  \State $\texttt{inner\_converged} \leftarrow$
    \textbf{true} if inner stopping criterion was met
  \State \textbf{[GPU, $n_c$]}\; Re-evaluate $\mathbf{c}(\mathbf{q})$
  \If{$\texttt{inner\_converged}$ \textbf{and} $n_c > 0$}
    \State \textbf{[GPU, $n_c$]}\;
           $\boldsymbol{\lambda} \leftarrow \boldsymbol{\lambda} + \rho\,\mathbf{c}(\mathbf{q})$
           \Comment{Dual ascent}
  \EndIf
  \State Transfer $\|\mathbf{c}\|$ to host;\;
    \textbf{if} $\texttt{inner\_converged}$ \textbf{and}
    $\|\mathbf{c}\| \le \varepsilon_{\mathrm{out}}$: \textbf{break}
    \Comment{Outer convergence}
\EndFor
\State \Return $\mathbf{q}_{n+1} \leftarrow \mathbf{q}$,\;
  $\boldsymbol{v}_{n+1} \leftarrow \boldsymbol{v}$
\end{algorithmic}
\end{algorithm}

\section{Performance Benchmarks}
\label{sec:performance_benchmark}
\setlength{\emergencystretch}{3em}

\begin{sloppypar}
We evaluate the performance of the GPU-accelerated Total Lagrangian FEA framework through two
complementary sets of experiments: scaling benchmarks for each of the three
supported element types using a canonical cantilever problem, and a
demonstration on three geometrically complex meshes.
All GPU experiments were conducted on an NVIDIA GeForce RTX~5090 (32\,GB
VRAM) under CUDA~12.8 (driver 570.195.03) with cuDSS~0.7.1.4.
CPU baselines were obtained on an Intel i7-13700KF;
FEniCS~\cite{Alnaes2015Fenics15} serves as the T10 CPU baseline and
Project Chrono~9.0~\cite{projectChronoWebSite} for ANCF element comparisons;
both use native MPI parallelization swept over $\{1,2,4,8,16\}$ ranks with
the best timing reported for each resolution.
\end{sloppypar}

\begin{sloppypar}
Throughout this section, the \emph{real-time factor} (RTF) denotes the ratio
of wall-clock runtime to total simulated time; lower RTF indicates faster
execution relative to real time. As a caveat, the GPU and CPU baselines reflect different points in the hardware design space, i.e., NVIDIA GeForce RTX 5090 vs. Intel i7-13700KF. These platforms are not directly comparable; one could choose a less or more capable GPU or CPU. Furthermore, the CPU solvers (FEniCS/PETSc and Project Chrono) employ different linear solver stacks and convergence criteria than the GPU implementation, so the measured RTF differences reflect a combination of hardware parallelism, solver architecture, and implementation maturity. The reported speedups should therefore be interpreted as indicative of the performance regime achievable with a fully GPU-resident implicit solver on current commodity hardware, not as a controlled hardware-normalized comparison.
\end{sloppypar}

For all GPU solvers, the inner loop terminates when the gradient norm
$\|\mathbf{g}\|_2$ of the augmented Lagrangian falls below $\varepsilon_{\mathrm{in}}$;
this criterion is identical for the Newton and AdamW solvers, so
$\varepsilon_{\mathrm{in}}$ is directly comparable across both.
The outer ALM loop terminates when the constraint residual satisfies
$\|\mathbf{c}\| \leq \varepsilon_{\mathrm{out}}$.
Because $\|\mathbf{g}\|_2$ is an un-normalized $\ell_2$ norm summed over all
degrees of freedom, it grows naturally with mesh size even at the same
per-DOF accuracy; a fixed absolute threshold therefore becomes increasingly
demanding at finer resolutions.
At the highest resolutions (RES32 for T10 and ANCF3243; RES8 and above for
ANCF3443), $\varepsilon_{\mathrm{in}}$ and $\varepsilon_{\mathrm{out}}$ are
relaxed from $10^{-4}$ to $10^{-3}$.
For the Newton solver this relaxation is a deliberate runtime trade-off: the
coarser tolerance remains well within the regime where tip-displacement results
are indistinguishable from those at $10^{-4}$.
For the AdamW solver, the first-order convergence rate causes the gradient norm
to plateau at large problem sizes within any practical iteration budget, making
the tighter tolerance unattainable without a prohibitive increase in iteration
count.
The ANCF3443 shell system requires the relaxation at a coarser mesh than the
solid or beam cases, reflecting the higher condition numbers typical of
thin-shell discretizations.
\begin{sloppypar}
FEniCS uses PETSc's Newton-with-line-search (\texttt{newtonls}) SNES solver
with the default convergence test (\texttt{SNESConvergedDefault})~\cite{petsc-user-ref},
which checks an un-normalized $\ell_2$ residual norm $\|\mathbf{F}\|_2$
against absolute (\texttt{snes\_atol}), relative (\texttt{snes\_rtol}), and
step (\texttt{snes\_stol}) tolerances; the absolute tolerance is therefore
subject to the same mesh-size dependence as $\varepsilon_{\mathrm{in}}$
and is not directly comparable in absolute terms.
Per-resolution values of $\varepsilon_{\mathrm{in}}$ and $\varepsilon_{\mathrm{out}}$
are reported in Tables~\ref{tab:t10_resolutions}, \ref{tab:ancf3243_resolutions},
and~\ref{tab:ancf3443_resolutions}.
\end{sloppypar}

\subsection{T10 Tetrahedral Element Scaling}
\label{sec:benchmark_t10}

\begin{sloppypar}
We discretize a 3\,m\,(L) $\times$ 2\,m\,(W) $\times$ 1\,m\,(H) beam with T10 tetrahedral elements and evaluate six mesh resolutions. One end of the beam is fully fixed via constraints, and a 5{,}000\,N load is applied at the opposite end, distributed uniformly across all nodes on that face. SVK and time-integration parameters are shared across all scaling benchmarks (Appendix~\ref{sec:appendix_benchmark_params}, Table~\ref{tab:scaling_sim_params}); Mooney--Rivlin parameters are given in Table~\ref{tab:t10_mr_params}. Table~\ref{tab:t10_resolutions} summarizes the mesh statistics for each resolution level, Table~\ref{tab:t10_rtf} reports the corresponding real-time factors (RTF) for the St.\ Venant--Kirchhoff (SVK) material model, and Table~\ref{tab:t10_rtf_mooney_rivlin} reports the corresponding RTFs for the Mooney--Rivlin material model.
\end{sloppypar}

\begin{table}[t]
  \centering
  \small
  \setlength{\tabcolsep}{5pt}
  \renewcommand{\arraystretch}{1.1}
  \begin{tabular}{@{}lrrrrrr@{}}
    \toprule
    Resolution & DOFs & Nodes & Elements & Constrained DOFs & $\varepsilon_{\mathrm{in}}$ & $\varepsilon_{\mathrm{out}}$\\
    \midrule
    RES0  &     315 &     105 &    36 &  45    & $10^{-4}$ & $10^{-4}$\\
    RES2  &   1,398 &     466 &   199 &  135   & $10^{-4}$ & $10^{-4}$\\
    RES4  &   6,039 &   2,013 &   936 &  459   & $10^{-4}$ & $10^{-4}$\\
    RES8  &  27,150 &   9,050 & 4,567 &  1,683 & $10^{-4}$ & $10^{-4}$\\
    RES16 & 118,326 &  39,442 & 20,829 & 6,435 & $10^{-4}$ & $10^{-4}$\\
    RES32 & 497,310 & 165,770 & 83,432 & 25,155 & $10^{-3}$ & $10^{-3}$\\
    \bottomrule
  \end{tabular}
  \caption{T10 beam mesh statistics and solver tolerances across six resolution levels.
    $\varepsilon_{\mathrm{in}}$: inner gradient-norm stopping criterion; $\varepsilon_{\mathrm{out}}$: outer ALM constraint-residual stopping criterion.}
  \label{tab:t10_resolutions}
\end{table}

\begin{table}[t]
  \centering
  \small
  \setlength{\tabcolsep}{5pt}
  \renewcommand{\arraystretch}{1.1}
  \begin{tabular}{@{}lrrr@{}}
    \toprule
    Resolution & Newton (GPU) & AdamW (GPU) & FEniCS (CPU) \\
    \midrule
    RES0  &   3.63 & 12.10 &    5.41 \\
    RES2  &   6.11 & 11.52 &   13.73 \\
    RES4  &  12.85 & 28.13 &   54.91 \\
    RES8  &  38.78 & 38.66 &  293.20 \\
    RES16 & 170.09 & 173.92 & 2509.94 \\
    RES32 & 1249.69 & 512.01 & 27596.69 \\
    \bottomrule
  \end{tabular}
  \caption{RTF for the T10 beam-sagging benchmark with the St.\ Venant--Kirchhoff
    material model across six resolution levels.
    Newton and AdamW run on GPU; FEniCS on CPU.}
  \label{tab:t10_rtf}
\end{table}

\begin{table}[t]
  \centering
  \small
  \setlength{\tabcolsep}{5pt}
  \renewcommand{\arraystretch}{1.1}
  \begin{tabular}{@{}lrrr@{}}
    \toprule
    Resolution & Newton (GPU) & AdamW (GPU) & FEniCS (CPU) \\
    \midrule
    RES0  & 4.02 & 14.04 & 5.64 \\
    RES2  & 6.66 & 13.28 & 14.10 \\
    RES4  & 13.23 & 31.93 & 57.50 \\
    RES8  & 39.40 & 51.14 & 293.91 \\
    RES16 & 171.23 & 212.66 & 2425.62 \\
    RES32 & 1255.48 & 619.19 & 30547.00 \\
    \bottomrule
  \end{tabular}
  \caption{RTF for the T10 beam-sagging benchmark with the Mooney--Rivlin material
    model across six resolution levels.
    Newton and AdamW run on GPU; FEniCS on CPU.}
  \label{tab:t10_rtf_mooney_rivlin}
\end{table}

\begin{figure}[t]
  \centering
  \maybeincludegraphics[width=\linewidth]{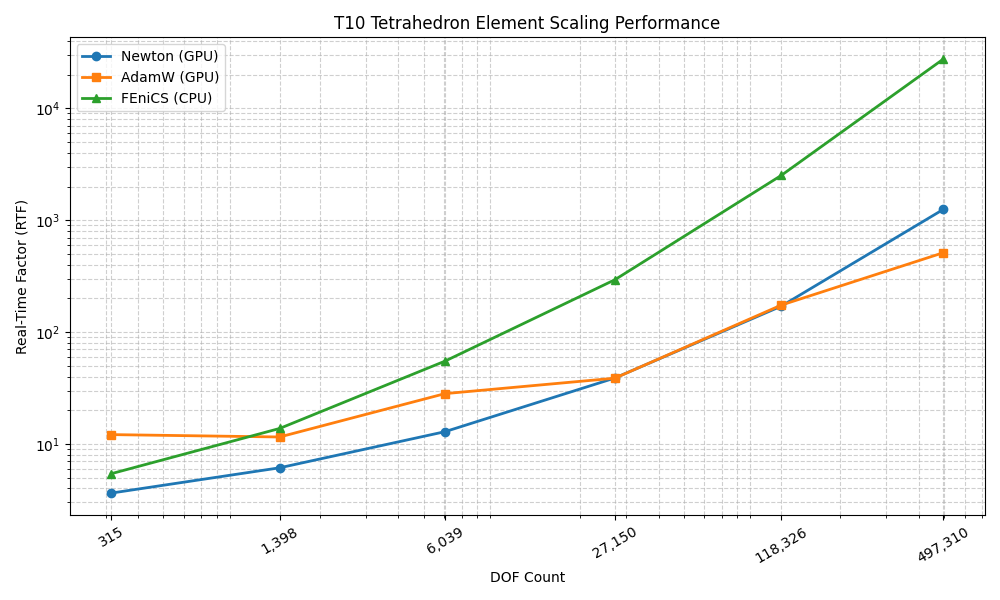}
  \caption{Time-to-solution comparison for the T10 beam-sagging benchmark across mesh resolutions. GPU solvers (AdamW and Newton) are compared against the FEniCS CPU baseline}
  \label{fig:beam_sag_timing}
\end{figure}

\begin{sloppypar}
Fig.~\ref{fig:beam_sag_timing} shows the time-to-solution for each solver across all six resolutions.
The Newton solver achieves the best RTF at coarse-to-mid resolutions, while the AdamW solver becomes competitive at RES8 and above; against the FEniCS CPU baseline, the Newton solver yields approximately a $22\times$ speedup at the largest resolution tested (RTF $1{,}250$ vs.\ $27{,}597$ at RES32, SVK material).
\end{sloppypar}

\subsection{ANCF3243 Beam Element Scaling}
\label{sec:benchmark_ancf3243}

\begin{sloppypar}
The beam is discretized as a one-dimensional chain of ANCF3243 elements with uniform element length \(L = 0.2\) and a constant rectangular cross section of width \(W = 0.1\) and height \(H = 0.1\), with consecutive nodes connected so that element \(e\) spans nodes \(e\) and \(e+1\) along the global \(x\) axis. A cantilever constraint clamps the leftmost four nodes throughout the simulation. A concentrated vertical tip force of magnitude \(F_z = 5{,}000\) is applied at the free end and held constant for the first 100 time steps, after which it is released and the beam evolves under inertia and elasticity alone.
\end{sloppypar}

\begin{table}[t]
  \centering
  \small
  \setlength{\tabcolsep}{6pt}
  \renewcommand{\arraystretch}{1.1}
  \begin{tabular}{@{}lrrrrrr@{}}
    \toprule
    Resolution & DOFs & Nodes & Elements & Constrained DOFs & $\varepsilon_{\mathrm{in}}$ & $\varepsilon_{\mathrm{out}}$\\
    \midrule
    RES0  & 12{,}012      & 1{,}001   & 1{,}000   & 12 & $10^{-4}$ & $10^{-4}$\\
    RES2  & 120{,}012     & 10{,}001  & 10{,}000  & 12 & $10^{-4}$ & $10^{-4}$\\
    RES4  & 600{,}012     & 50{,}001  & 50{,}000  & 12 & $10^{-4}$ & $10^{-4}$\\
    RES8  & 1{,}200{,}012 & 100{,}001 & 100{,}000 & 12 & $10^{-4}$ & $10^{-4}$\\
    RES16 & 2{,}400{,}012 & 200{,}001 & 200{,}000 & 12 & $10^{-4}$ & $10^{-4}$\\
    RES32 & 6{,}000{,}012 & 500{,}001 & 500{,}000 & 12 & $10^{-3}$ & $10^{-3}$\\
    \bottomrule
  \end{tabular}
  \caption{ANCF3243 beam mesh statistics across six resolution levels.
    $\varepsilon_{\mathrm{in}}$: inner gradient-norm stopping criterion; $\varepsilon_{\mathrm{out}}$: outer ALM constraint-residual stopping criterion.}
  \label{tab:ancf3243_resolutions}
\end{table}

\begin{table}[t]
  \centering
  \small
  \setlength{\tabcolsep}{5pt}
  \renewcommand{\arraystretch}{1.1}
  \begin{tabular}{@{}lrrr@{}}
    \toprule
    Resolution & Newton (GPU) & AdamW (GPU) & Project Chrono (CPU) \\
    \midrule
    RES0  &   4.85 & 14.21 &   93.61 \\
    RES2  &   16.57 & 44.29 &  1151.64 \\
    RES4  &  74.38 & 222.83 &   6571.27 \\
    RES8  &  146.22 & 444.42 &  13018.00 \\
    RES16 & 297.35 & 891.52 & 25844.58 \\
    RES32 & 530.05 & 1769.61 & 66806.46 \\
    \bottomrule
  \end{tabular}
  \caption{RTF for the ANCF3243 beam-sagging benchmark across six resolution levels.
    Newton and AdamW are run on the GPU; Project Chrono on the CPU.
    Project Chrono enforces constraints via a different formulation (penalty/Lagrange multiplier
    at the rigid-body level) and uses its own internal convergence criterion,
    which is not directly comparable to the ALM criteria $\varepsilon_{\mathrm{in}}$/$\varepsilon_{\mathrm{out}}$.}
  \label{tab:ancf3243_rtf}
\end{table}

\begin{figure}[t]
  \centering
  \maybeincludegraphics[width=\linewidth]{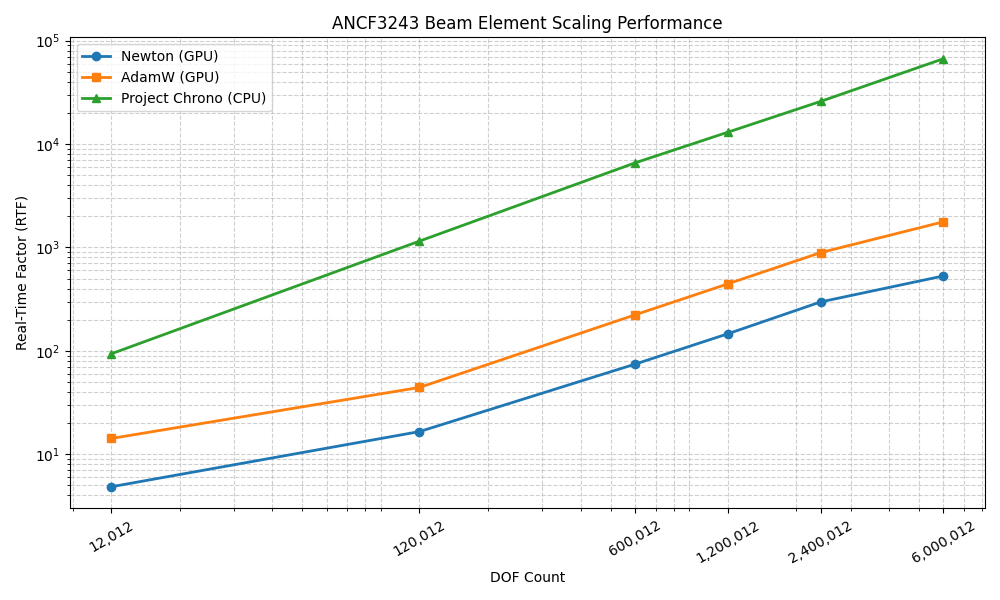}
  \caption{Time-to-solution comparison for the ANCF3243 beam-sagging benchmark across mesh resolutions. GPU solvers (AdamW and Newton) are compared against the Project Chrono CPU baseline}
  \label{fig:beam_sag_timing_ancf3243}
\end{figure}

Fig.~\ref{fig:beam_sag_timing_ancf3243} shows the timing comparison across resolutions.

\subsection{ANCF3443 Shell Element Scaling}
\label{sec:benchmark_ancf3443}

\begin{sloppypar}
We discretize a rectangular ANCF3443 shell with in-plane dimensions \(x_{\mathrm{size}} = 4\) and \(y_{\mathrm{size}} = 2\) and thickness \(H = 0.1\) using a structured mesh of \(n_x \times n_y\) elements, where \(n_x = n_y \in \{10, 20, 50, 100, 200, 400\}\) defines six resolution levels with uniform element sizes \(L = x_{\mathrm{size}}/n_x\) and \(W = y_{\mathrm{size}}/n_y\). A cantilever boundary condition is imposed by clamping all nodes on the edge \(x = 0\). A total vertical force \(F_z = -500\) is applied uniformly to the opposite free edge at \(x = x_{\mathrm{size}}\), held constant for the first 100 time steps and then released. Material and time-integration parameters are given in Appendix~\ref{sec:appendix_benchmark_params} (Table~\ref{tab:scaling_sim_params}). Table~\ref{tab:ancf3443_resolutions} summarizes the mesh statistics and Table~\ref{tab:ancf3443_rtf} reports the corresponding real-time factors.
\end{sloppypar}

\begin{table}[t]
  \centering
  \small
  \setlength{\tabcolsep}{6pt}
  \renewcommand{\arraystretch}{1.1}
  \begin{tabular}{@{}lrrrrrr@{}}
    \toprule
    Resolution & DOFs & Nodes & Elements & Constrained DOFs & $\varepsilon_{\mathrm{in}}$ & $\varepsilon_{\mathrm{out}}$\\
    \midrule
    RES0  & 1{,}452   & 121      & 100      & 132     & $10^{-4}$ & $10^{-4}$\\
    RES2  & 5{,}292   & 441      & 400      & 252     & $10^{-4}$ & $10^{-4}$\\
    RES4  & 31{,}212  & 2{,}601  & 2{,}500  & 612     & $10^{-4}$ & $10^{-4}$\\
    RES8  & 122{,}412 & 10{,}201 & 10{,}000 & 1{,}212 & $10^{-3}$ & $10^{-3}$\\
    RES16 & 273{,}612 & 22{,}801 & 22{,}500 & 1{,}812 & $10^{-3}$ & $10^{-3}$\\
    RES32 & 484{,}812 & 40{,}401 & 40{,}000 & 2{,}412 & $10^{-3}$ & $10^{-3}$\\
    \bottomrule
  \end{tabular}
  \caption{ANCF3443 shell mesh statistics across six resolution levels.
    $\varepsilon_{\mathrm{in}}$: inner gradient-norm stopping criterion; $\varepsilon_{\mathrm{out}}$: outer ALM constraint-residual stopping criterion.}
  \label{tab:ancf3443_resolutions}
\end{table}

\begin{table}[t]
  \centering
  \small
  \setlength{\tabcolsep}{5pt}
  \renewcommand{\arraystretch}{1.1}
  \begin{tabular}{@{}lrrr@{}}
    \toprule
    Resolution & Newton (GPU) & AdamW (GPU) & Project Chrono (CPU) \\
    \midrule
    RES0  & 8.11 & 38.92 & 25.22 \\
    RES2  & 12.89 & 63.79 & 110.66 \\
    RES4  & 46.19 & 229.57 & 840.84 \\
    RES8  & 188.26 & 835.40 & 4476.33 \\
    RES16 & 432.60 & 2021.65 & 11931.25 \\
    RES32 & 967.82 & 6202.02 & 23709.70 \\
    \bottomrule
  \end{tabular}
  \caption{RTF for the ANCF3443 shell cantilever benchmark across six resolution levels.
    Newton and AdamW are run on the GPU; Project Chrono on the CPU.
    Project Chrono enforces constraints via a different formulation (penalty/Lagrange multiplier
    at the rigid-body level) and uses its own internal convergence criterion,
    which is not directly comparable to the ALM criteria $\varepsilon_{\mathrm{in}}$/$\varepsilon_{\mathrm{out}}$.}
  \label{tab:ancf3443_rtf}
\end{table}

\begin{figure}[t]
  \centering
  \maybeincludegraphics[width=\linewidth]{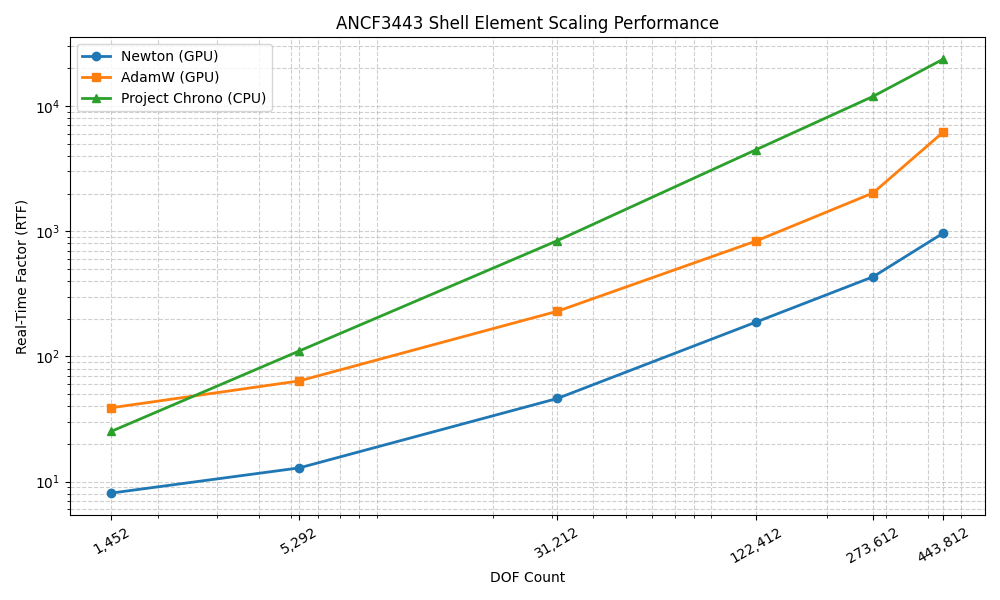}
  \caption{Time-to-solution comparison for the ANCF3443 shell beam-sagging benchmark across mesh resolutions. GPU solvers (AdamW and Newton) are compared against the Project Chrono CPU baseline}
  \label{fig:beam_sag_timing_ancf3443}
\end{figure}

Fig.~\ref{fig:beam_sag_timing_ancf3443} presents the timing comparison across resolutions.

\subsection{Demonstration on Geometrically Complex Meshes}
\label{sec:benchmark_complex}

\begin{figure}[t]
  \centering
  \maybeincludegraphics[width=0.30\linewidth,valign=c]{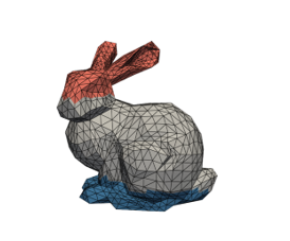}%
  \hfill%
  \maybeincludegraphics[width=0.67\linewidth,valign=c]{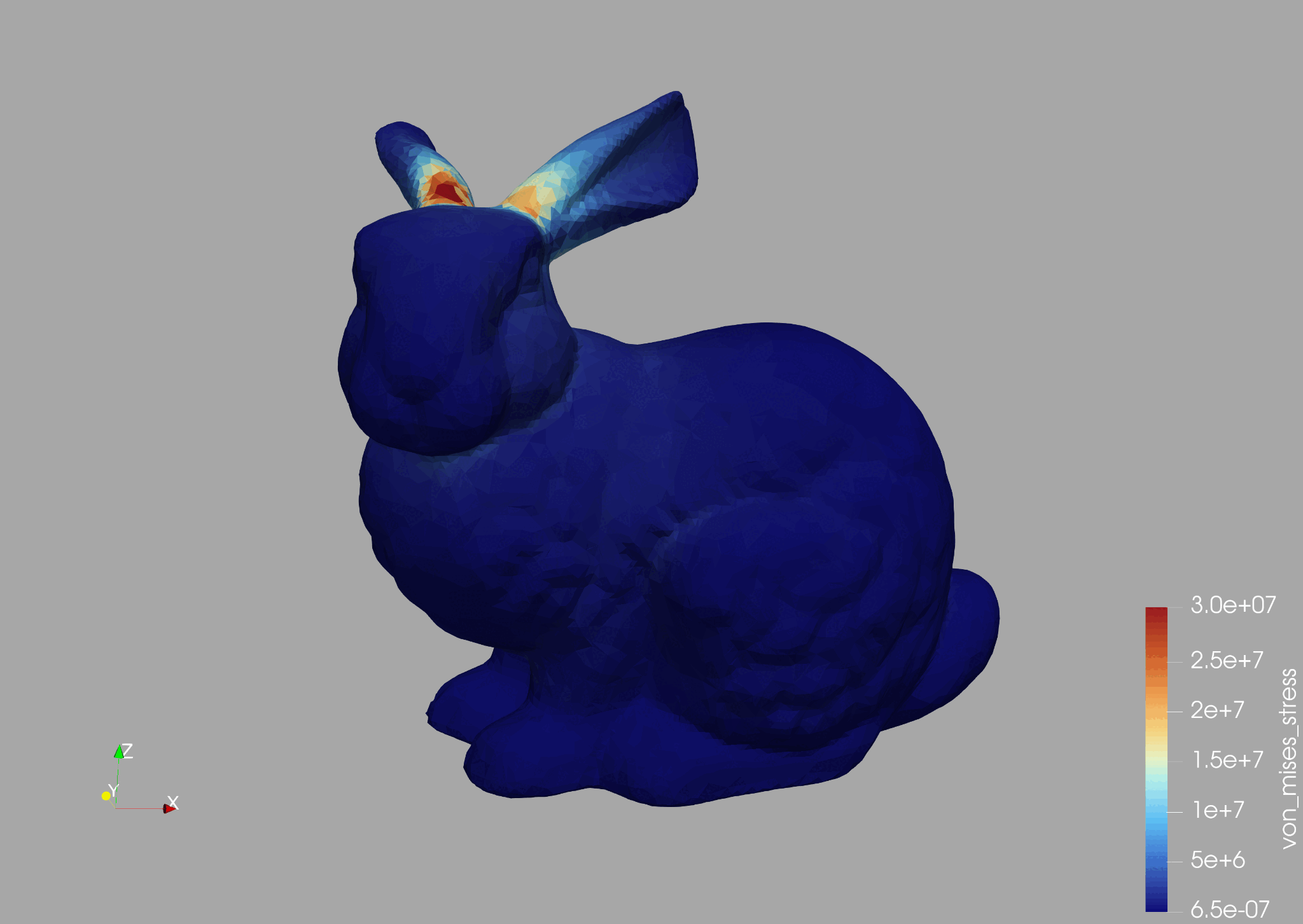}\\[6pt]
  \maybeincludegraphics[width=0.30\linewidth,valign=c]{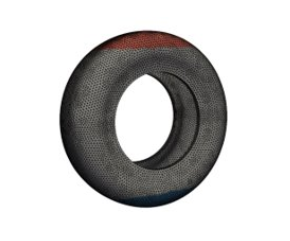}%
  \hfill%
  \maybeincludegraphics[width=0.67\linewidth,valign=c]{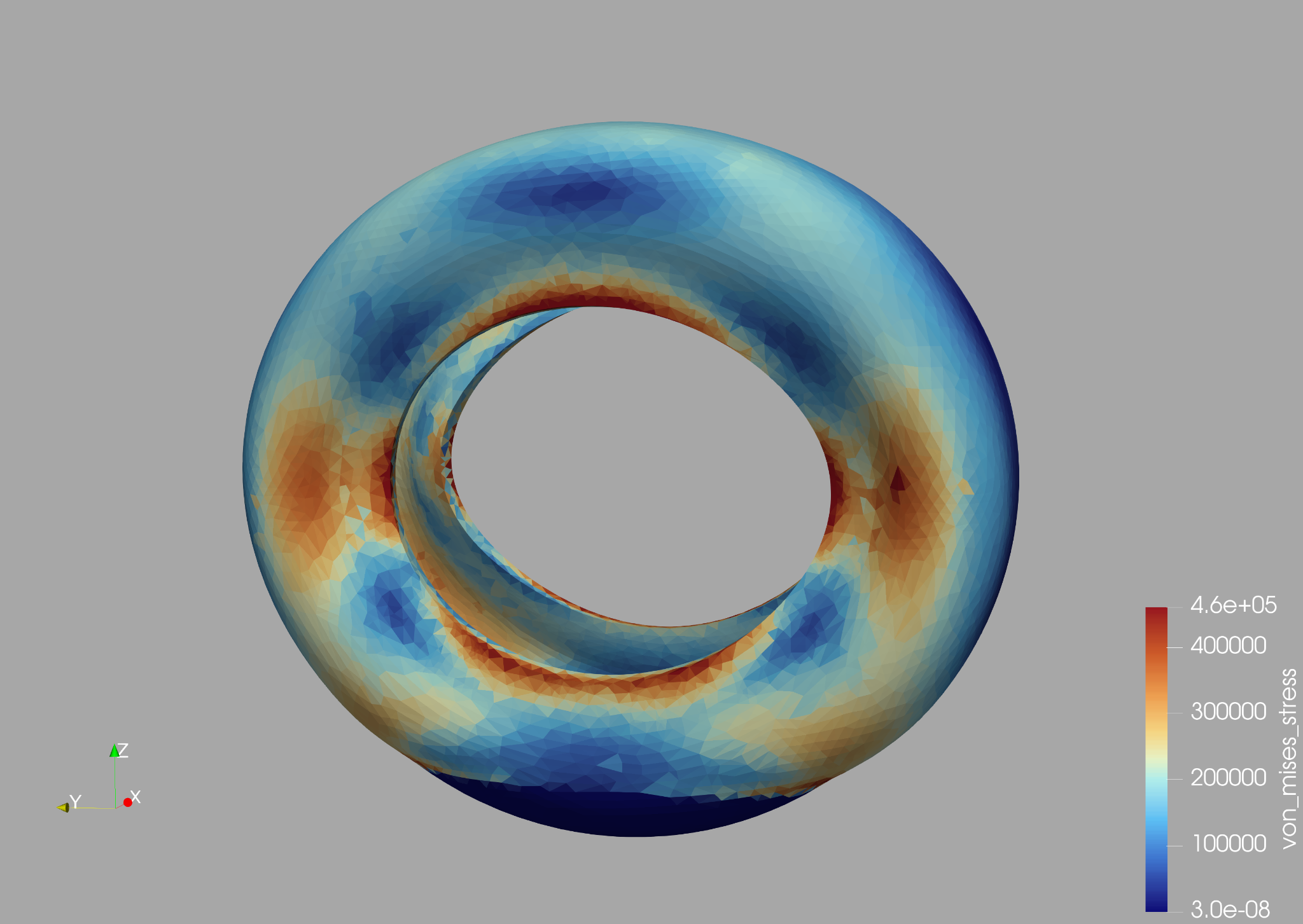}
  \caption{Loading configurations and von Mises stress fields (Pa) at $t = 0.01\,\mathrm{s}$
    for the two geometrically complex T10 benchmarks.
    \textit{Left column}: mesh colored by boundary condition (red: applied force nodes;
    blue: clamped nodes).
    \textit{Right column}: resulting von Mises stress field.
    \textit{Top row} --- Stanford Bunny ($2{,}095$ nodes, $1{,}066$ elements):
    stress concentrates at the thin ear roots, consistent with the cross-sectional minima
    of a loaded cantilever-like feature.
    \textit{Bottom row} --- Deformable tire ($34{,}414$ nodes, $17{,}167$ elements):
    a characteristic stress band forms along the inner radius with secondary peaks at
    the contact arc, consistent with a rim-clamped toroidal shell under axial compression}
  \label{fig:bunny_deformation}
  \label{fig:tire_deformation}
\end{figure}

\begin{sloppypar}
Beyond the structured beam-sagging case used for scaling analysis, we demonstrate the framework on three geometrically complex meshes representative of shapes arising in practical engineering and computer graphics applications: a deformable tire, a Stanford Bunny, and a Utah Teapot. These geometries feature irregular mesh topology, varying element valence, and non-trivial curvature distributions, providing a more demanding test of the GPU implementation than a regular structured grid. All three cases use T10 tetrahedral elements with the Newton solver. Fig.~\ref{fig:bunny_deformation} shows the loading configuration and resulting von Mises stress field for the Stanford Bunny and deformable tire cases; performance data for all three cases are summarized in Tables~\ref{tab:complex_materials} and~\ref{tab:complex_rtf}.
\end{sloppypar}

\paragraph{Utah Teapot}
\begin{sloppypar}
The Utah Teapot mesh comprises 23{,}006 nodes (69{,}018 DOFs) and 12{,}280 T10 tetrahedral elements, with overall dimensions of approximately 1.0053\,m (height) $\times$ 1.2753\,m (width) $\times$ 2.0529\,m (length). The bottom 20\% of nodes (5{,}444 nodes) are clamped, and a total upward force of 1{,}000\,N is applied to the top 20\% (1{,}873 nodes). The applied load is intentionally moderate; this case primarily serves to demonstrate the solver's ability to handle geometrically complex meshes with non-manifold-like junctions (spout, handle, and lid), and its performance relative to the FEniCS CPU baseline is reported in Table~\ref{tab:complex_rtf}.
\end{sloppypar}

\paragraph{Stanford Bunny}
\begin{sloppypar}
The Stanford Bunny mesh comprises 2{,}095 nodes (6{,}285 DOFs) and 1{,}066 T10 tetrahedral elements, spanning approximately 1.5413\,m (height) $\times$ 1.2083\,m (width) $\times$ 1.5594\,m (length). All nodes with $z < -4\,\mathrm{m}$ are clamped (545 nodes), and a total downward force of 35{,}000\,N is distributed over all nodes with $z > 4\,\mathrm{m}$ (582 nodes). Fig.~\ref{fig:bunny_deformation} shows the loading configuration and the resulting von Mises stress field.
\end{sloppypar}

Material parameters for all three cases are given in Appendix~\ref{sec:appendix_benchmark_params} (Table~\ref{tab:complex_materials}).

\begin{table}[t]
  \centering
  \small
  \setlength{\tabcolsep}{5pt}
  \renewcommand{\arraystretch}{1.1}
  \begin{tabular}{@{}lrr@{}}
    \toprule
    Case & Newton (GPU) & FEniCS (CPU) \\
    \midrule
    Utah Teapot     & 100.83 & 2348.78 \\
    Stanford Bunny  & 13.87 & 108.61 \\
    Deformable Tire & 148.36 & 4141.85 \\
    \bottomrule
  \end{tabular}
  \caption{RTF for the three geometrically complex mesh benchmarks with the Newton solver.
    All GPU cases use $\varepsilon_{\mathrm{in}} = \varepsilon_{\mathrm{out}} = 10^{-4}$.
    GPU: NVIDIA RTX~5090; CPU baseline: FEniCS on Intel i7-13700KF.
    FEniCS enforces Dirichlet conditions by direct matrix modification (strong form)
    and terminates on force-residual norm; its tolerance is not directly comparable
    to the ALM criteria used by the GPU solver.}
  \label{tab:complex_rtf}
\end{table}

\subsection{Roofline Analysis}
\label{sec:roofline}
We next examine the dominant GPU kernels to identify the main hardware limits on per-kernel throughput. The structured beam-sagging benchmarks are used for this analysis because they expose the core GPU kernels without the additional variability associated with contact detection and irregular mesh topology. The goal is to determine which kernels are limited primarily by DRAM bandwidth, which are limited by occupancy, and which are affected by atomic contention in sparse assembly.

\begin{figure}[!htbp]
  \centering
  \maybeincludegraphics[width=\linewidth]{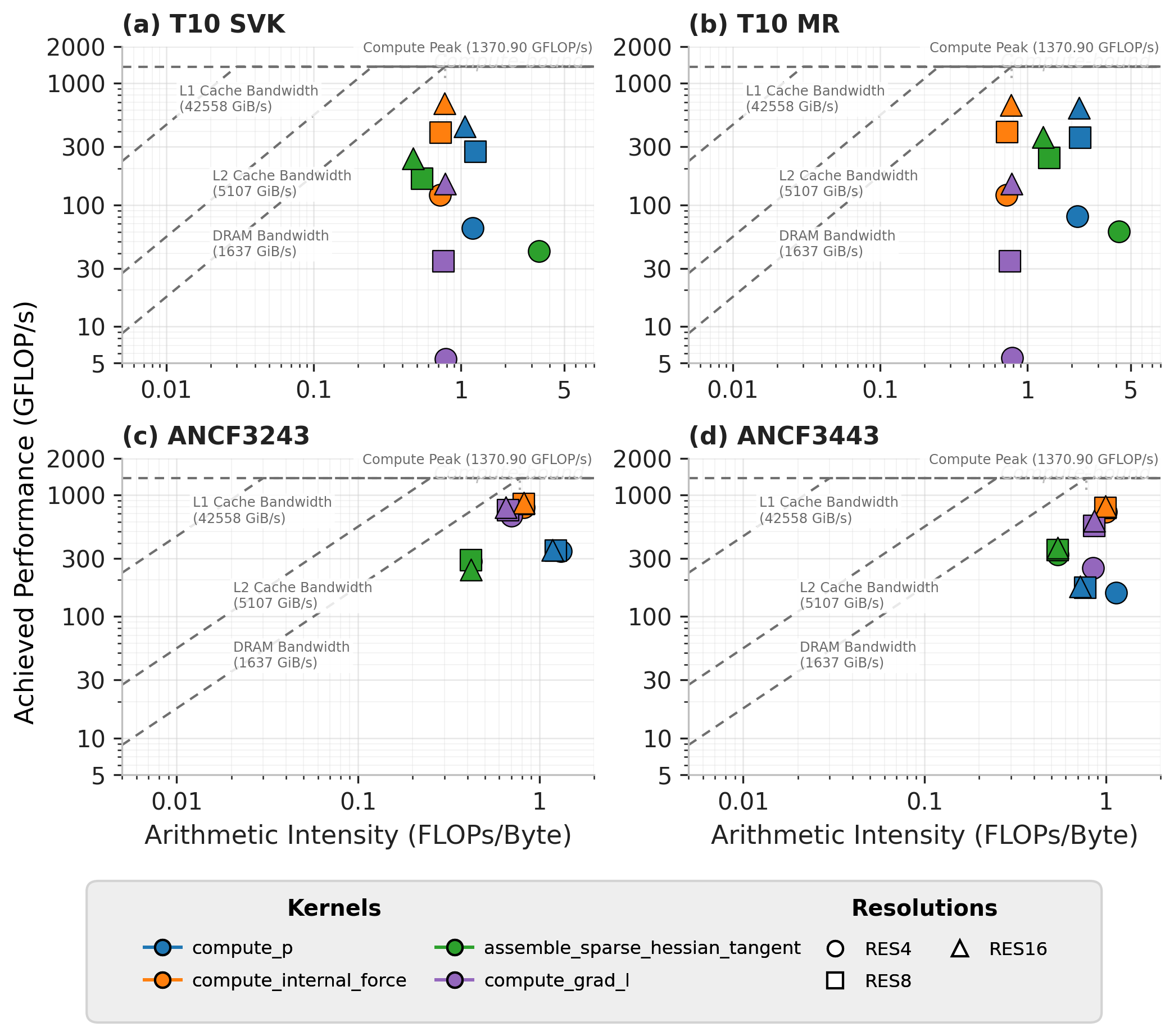}
  \caption{Roofline plots for the dominant kernels across all profiled
    element types and material models
    (RTX~5090; FP64 peak: 1{,}370.90 GFLOP/s;
    DRAM bandwidth: 1{,}637~GiB/s;
    ridge point: $\approx 0.78$ FLOP/byte).
    Each point corresponds to one kernel at one mesh resolution
    (RES4: circle, RES8: square, RES16: triangle).
    Dashed lines indicate the L1/L2/DRAM bandwidth ceilings and the FP64
    compute ceiling}
  \label{fig:roofline_combined}
\end{figure}

\begin{sloppypar}
Four GPU kernels are profiled using NVIDIA Nsight Compute on the RTX~5090 at RES4, RES8, and RES16: first Piola--Kirchhoff stress evaluation (\texttt{compute\_p}, Section~\ref{sec:parallel_fint}); internal-force assembly (\texttt{compute\_internal\_force}, Section~\ref{sec:parallel_fint}); sparse Hessian assembly (\texttt{assemble\_sparse\_hessian\_tangent}, Section~\ref{sec:hessian_assembly}); and augmented-Lagrangian gradient evaluation (\texttt{compute\_grad\_l}, Section~\ref{sec:gradient_eval}). Arithmetic intensity is defined as the ratio of executed FP64 operations to measured DRAM traffic, obtained from hardware performance counters~\cite{roofline2009}. For the RTX~5090, the FP64 peak is 1{,}370.90 GFLOP/s and the sustained DRAM bandwidth measured by Nsight is 1{,}637~GiB/s, yielding a ridge point of approximately 0.78 FLOP/byte. The resulting roofline plots are shown in Fig.~\ref{fig:roofline_combined}.
\end{sloppypar}

\paragraph{ANCF elements}
\begin{sloppypar}
For ANCF elements, \texttt{compute\_internal\_force} delivers the highest throughput among the kernels profiled. ANCF3243 reaches 788--856 GFLOP/s at an arithmetic intensity of approximately 0.82 FLOP/byte, placing it just to the right of the ridge point and at roughly 57--62\% of the FP64 peak. ANCF3443 reaches 725--802 GFLOP/s near 1.0 FLOP/byte, or roughly 53--59\% of peak. For context, Taylor~\cite{mikeANCF-implementationAspects2023} reports CPU arithmetic intensities of 0.106--0.136 FLOP/byte for ANCF3243 and 0.113--0.146 FLOP/byte for ANCF3443, together with achieved rates no higher than approximately 20 GFLOP/s for force evaluation and 40 GFLOP/s for Jacobian evaluation. These CPU numbers provide useful reference points, but they are not directly comparable to the GPU measurements reported here because the hardware, measurement methodology, and implementation details differ.

The gradient kernel \texttt{compute\_grad\_l} reaches 673--786 GFLOP/s for ANCF3243 at arithmetic intensities of approximately 0.65--0.70 FLOP/byte, placing it near the DRAM ceiling at those intensities. For ANCF3443, the corresponding throughput ranges from 250 to 610 GFLOP/s. In both ANCF families, throughput is relatively flat across RES4--RES16, indicating that even the coarsest ANCF problems considered here are already large enough to occupy the GPU effectively.

The main exception is \texttt{compute\_p} for ANCF3443, which achieves only 157--176 GFLOP/s despite arithmetic intensities in the range 0.72--1.14 FLOP/byte. This behavior is consistent with reduced occupancy caused by the elevated register pressure associated with the full three-dimensional shell kinematics. The Hessian kernel is more regular for ANCF, with arithmetic intensities of 0.42--0.54 FLOP/byte and throughputs of 241--367 GFLOP/s, likely because the larger element counts reduce per-entry atomic contention even at RES4.
\end{sloppypar}

\paragraph{T10 element}
\begin{sloppypar}
For the T10 element, \texttt{compute\_internal\_force} remains at a nearly constant arithmetic intensity of 0.73--0.74 FLOP/byte across all resolutions and both material models, placing it just below the ridge point and therefore in the bandwidth-bound regime. Its throughput increases from approximately 120 GFLOP/s at RES4 to approximately 695 GFLOP/s at RES16, consistent with improved SM occupancy as the problem size grows.

The stress-evaluation kernel \texttt{compute\_p} lies to the right of the ridge for all T10 cases (SVK: 1.0--1.3 FLOP/byte; MR: 2.2--2.3 FLOP/byte), placing it in the compute-bound region of the roofline. Nevertheless, it reaches only 430--606 GFLOP/s at RES16, or about 31--44\% of the FP64 peak, indicating that throughput is limited by occupancy rather than by raw arithmetic capability. The gradient kernel \texttt{compute\_grad\_l} achieves only about 5.5 GFLOP/s at T10 RES4; with 18{,}117 DOFs, the kernel launches too little work to occupy the 170 streaming multiprocessors of the RTX~5090 efficiently. Its throughput rises to approximately 150 GFLOP/s at RES16 as the problem size grows.

The Hessian kernel exhibits the clearest small-problem bottleneck. At T10 RES4 it has an arithmetic intensity of approximately 3.25 FLOP/byte, which places it well inside the compute-bound region, yet it achieves only about 41 GFLOP/s, i.e., roughly 3\% of the FP64 peak. This behavior is consistent with the use of atomic scatter operations into the CSR matrix, where many threads target a relatively small set of entries at only 2{,}013 elements. As the mesh is refined, contention becomes less severe and throughput rises to approximately 247 GFLOP/s at RES16.
\end{sloppypar}

\paragraph{Summary}
\begin{sloppypar}
The roofline analysis in Fig.~\ref{fig:roofline_combined} indicates that DRAM bandwidth is the main throughput limiter for most kernels in this framework. Kernels that lie in the nominally compute-bound region, most notably \texttt{compute\_p} and the T10 Hessian kernel at RES4, still remain well below the FP64 ceiling, consistent with additional limitations from occupancy loss and, in the case of Hessian assembly, atomic contention. These kernel-level observations are consistent with the end-to-end scaling trends reported earlier for the T10, ANCF3243, and ANCF3443 benchmarks, as well as for the Stanford Bunny case. In particular, the small-to-mid-resolution behavior of the T10 Newton runs is consistent with low occupancy and atomic contention in Hessian assembly, whereas the relatively flat ANCF scaling is consistent with internal-force kernels that already operate near the DRAM ceiling at coarse resolutions. Sparse Hessian assembly therefore appears to be the most promising optimization target. Element coloring, blockwise accumulation, or two-pass scatter/reduction strategies could reduce the atomic bottleneck, particularly for the smaller T10 problems, while the ANCF path already appears to be close to limits imposed by bandwidth and register pressure.
\end{sloppypar}

\section{Unit Tests}
\label{sec:experiments}

This section validates the frictional contact model and the bilateral constraint enforcement. The first two subsections cover contact; the third validates the joint constraint machinery formulated in Part~I~\cite{json-ganesh-danTLFEA-1-2026}, Section~2. The frictional contact model is validated through two complementary unit tests targeting distinct physical regimes. The first (Section~\ref{sec:brick_slope}) examines the quasi-static stick--slip transition of a block on an inclined plane, where the critical sliding angle admits a closed-form prediction from Coulomb theory. The second (Section~\ref{sec:oblique_impact}) targets dynamic oblique impact of a sphere against a flat surface under microgravity, where the tangential coefficient of restitution and post-impact angular velocity furnish quantitative comparisons against rigid-body analytical solutions across a systematic sweep of impact angles. Together, the two tests bracket the quasi-static and impulsive limits of frictional contact, providing complementary coverage of the implemented model.

\subsection{Brick Sliding on a Slope}
\label{sec:brick_slope}

We consider a rectangular prism (the ``brick'') resting on a flat surface tilted to a fixed slope angle $\alpha$. This configuration isolates the quasi-static stick--slip transition governed by Coulomb friction: for slope angles below the critical angle $\alpha_c$, the frictional force is sufficient to prevent sliding and the brick remains stationary; above $\alpha_c$, the brick accelerates down the slope under kinetic friction. The test thereby exercises both branches of the Coulomb friction law and verifies that the contact model correctly captures the stick--slip threshold and subsequent sliding kinematics.

For validation purposes, we compare against the rigid-body analytical solution, which provides a well-defined reference for the onset of sliding and the subsequent acceleration. Deformability introduces transient elastic oscillations, but these do not alter the steady-state sliding behavior. For a rigid block on an inclined plane, the critical angle at which sliding initiates satisfies:
\begin{equation}
    \alpha_c = \arctan(\mu_s),
\end{equation}
where $\mu_s$ is the static friction coefficient. Once sliding occurs, the acceleration of the center of mass (COM) along the slope is:
\begin{equation}
    a_{\mathrm{COM},\parallel} = g\left(\sin\alpha - \mu_k\cos\alpha\right),
\end{equation}
where $g$ is the gravitational acceleration and $\mu_k$ is the kinetic friction coefficient. If sliding initiates from rest at $t = 0$, the COM velocity and displacement along the slope follow $v_{\parallel}(t) = a_{\mathrm{COM},\parallel}\,t$ and $s_{\parallel}(t) = \tfrac{1}{2}a_{\mathrm{COM},\parallel}\,t^2$, respectively.

To bracket the stick--slip transition systematically, four slope configurations are tested using $\mu_k = 0.2$ and $\mu_s = 0.25$, yielding $\alpha_c = \arctan(0.25) \approx 0.2450\,\mathrm{rad}$: Slope~1 at $\alpha = 0.18\,\mathrm{rad}$, well below $\arctan(\mu_k)$ and firmly in the stick regime; Slope~2 at $\alpha = \arctan(\mu_k) \approx 0.1974\,\mathrm{rad}$, at the kinetic-friction threshold but still below $\alpha_c$ and therefore also in stick; Slope~3 at $\alpha = \alpha_c \approx 0.2450\,\mathrm{rad}$, the marginal case at the static threshold; and Slope~4 at $\alpha = 0.25\,\mathrm{rad}$, marginally above $\alpha_c$ and expected to slide. For Slope~4, the analytical prediction gives $a_{\mathrm{COM},\parallel} = 9.81\,(\sin 0.25 - 0.2\cos 0.25) \approx 0.526\,\mathrm{m/s^2}$, with Cartesian projections $a_x = -a_{\mathrm{COM},\parallel}\cos\alpha \approx -0.510\,\mathrm{m/s^2}$ and $a_z = -a_{\mathrm{COM},\parallel}\sin\alpha \approx -0.130\,\mathrm{m/s^2}$. The solver, contact, and material parameters are listed in Table~\ref{tab:brick_slope_params}.

\begin{table}[!htbp]
    \centering
    \footnotesize
    \setlength{\tabcolsep}{4pt}
    \renewcommand{\arraystretch}{0.9}
    \begin{tabular}{@{}lr@{}}
        \toprule
        Parameter & Value \\
        \midrule
        Time step, $\Delta t$ & $5\times10^{-4}\,\mathrm{s}$ \\
        Inner/outer abs. tolerance & $10^{-6}\,(-)$ \\
        ALM $\rho$ & $10^{12}\,(-)$ \\
        Max outer iterations & $3\,(-)$ \\
        Max inner iterations & $10\,(-)$ \\
        Gravity & $9.81\,\mathrm{m/s^2}$ \\
        Material Young's modulus, $E$ & $1\times10^{7}\,\mathrm{Pa}$ \\
        Coefficient of restitution, $e$ & $0.0\,(-)$ \\
        Static friction coefficient, $\mu_s$ & $0.25\,(-)$ \\
        Kinetic friction coefficient, $\mu_k$ & $0.2\,(-)$ \\
        Contact stiffness, $k_n^{\mathrm{area}}$ & $1\times10^{8}\,\mathrm{Pa/m}$ \\
        \bottomrule
    \end{tabular}
    \caption{Solver, contact, and material parameters for the brick sliding on a slope test}
    \label{tab:brick_slope_params}
\end{table}

\begin{figure}[t]
    \centering
    \maybeincludegraphics[width=\linewidth]{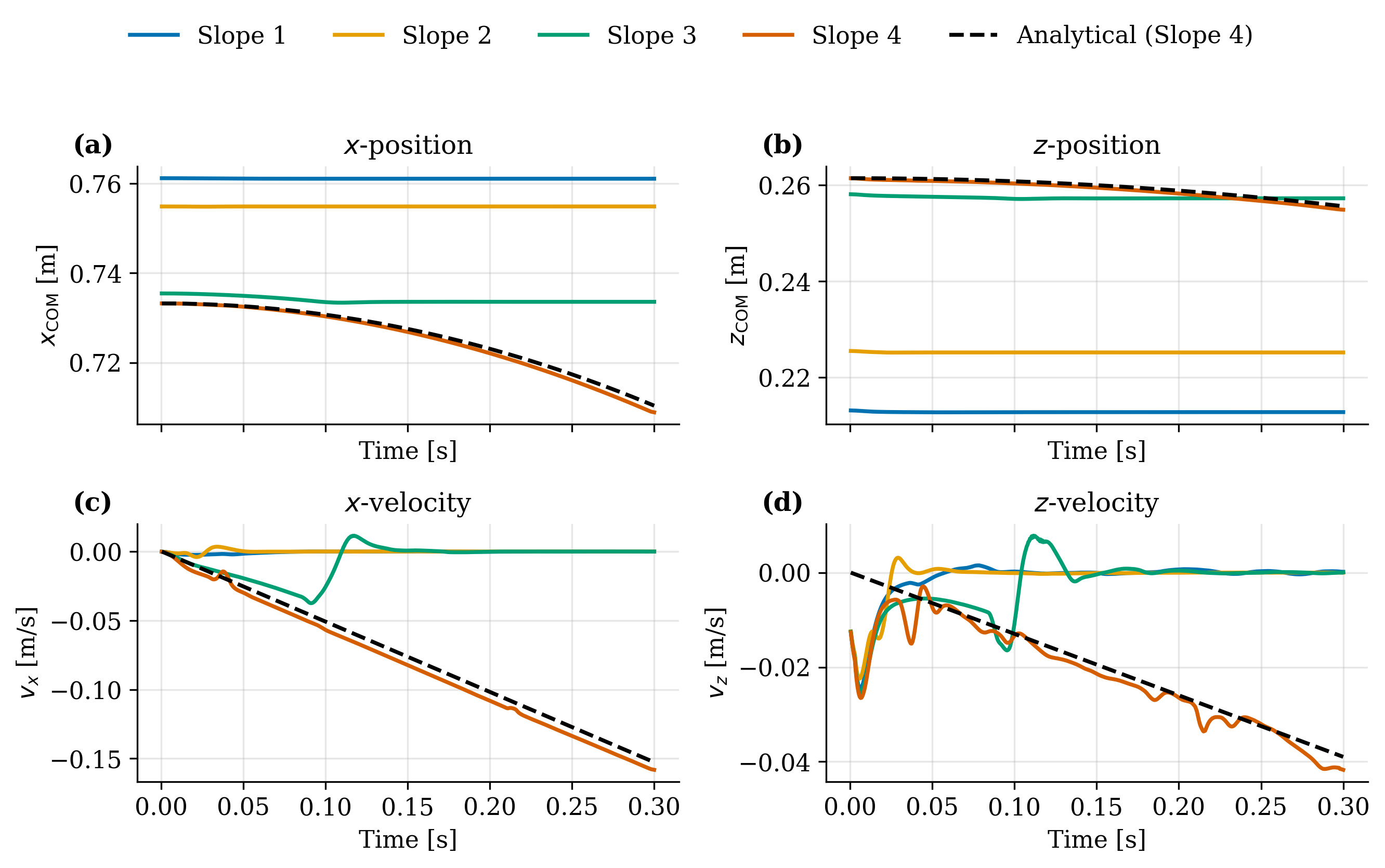}
    \caption{COM position and velocity histories for the brick-sliding validation test ($\mu_s = 0.25$, $\mu_k = 0.2$, $\alpha_c \approx 0.2450\,\mathrm{rad}$). Slope~1: $\alpha = 0.18\,\mathrm{rad}$ (stick); Slope~2: $\alpha \approx 0.1974\,\mathrm{rad} = \arctan(\mu_k)$ (stick); Slope~3: $\alpha = \alpha_c \approx 0.2450\,\mathrm{rad}$ (marginal, brief slide then arrest); Slope~4: $\alpha = 0.25\,\mathrm{rad}$ (continuous sliding). The dashed line is the rigid-body analytical prediction for Slope~4, with along-slope acceleration $a_{\mathrm{COM},\parallel} \approx 0.526\,\mathrm{m/s^2}$ projected onto the $x$- and $z$-axes. Panels~(a) and~(c) show position and velocity in the $x$-direction; panels~(b) and~(d) show the $z$-direction}
    \label{fig:brick_sliding_COM}
\end{figure}

Fig.~\ref{fig:brick_sliding_COM} shows the COM position and velocity histories for all four cases, with the rigid-body analytical prediction overlaid for Slope~4. Slopes~1 and~2 remain stationary throughout: $x_\mathrm{COM}$ and $z_\mathrm{COM}$ are flat in panels~(a) and~(b), and both velocity components stay at zero in panels~(c) and~(d), confirming correct enforcement of static friction below $\alpha_c$. Slope~3, at the marginal static threshold, exhibits a brief bidirectional velocity transient in panels~(c) and~(d) — including a momentary positive-$v_x$ excursion as the elastic body rebounds against the contact — before arresting, a consequence of elastic wave dynamics that locally redistribute contact forces below the sustained kinetic threshold. Slope~4 undergoes continuous sliding: panels~(a) and~(c) show $x_\mathrm{COM}$ and $v_x$ tracking the analytical parabola and linear ramp closely after the initial elastic settling, confirming that the inferred along-slope acceleration is consistent with the predicted $0.526\,\mathrm{m/s^2}$. The $z$-velocity in panel~(d) converges to the analytical slope after $t \approx 0.15\,\mathrm{s}$, with the early oscillations attributable to elastic wave reflections in the deformable brick.

\subsection{Oblique Impact}
\label{sec:oblique_impact}

\begin{figure}[t]
    \centering
    \maybeincludegraphics[width=0.7\linewidth]{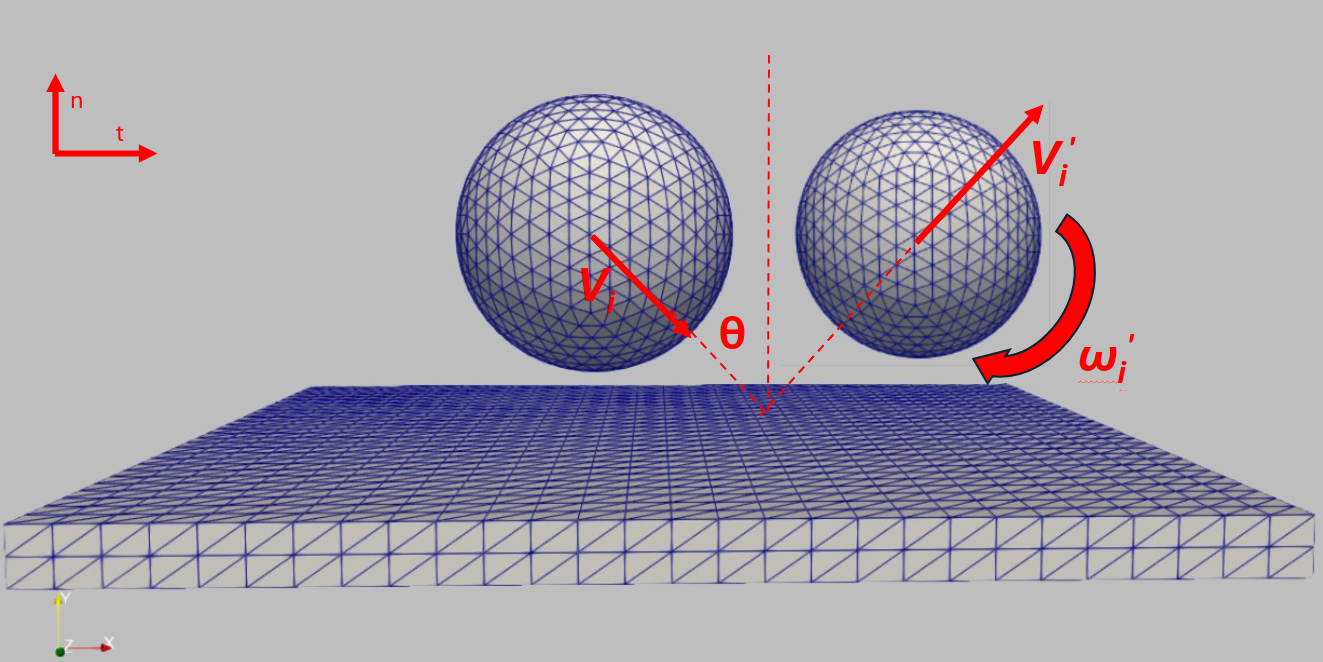}
    \caption{Oblique impact configuration. A sphere impacts a flat surface at impact angle $\theta$ with initial velocity $\mathbf{v}_i$; following rebound, the post-collision velocity $\mathbf{v}'_i$ and angular velocity $\omega'_i$ are recorded. The surface-normal ($n$) and tangential ($t$) directions are defined relative to the contact surface}
    \label{fig:oblique_setup}
\end{figure}

To validate the contact model in the dynamic impulsive regime, we perform a series of oblique impact simulations under microgravity conditions. A sphere is launched with an initial velocity $\mathbf{v}_i = \mathbf{v}_{i,n} + \mathbf{v}_{i,t}$ directed at an impact angle $\theta$ to the contact surface, where $n$ and $t$ denote the surface-normal and tangential directions, respectively. Following impact and rebound, the post-collision linear velocity $\mathbf{v}'_i = \mathbf{v}'_{i,n} + \mathbf{v}'_{i,t}$ and angular velocity $\omega'_i$ are recorded. A visualization of the configuration is shown in Fig.~\ref{fig:oblique_setup}.

The primary observable for assessing tangential contact fidelity is the tangential coefficient of restitution (COR), defined as the ratio of post- to pre-collision tangential speed:

\begin{equation}
    e_t = \frac{|\mathbf{v}'_{i,t}|}{|\mathbf{v}_{i,t}|}
    \label{eq:tangential_cor}
\end{equation}

When the contact is in the gross sliding regime ($|F_t| = \mu|F_n|$), the post-collision kinematics admit a closed-form solution from rigid-body theory~\cite{wu2003coefficients}. The theoretical tangential COR and post-impact angular velocity are, respectively:

\begin{equation}
    e_t = 1 - \frac{\mu(1+e)}{\tan\theta},
    \label{eq:et_theory_sliding}
\end{equation}
\begin{equation}
    |\omega'_i| = \frac{5}{2}\frac{\mu(1+e)|v_{i,n}|}{R}.
    \label{eq:omega_post_sliding}
\end{equation}

The transition between the sliding and sticking regimes is governed by a critical impact angle~\cite{yu2017impulse}:

\begin{equation}
    \theta^* = \arctan \left[ \frac{7}{2} \mu_s (1 + e) \right].
    \label{eq:critical_impact_angle}
\end{equation}

For $\theta > \theta^*$ the contact remains in sustained sliding throughout the collision and Eqs.~\eqref{eq:et_theory_sliding}--\eqref{eq:omega_post_sliding} apply; for $\theta < \theta^*$ the contact enters the sticking regime and these expressions no longer hold. In both experiments a single friction coefficient $\mu$ is used (no static/kinetic distinction), so $\mu_s = \mu$ in Eq.~\eqref{eq:critical_impact_angle}.

The sphere mesh comprises 5883 vertices and 3438 tetrahedral elements, with a radius of $0.15\,\mathrm{m}$. The floor is discretized by a tetrahedral mesh of 12\,645 vertices and 9408 elements. The Newton solver parameters are summarized in Table~\ref{tab:oblique_impact_solver_params}.

\begin{table}[!htbp]
    \centering
    \footnotesize
    \setlength{\tabcolsep}{4pt}
    \renewcommand{\arraystretch}{0.9}
    \begin{tabular}{@{}lr@{}}
        \toprule
        Parameter & Value \\
        \midrule
        Time step, $\Delta t$ & $1\times 10^{-4}\,\mathrm{s}$ \\
        Inner/outer abs. tolerance & $10^{-6}$ \\
        ALM $\rho$ & $10^{12}$ \\
        Max outer iterations & $3$ \\
        Max inner iterations & $10$ \\
        Gravity & $0\,\mathrm{m/s^2}$ \\
        \bottomrule
    \end{tabular}
    \caption{Solver parameters for the oblique impact test}
    \label{tab:oblique_impact_solver_params}
\end{table}

\begin{figure}[t]
    \centering
    \maybeincludegraphics[width=\linewidth]{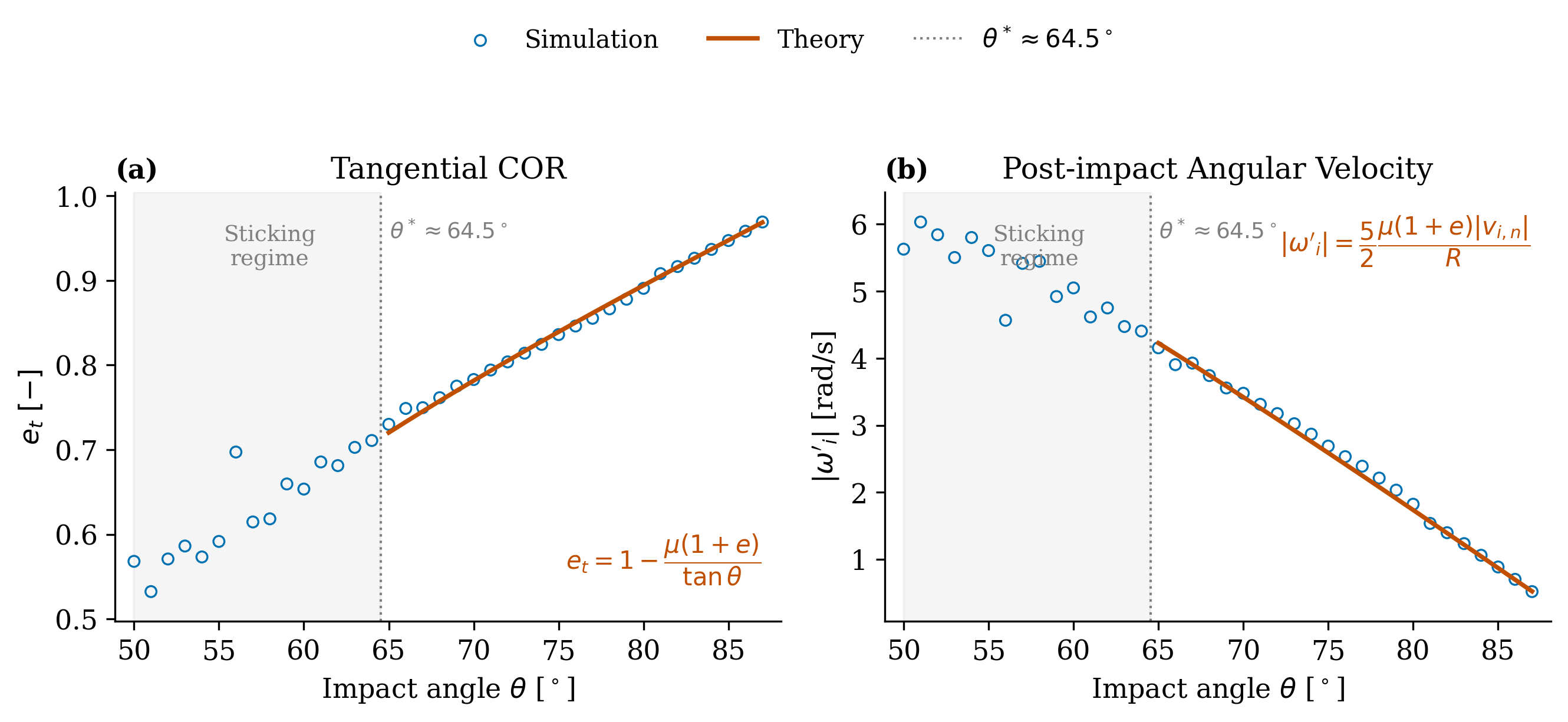}
    \caption{Oblique impact results for Experiment~1 ($\mu = 0.3$, $e = 1.0$, $\theta^* \approx 64.5^\circ$). Panel~(a): tangential COR $e_t$ vs.\ impact angle $\theta$. Panel~(b): post-impact angular velocity $|\omega'_i|$ vs.\ $\theta$. Solid line: rigid-body analytical prediction, valid for $\theta > \theta^*$ (sliding regime). Shaded region: sticking regime where the analytical expressions do not apply. Simulated $|\omega'_i|$ in panel~(b) lies slightly above the rigid-body prediction due to elastic energy released as additional spin upon rebound}
    \label{fig:oblique_plots_exp1}
\end{figure}

\begin{figure}[t]
    \centering
    \maybeincludegraphics[width=\linewidth]{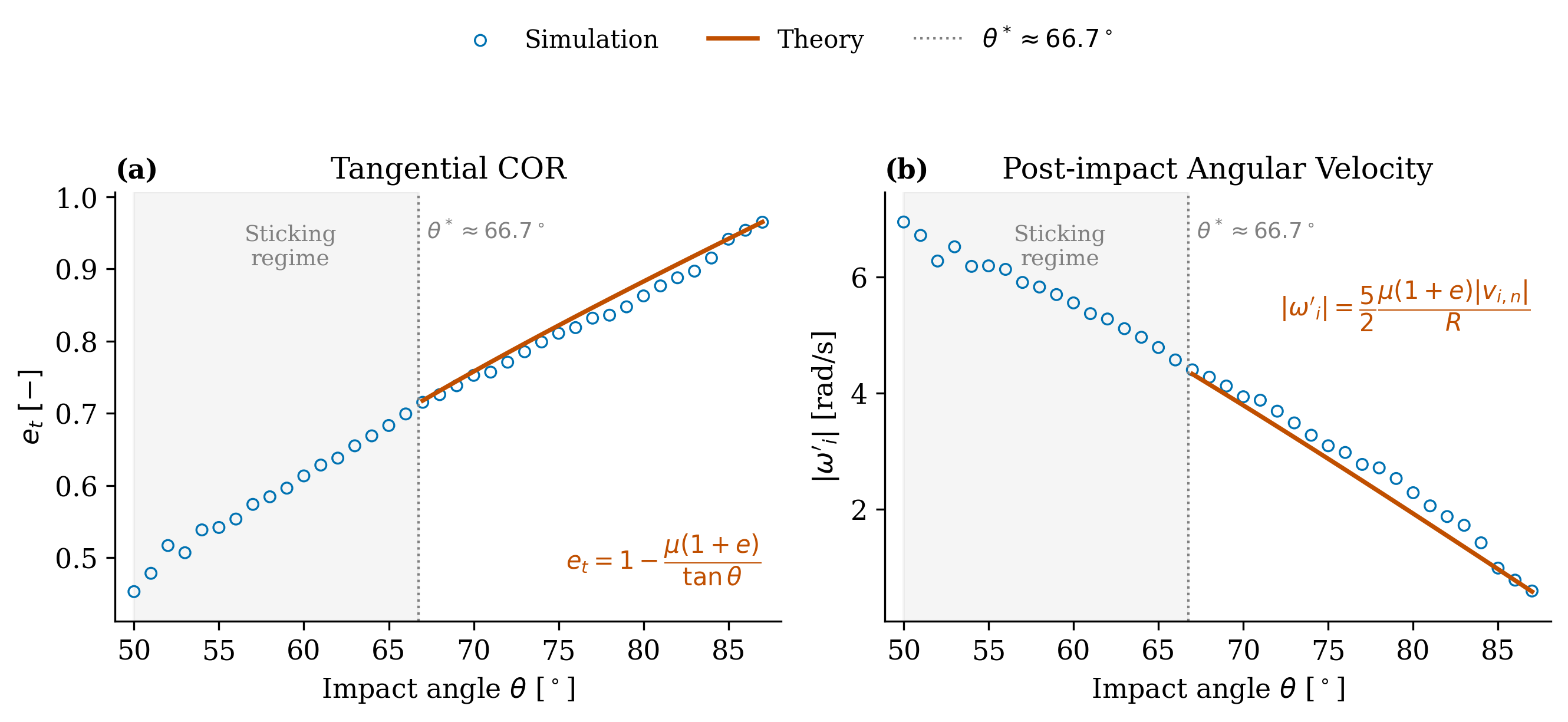}
    \caption{Oblique impact results for Experiment~2 ($\mu = 0.35$, $e = 0.9$, $\theta^* \approx 66.7^\circ$). Panel~(a): tangential COR $e_t$ vs.\ impact angle $\theta$. Panel~(b): post-impact angular velocity $|\omega'_i|$ vs.\ $\theta$. Higher material damping suppresses elastic rebound effects; simulated values agree with the rigid-body analytical prediction to within data scatter for $\theta > \theta^*$}
    \label{fig:oblique_plots_exp2}
\end{figure}

Two parameter sweeps are performed with impact angle $\theta$ ranging from $50^\circ$ to $87^\circ$ in $1^\circ$ increments; the material and contact properties are given in Table~\ref{tab:oblique_impact_material_props}. The density $\rho = 70.7355\,\mathrm{kg/m^3}$ is chosen such that the discretized sphere mesh has a total mass of $1\,\mathrm{kg}$. Results for experiments~1 and~2 are shown in Figs.~\ref{fig:oblique_plots_exp1} and~\ref{fig:oblique_plots_exp2}, respectively, with the rigid-body analytical predictions of Eqs.~\eqref{eq:et_theory_sliding}--\eqref{eq:omega_post_sliding} overlaid for $\theta > \theta^*$. The shaded region marks the sticking regime ($\theta < \theta^*$) where the analytical expressions do not apply.

In Experiment~1 ($\mu = 0.3$, $e = 1.0$, $\theta^* \approx 64.5^\circ$), the tangential COR in panel~(a) rises from approximately $0.45$ at $\theta = 50^\circ$ toward unity as $\theta \to 87^\circ$, and follows the analytical prediction closely for $\theta > \theta^*$. The post-impact angular velocity in panel~(b) decreases monotonically with $\theta$ as expected from Eq.~\eqref{eq:omega_post_sliding}, but the simulated values lie systematically above the rigid-body prediction; this offset reflects elastic energy stored in the deformable sphere during contact and partially released as additional spin upon rebound, an effect absent from the rigid-body model.

In Experiment~2 ($\mu = 0.35$, $e = 0.9$, $\theta^* \approx 66.7^\circ$), the higher damping ($\eta_{\mathrm{damp}} = \lambda_{\mathrm{damp}} = 5\times10^2$\,Pa$\cdot$s) suppresses elastic oscillations, and both $e_t$ and $|\omega'_i|$ agree with the analytical predictions to within the scatter of the data for $\theta > \theta^*$. This near-exact agreement constitutes the stronger validation result: once material damping is sufficient to remove elastic rebound effects, the frictional contact model recovers the rigid-body impulse response with high fidelity.

\begin{table}[!htbp]
    \centering
    \footnotesize
    \setlength{\tabcolsep}{4pt}
    \renewcommand{\arraystretch}{0.9}
    \begin{tabular}{@{}lcc@{}}
        \toprule
        Parameter & Experiment 1 & Experiment 2 \\
        \midrule
        Material model & SVK & SVK \\
        Young's modulus, $E$ & $1\times 10^{7}$ & $1\times 10^{7}$ \\
        Poisson's ratio, $\nu$ & $0.3$ & $0.3$ \\
        Density, $\rho$ & $70.7355$ & $70.7355$ \\
        Damping, $\eta_{\mathrm{damp}}$ & $5\times 10^{1}$ & $5\times 10^{2}$ \\
        Damping, $\lambda_{\mathrm{damp}}$ & $5\times 10^{1}$ & $5\times 10^{2}$ \\
        Contact friction, $\mu$ & $0.3$ & $0.35$ \\
        Normal COR, $e$ & $1.0$ & $0.9$ \\
        \bottomrule
    \end{tabular}
    \caption{Material and contact properties for the oblique impact parameter sweeps}
    \label{tab:oblique_impact_material_props}
\end{table}

\subsection{Joint Constraint Validation}
\label{sec:joint_validation}

The bilateral constraint enforcement described in the Augmented Lagrangian Formulation section is validated through double-pendulum tests using revolute and spherical joints. Each joint is composed of coordinate-difference (CD) and dot-product (DP1) primitives as defined in Part~I~\cite{json-ganesh-danTLFEA-1-2026}: a revolute joint enforces three CD constraints (point coincidence) and two DP1 constraints (axis alignment), while a spherical joint enforces only three CD constraints. The tests below verify constraint residual magnitudes, consistency of recovered joint forces with analytical quasi-static predictions, and dynamic plausibility against a Project Chrono~\cite{projectChronoWebSite} rigid-body reference. Simulation and material parameters are collected in Appendix~\ref{sec:appendix_benchmark_params}.

\subsubsection{Revolute Joint: Two-Link Pendulum}
\label{sec:rev_pendulum_motion}

Two identical T10 deformable beams ($0.5\,\mathrm{m}\times 0.04\,\mathrm{m}\times 0.04\,\mathrm{m}$, SVK material) are chained by two revolute joints whose hinge axes are aligned with the global $y$-direction, restricting the motion predominantly to the $x$--$z$ plane. The upper beam is anchored to the world frame at $\mathbf{p}_{\mathrm{top}} = (0,\,0,\,0.7)\,\mathrm{m}$; the initial configuration places the upper and lower links at $35^\circ$ and $-25^\circ$ from the downward vertical, respectively. Four Kelvin--Voigt damping levels ($\eta_{\mathrm{damp}} = \lambda_{\mathrm{damp}} \in \{0,\,10^2,\,10^3,\,10^4\}$\,Pa$\cdot$s) are swept while all other parameters are held fixed. Each case is advanced for 5000 steps at $\Delta t = 5\times10^{-4}\,\mathrm{s}$ (2.5\,s total).

\begin{figure}[t]
    \centering
    \maybeincludegraphics[width=\linewidth]{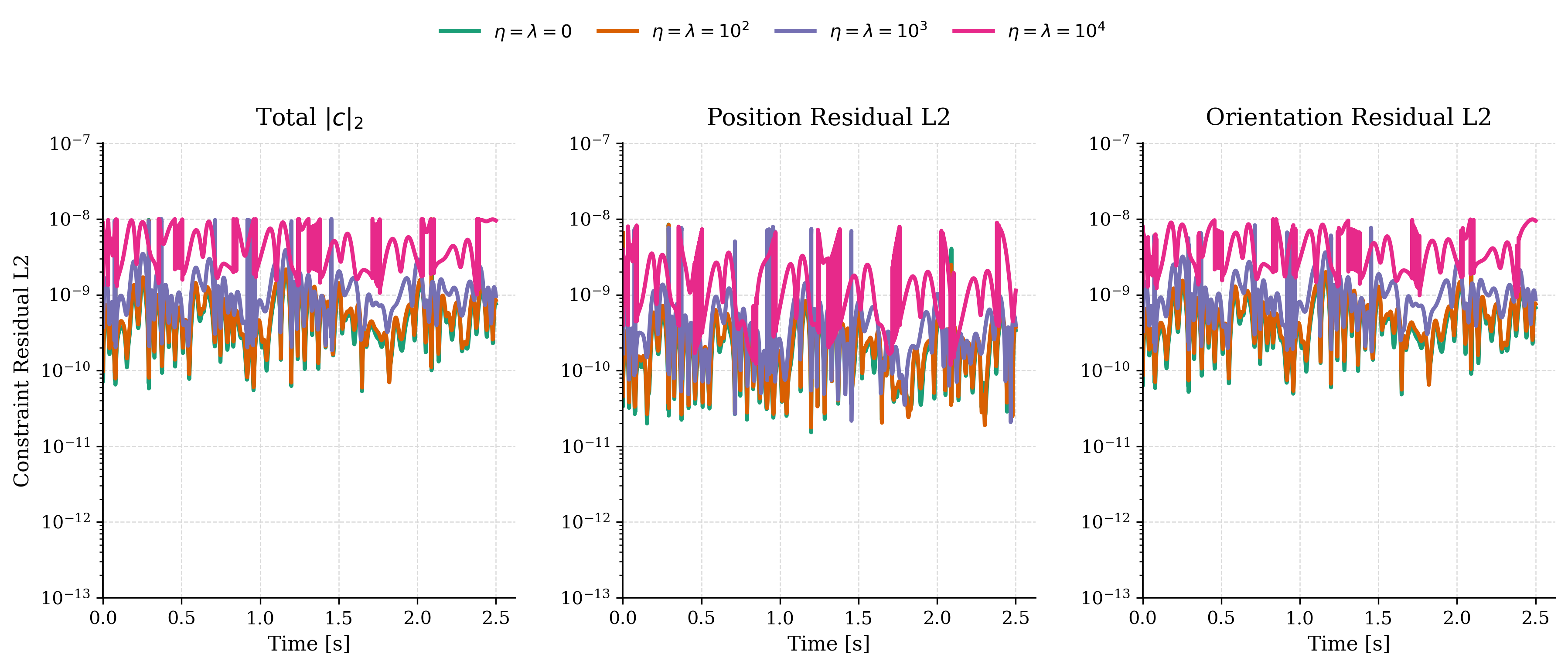}
    \caption{Revolute-joint constraint residual histories for the double-pendulum motion test under four damping levels ($\eta_{\mathrm{damp}} = \lambda_{\mathrm{damp}} \in \{0,10^2,10^3,10^4\}$\,Pa$\cdot$s). Left: total residual $\|\mathbf{c}\|_2$. Center: position residual (CD rows). Right: orientation residual (DP1 rows). All components remain within $10^{-10}$--$10^{-8}$ with no drift over 2.5\,s}
    \label{fig:rev_pend_constraint_summary}
\end{figure}

\begin{figure}[t]
    \centering
    \maybeincludegraphics[width=\linewidth]{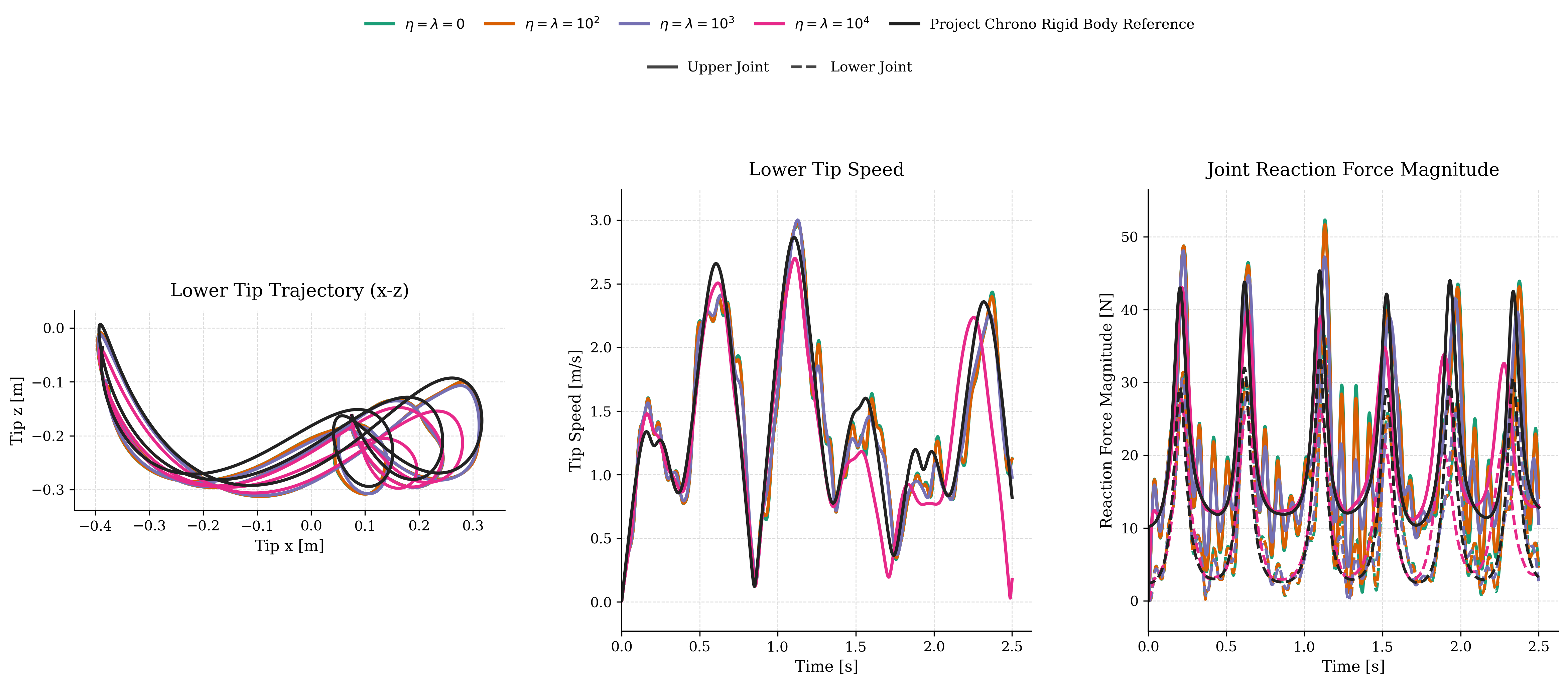}
    \caption{Motion and reaction summary for the revolute-joint double-pendulum under four damping configurations, with a Project Chrono rigid-body reference. Left: lower-tip trajectory in the $x$--$z$ plane. Center: lower-tip speed histories. Right: joint reaction-force magnitudes at the upper (solid) and lower (dashed) revolute joints}
    \label{fig:rev_pend_motion_summary}
\end{figure}

Fig.~\ref{fig:rev_pend_constraint_summary} shows the revolute-joint constraint residuals. The total, position, and orientation components remain bounded within $10^{-10}$--$10^{-8}$ throughout the simulation with no cumulative drift, confirming that the ALM outer loop drives hinge-point coincidence (CD rows) and revolute-axis alignment (DP1 rows) to negligible levels across all damping configurations. Fig.~\ref{fig:rev_pend_motion_summary} presents the lower-tip trajectory, tip speed, and recovered joint reaction-force magnitudes against the Project Chrono rigid-body reference. The lightly damped flexible cases track the rigid-body trajectories closely; deviations grow with increasing damping as the motion envelope shrinks. The upper-joint reaction consistently exceeds the lower-joint reaction, as expected because the upper joint transmits the combined inertial and gravitational load of both links.

\subsubsection{Revolute Joint: Vertical Force Recovery}
\label{sec:rev_pendulum_pull}

\begin{figure}[t]
    \centering
    \maybeincludegraphics[width=\linewidth]{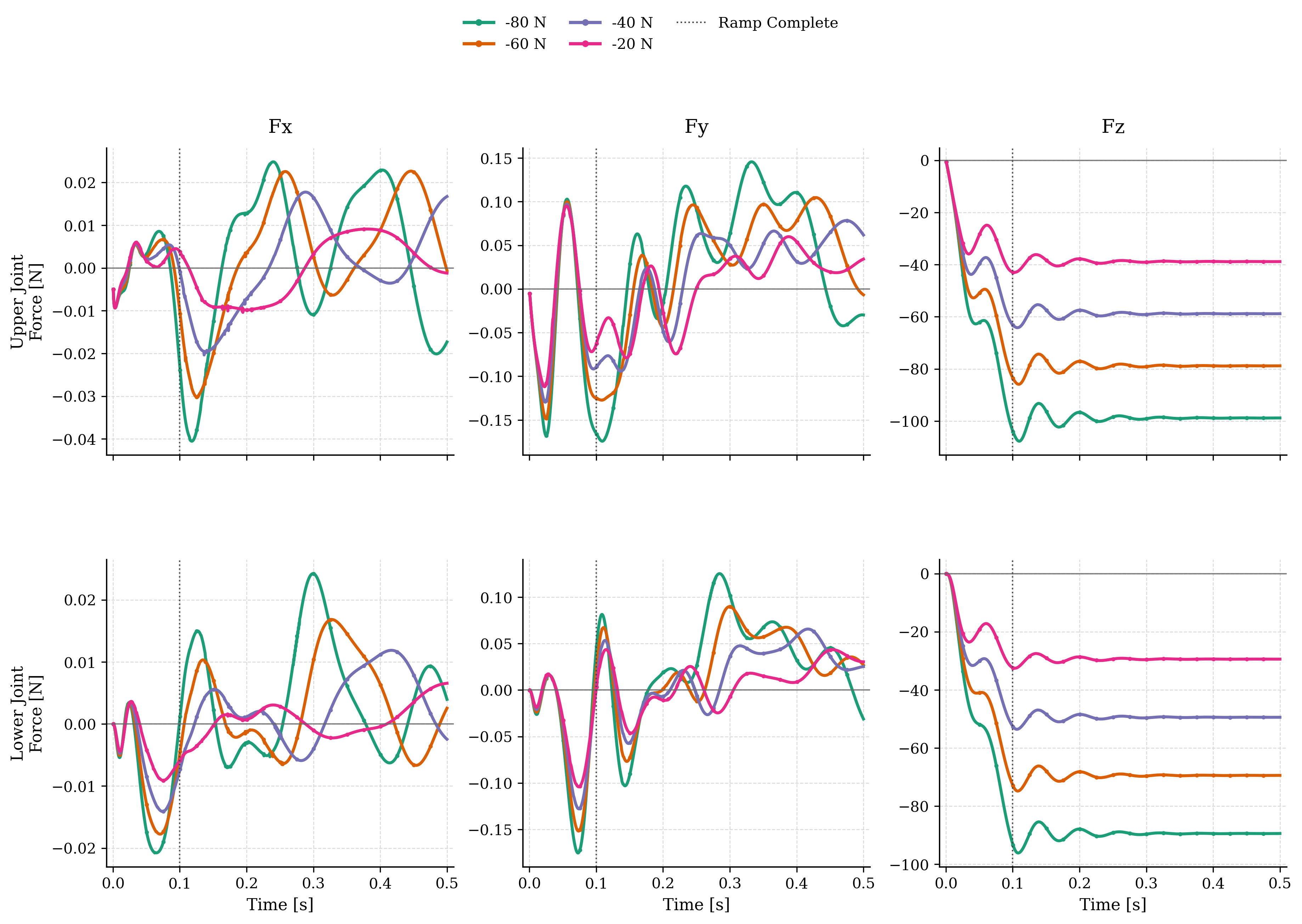}\\[0.5ex]
    \maybeincludegraphics[width=\linewidth]{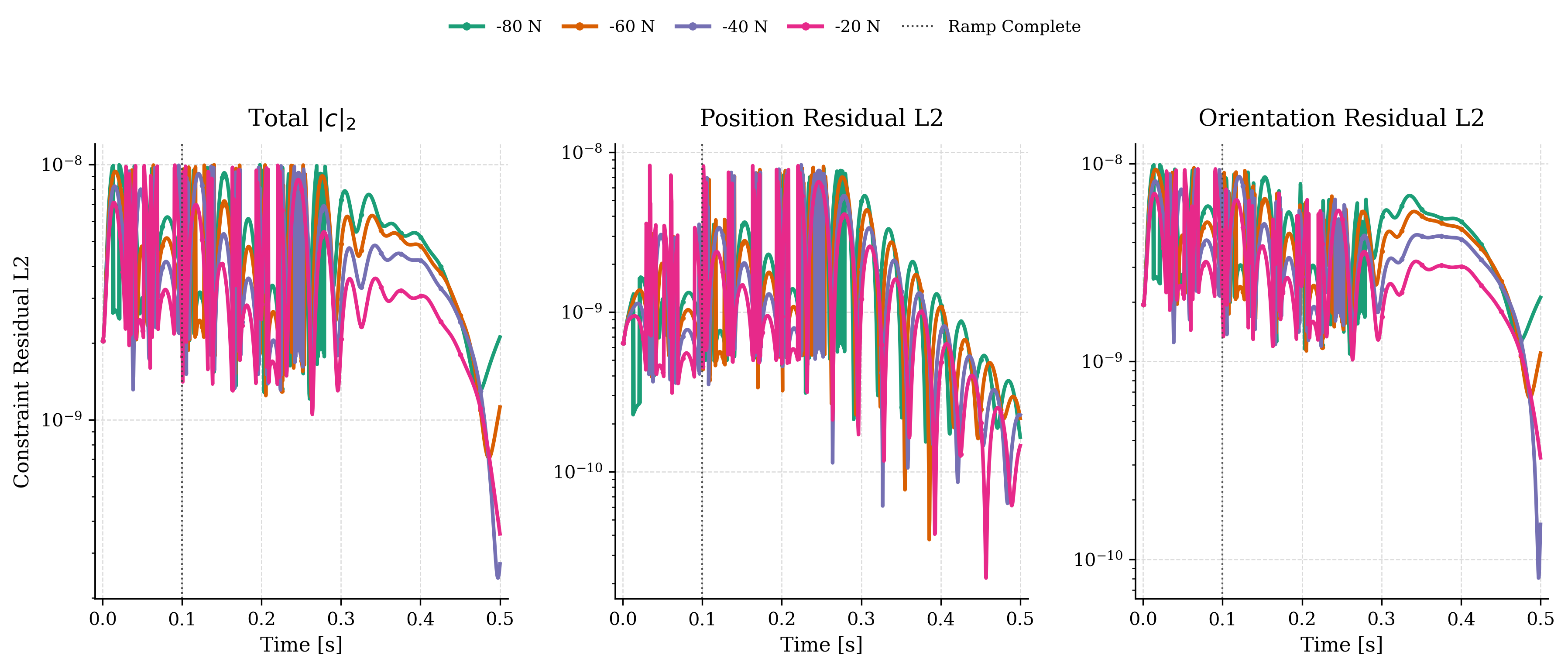}
    \caption{Revolute-joint vertical pulling test for four loading cases ($-20$, $-40$, $-60$, $-80\,\mathrm{N}$). Top: Cartesian reaction-force histories at the upper (top row) and lower (bottom row) joints; $F_z$ dominates while $F_x$, $F_y$ remain near zero. Bottom: constraint violation histories; all residuals remain below $10^{-8}$}
    \label{fig:rev_pend_pull_results}
\end{figure}

The same two-link geometry is placed in a vertical hanging configuration and subjected to four downward distributed loads ($-20$, $-40$, $-60$, $-80\,\mathrm{N}$) applied to the lower tip region and ramped linearly over the first 200 steps. Under quasi-static vertical loading the expected joint reactions are: lower joint $F_z = -(m_b g + |F_{\mathrm{pull}}|)$, upper joint $F_z = -(2\,m_b g + |F_{\mathrm{pull}}|)$, with beam mass $m_b = 0.96\,\mathrm{kg}$. Fig.~\ref{fig:rev_pend_pull_results} shows the Cartesian reaction-force and constraint-violation histories; $F_z$ dominates in all cases while $F_x$ and $F_y$ remain near zero, consistent with the vertical loading and boundary conditions. Constraint residuals remain below $10^{-8}$ throughout. Table~\ref{tab:rev_pend_pull_validation} compares the simulated reactions at step~1{,}000 against the analytical reference; the errors grow modestly with increasing applied load due to residual transient inertia in the flexible simulation but remain small relative to the total support force.

\begin{table}[!htbp]
    \centering
    \footnotesize
    \setlength{\tabcolsep}{4pt}
    \renewcommand{\arraystretch}{0.9}
    \begin{tabular}{@{}r rr r rr r@{}}
        \toprule
        Pull & \multicolumn{3}{c}{Upper Joint $F_z$ (N)} & \multicolumn{3}{c}{Lower Joint $F_z$ (N)} \\
        \cmidrule(lr){2-4} \cmidrule(lr){5-7}
        (N) & Simulated & Expected & Rel.\ Err.\ (\%) & Simulated & Expected & Rel.\ Err.\ (\%) \\
        \midrule
        $-20$ & $-37.78$ & $-38.84$ & $2.73$ & $-28.66$ & $-29.42$ & $2.58$ \\
        $-40$ & $-57.47$ & $-58.84$ & $2.33$ & $-48.44$ & $-49.42$ & $1.98$ \\
        $-60$ & $-77.08$ & $-78.84$ & $2.23$ & $-68.16$ & $-69.42$ & $1.82$ \\
        $-80$ & $-96.66$ & $-98.84$ & $2.21$ & $-87.84$ & $-89.42$ & $1.77$ \\
        \bottomrule
    \end{tabular}
    \caption{Simulated vs.\ expected vertical joint reactions at step~1{,}000 for the revolute-joint vertical pulling test. Expected values: lower joint $F_z = -(m_b g + |F_{\mathrm{pull}}|)$, upper joint $F_z = -(2\,m_b g + |F_{\mathrm{pull}}|)$, with $m_b = 0.96\,\mathrm{kg}$ and $g = 9.81\,\mathrm{m/s^2}$. Relative error $= |F_{\mathrm{sim}} - F_{\mathrm{exp}}| / |F_{\mathrm{exp}}| \times 100\%$}
    \label{tab:rev_pend_pull_validation}
\end{table}

\subsubsection{Spherical Joint: Double Pendulum}
\label{sec:sph_pendulum}

\begin{figure}[t]
    \centering
    \maybeincludegraphics[width=\linewidth]{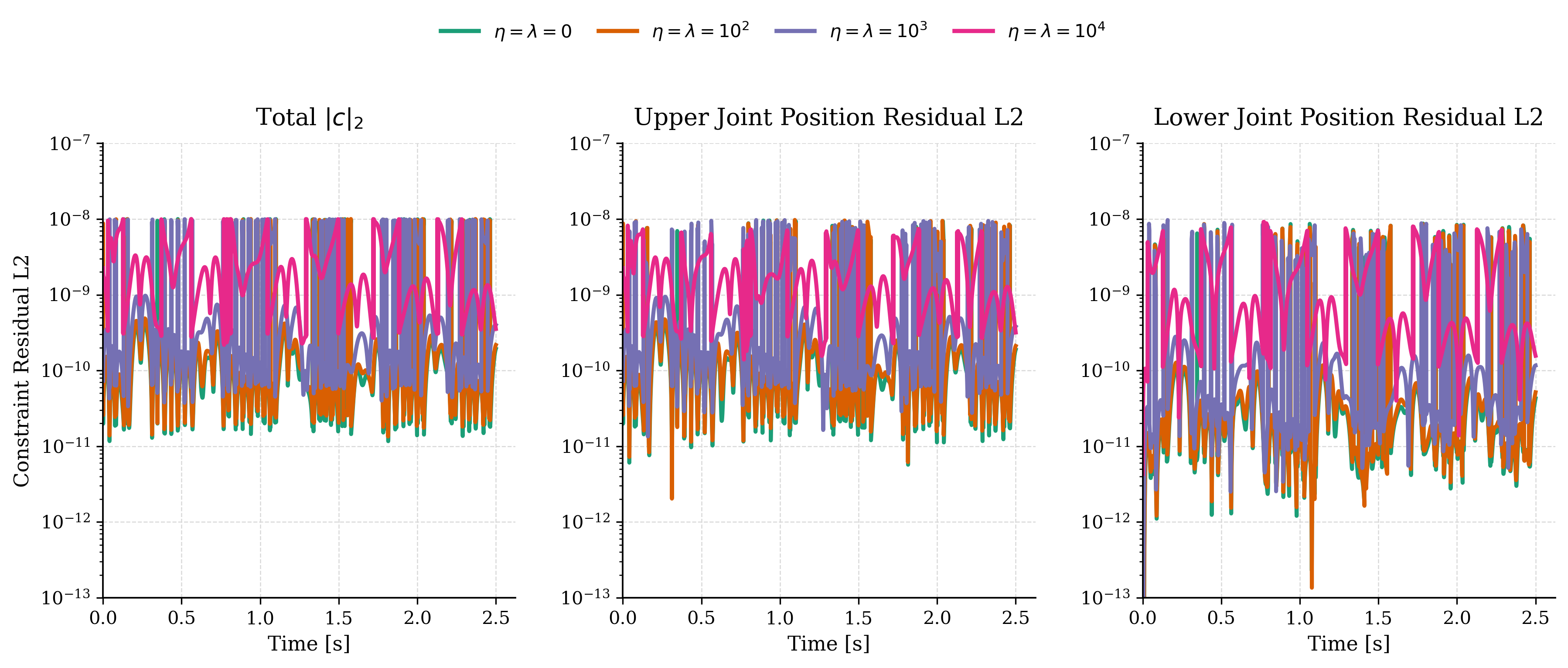}
    \caption{Spherical-joint constraint residual histories for the double-pendulum motion test under four damping levels. Left: total residual $\|\mathbf{c}\|_2$. Center: upper-joint position residual. Right: lower-joint position residual. Residuals remain within $10^{-11}$--$10^{-8}$ with no cumulative drift}
    \label{fig:sph_pend_constraint_summary}
\end{figure}

\begin{figure}[t]
    \centering
    \maybeincludegraphics[width=\linewidth]{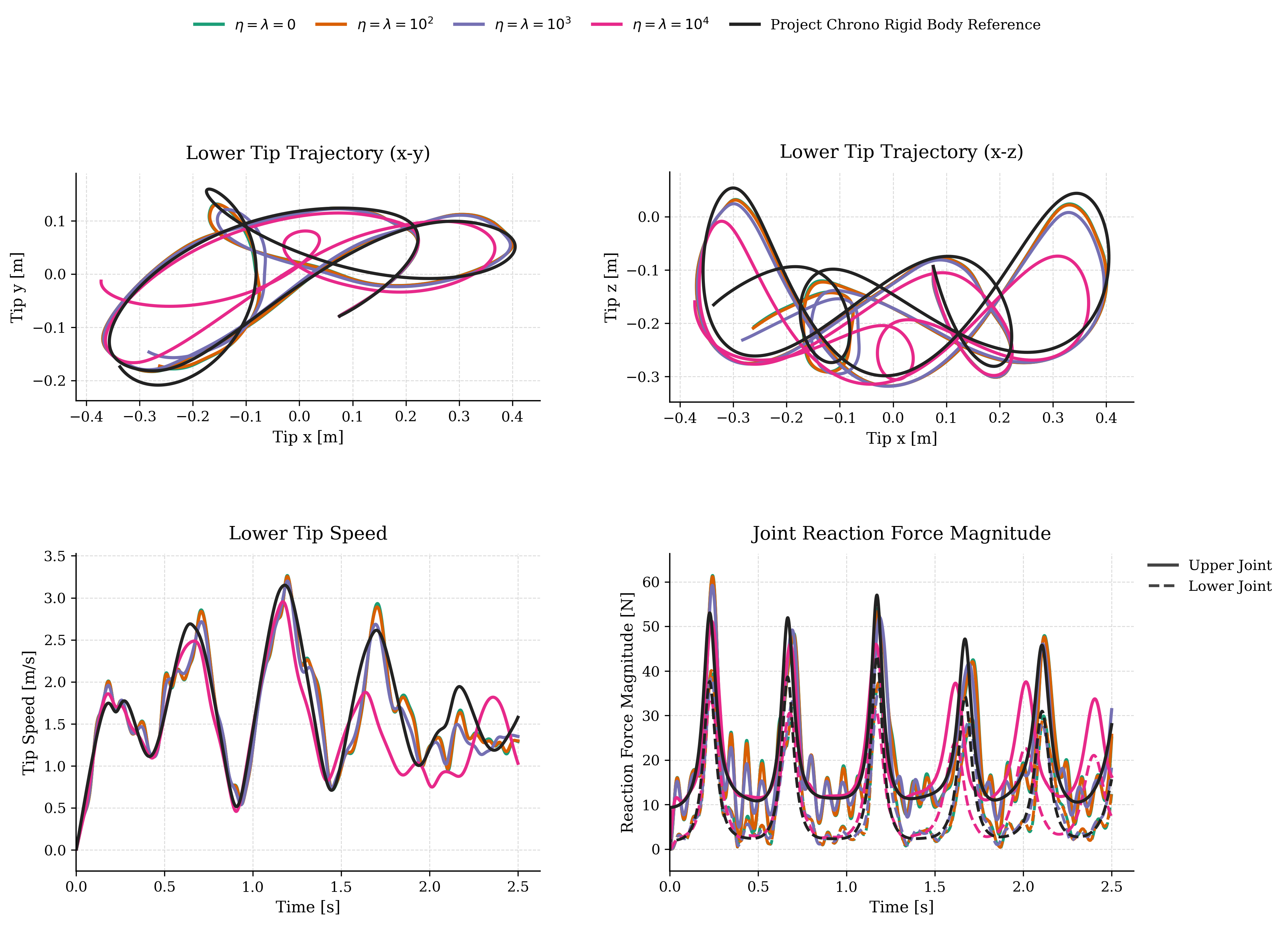}
    \caption{Motion and reaction summary for the spherical-joint double-pendulum under four damping configurations, compared against a Project Chrono ball-joint reference. Top: lower-tip trajectories in the $x$--$y$ and $x$--$z$ planes. Bottom left: lower-tip speed histories. Bottom right: joint reaction-force magnitudes at the upper (solid) and lower (dashed) spherical joints}
    \label{fig:sph_pend_motion_summary}
\end{figure}

The same protocol is applied with spherical joints (three CD constraints per joint, no DP1). The motion test uses a fully three-dimensional initial configuration: the upper and lower links are rotated $35^\circ$/$-25^\circ$ in the pendulum plane and additionally tilted $18^\circ$/$-27^\circ$ out of plane, so that all three translational degrees of freedom at each joint are exercised. Figs.~\ref{fig:sph_pend_constraint_summary} and~\ref{fig:sph_pend_motion_summary} show the constraint residuals and lower-tip trajectories under four damping levels, compared against a Project Chrono ball-joint reference. Position residuals remain within $10^{-11}$--$10^{-8}$ with no drift, and the flexible trajectories exhibit the expected three-dimensional precession absent from the planar revolute case. For the vertical force-recovery test (same four loading cases as the revolute variant), Table~\ref{tab:sph_pend_pull_validation} shows that the recovered reactions agree with the analytical predictions to within 0.10\% -- an order of magnitude better than the revolute case owing to the absence of orientation-constraint coupling.

\begin{table}[!htbp]
    \centering
    \footnotesize
    \setlength{\tabcolsep}{4pt}
    \renewcommand{\arraystretch}{0.9}
    \begin{tabular}{@{}r rr r rr r@{}}
        \toprule
        Pull & \multicolumn{3}{c}{Upper Joint $F_z$ (N)} & \multicolumn{3}{c}{Lower Joint $F_z$ (N)} \\
        \cmidrule(lr){2-4} \cmidrule(lr){5-7}
        (N) & Simulated & Expected & Rel.\ Err.\ (\%) & Simulated & Expected & Rel.\ Err.\ (\%) \\
        \midrule
        $-20$ & $-38.88$ & $-38.84$ & $0.10$ & $-29.45$ & $-29.42$ & $0.10$ \\
        $-40$ & $-58.84$ & $-58.84$ & $0.02$ & $-49.42$ & $-49.42$ & $0.00$ \\
        $-60$ & $-78.81$ & $-78.84$ & $0.04$ & $-69.40$ & $-69.42$ & $0.03$ \\
        $-80$ & $-98.80$ & $-98.84$ & $0.03$ & $-89.39$ & $-89.42$ & $0.02$ \\
        \bottomrule
    \end{tabular}
    \caption{Simulated vs.\ expected vertical joint reactions at step~1{,}000 for the spherical-joint vertical pulling test. Expected values use the same static equilibrium formula as Table~\ref{tab:rev_pend_pull_validation}. Relative error $= |F_{\mathrm{sim}} - F_{\mathrm{exp}}| / |F_{\mathrm{exp}}| \times 100\%$}
    \label{tab:sph_pend_pull_validation}
\end{table}

\section{Large-Scale Tests}
\label{sec:large_scale}

\subsection{Vase Dropping on Foam}
\label{sec:vase_foam}

This test examines the transient dynamics of a brittle ceramic vase interacting with a protective foam insert under gravity and subsequent lateral excitation. The scenario is representative of transport of fragile objects, where insert compliance governs the stress transmitted to the enclosed body~\cite{MILHDBK304B1978}. Both bodies are discretized with T10 quadratic
tetrahedral elements. Surface triangle meshes extracted from the volumetric
discretizations feed the collision pipeline exclusively, decoupling
the higher-order volumetric interpolation from the contact detection layer.

The vase is modeled as a Saint Venant--Kirchhoff elastic solid
($E=50\,\mathrm{GPa}$, $\nu=0.25$, $\rho_0=2400\,\mathrm{kg/m^3}$),
representative of a dense fired ceramic~\cite{Hulan2020IlliticClays,Kojima2024Poisson}.
The foam insert is modeled with a compressible two-parameter Mooney--Rivlin
law, which supports large contact-induced shape changes while remaining
numerically tractable in near-incompressible regimes~\cite{Abaqus2024Hyperelastic}.
Five material presets spanning neoprene, polyurethane, and EVA foams of
two densities are tested; their parameters are listed in
Table~\ref{tab:vase_foam_presets}. Neoprene and polyurethane parameters are
taken from published characterizations~\cite{HanChe2021Neoprene,Mamashli2026Polyurethane}.
EVA parameters are study-defined compressible Mooney--Rivlin surrogate
calibrations that preserve the experimentally observed density-dependent
stiffness ordering reported by Chen et al.~\cite{Chen2024EVA}.
Rayleigh-type viscous damping ($\eta_{\mathrm{damp}}=\lambda_{\mathrm{damp}}=5\times10^3\,\mathrm{Pa\cdot s}$)
is applied to the insert only.

\begin{table}[t]
\centering
\footnotesize
\renewcommand{\arraystretch}{1.05}
\setlength{\tabcolsep}{4pt}
\caption{Foam material presets for the vase--insert benchmark. $^\ast$~Study-defined compressible Mooney--Rivlin surrogate calibrations based on EVA compression data; not reported directly by Chen et~al.}
\label{tab:vase_foam_presets}
\begin{tabular}{lrrC{2.6cm}rr}
\toprule
Material & $C_{10}$ (MPa) & $C_{01}$ (MPa) & Vol.\ param. & $\rho_0$ (kg/m$^3$) & Ref. \\
\midrule
Neoprene 50A     & 0.302        & 0.076        & $\nu = 0.490$                      & 1350 & \cite{HanChe2021Neoprene,TRCNeoprene50A} \\
Neoprene 60A     & 0.382        & 0.096        & $\nu = 0.490$                      & 1400 & \cite{HanChe2021Neoprene,DeltaRubberNeoprene60A} \\
Polyurethane 50A & 0.302        & 0.076        & $\nu = 0.499$                      & 2000 & \cite{Mamashli2026Polyurethane} \\
EVA 80           & $0.417^{\ast}$ & $0.104^{\ast}$ & $D_1\!=\!0.736\,\mathrm{MPa}^{-1}{}^{\ast}$ & 80   & \cite{Chen2024EVA} \\
EVA 95           & $0.641^{\ast}$ & $0.160^{\ast}$ & $D_1\!=\!0.478\,\mathrm{MPa}^{-1}{}^{\ast}$ & 95   & \cite{Chen2024EVA} \\
\bottomrule
\end{tabular}
\end{table}

The foam base nodes ($z < -0.09\,\mathrm{m}$) are fixed throughout; the vase is
unconstrained except for gravity and contact. Inter-body contact uses a penalty
model with stiffness $E_c=8\times10^6\,\mathrm{Pa}$, restitution coefficient
$0.1$, and friction coefficients $\mu_s=0.6$, $\mu_k=0.5$. The coupled system
is advanced with an implicit backward-Euler integrator at
$\Delta t = 1\times10^{-4}\,\mathrm{s}$ (6000 steps, $0.6\,\mathrm{s}$ total)
using the multi-block augmented Lagrangian Newton solver (inner tolerances
$10^{-3}$ absolute / $10^{-4}$ relative; outer tolerance $10^{-6}$;
$\rho=10^{12}$; up to 5 outer and 10 inner iterations; line search enabled);
mean RTF is approximately 3{,}670 during settling and 5{,}930 during shaking,
reflecting the increased per-step cost once frictional contact is active.
The loading protocol consists of three phases:
\begin{enumerate}
  \item \textbf{Settling} ($t\in[0, 0.1]\,\mathrm{s}$): gravity only; the vase
    descends into the cavity.
  \item \textbf{Shaking} ($t\in[0.1, 0.5]\,\mathrm{s}$): the foam base follows
    a triangular velocity waveform in $x$ with amplitude $A_x=5\,\mathrm{mm}$
    and peak speed $v_x=0.2\,\mathrm{m/s}$ ($T=0.1\,\mathrm{s}$), inducing
    repeated sliding, separation, and re-contact at the vase--cavity interface.
  \item \textbf{Relaxation} ($t\in[0.5, 0.6]\,\mathrm{s}$): base motion
    removed; system returns to gravity-only equilibrium.
\end{enumerate}

\begin{figure}[t]
  \centering
  \setlength{\tabcolsep}{2pt}
  \renewcommand{\arraystretch}{0.9}
  \small
  \begin{tabular}{@{}ccc@{}}
    $t = 0.08\,\mathrm{s}$ & $t = 0.10\,\mathrm{s}$ & $t = 0.25\,\mathrm{s}$ \\[2pt]
    \maybeincludegraphics[width=0.315\linewidth]{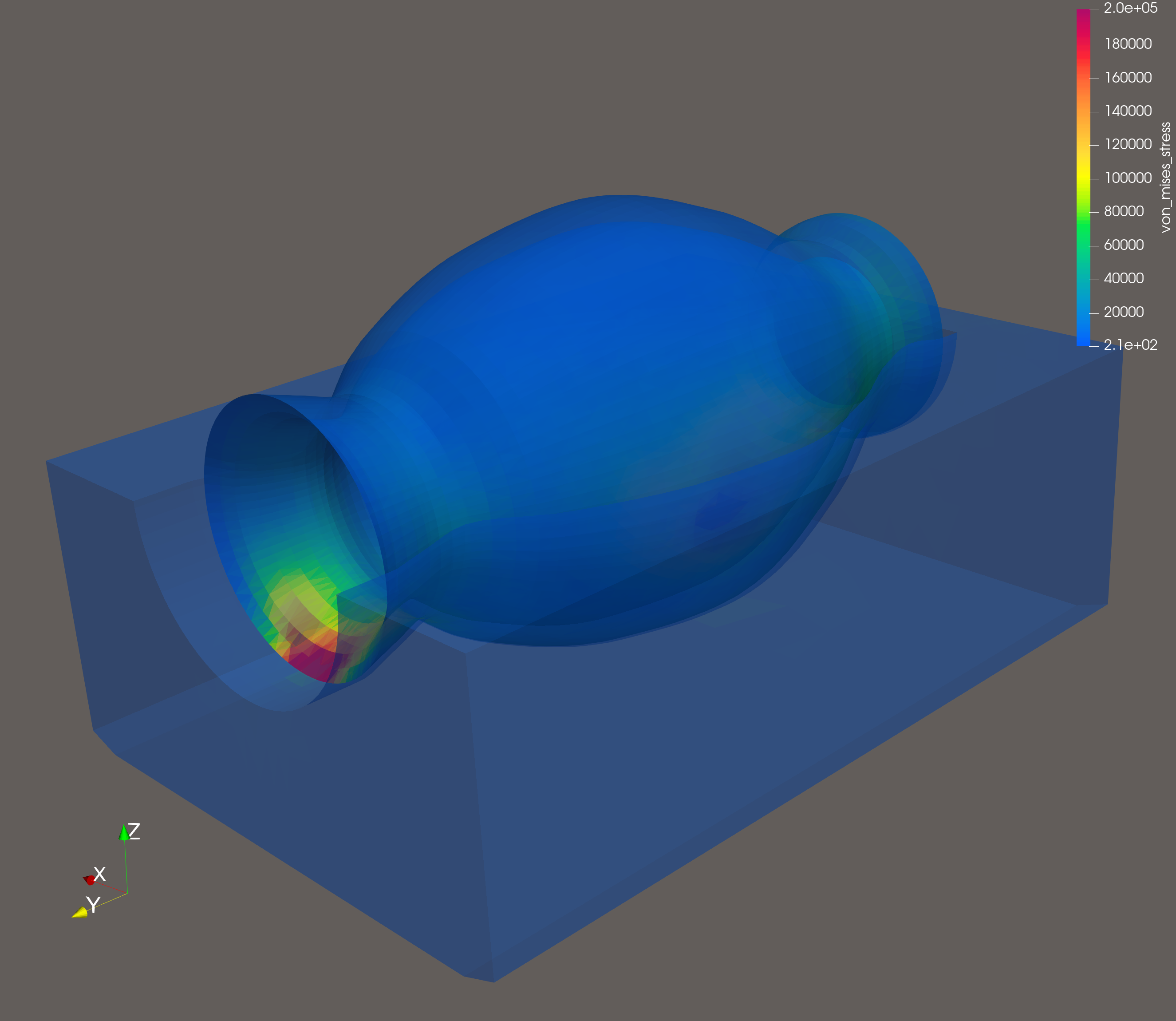} &
    \maybeincludegraphics[width=0.315\linewidth]{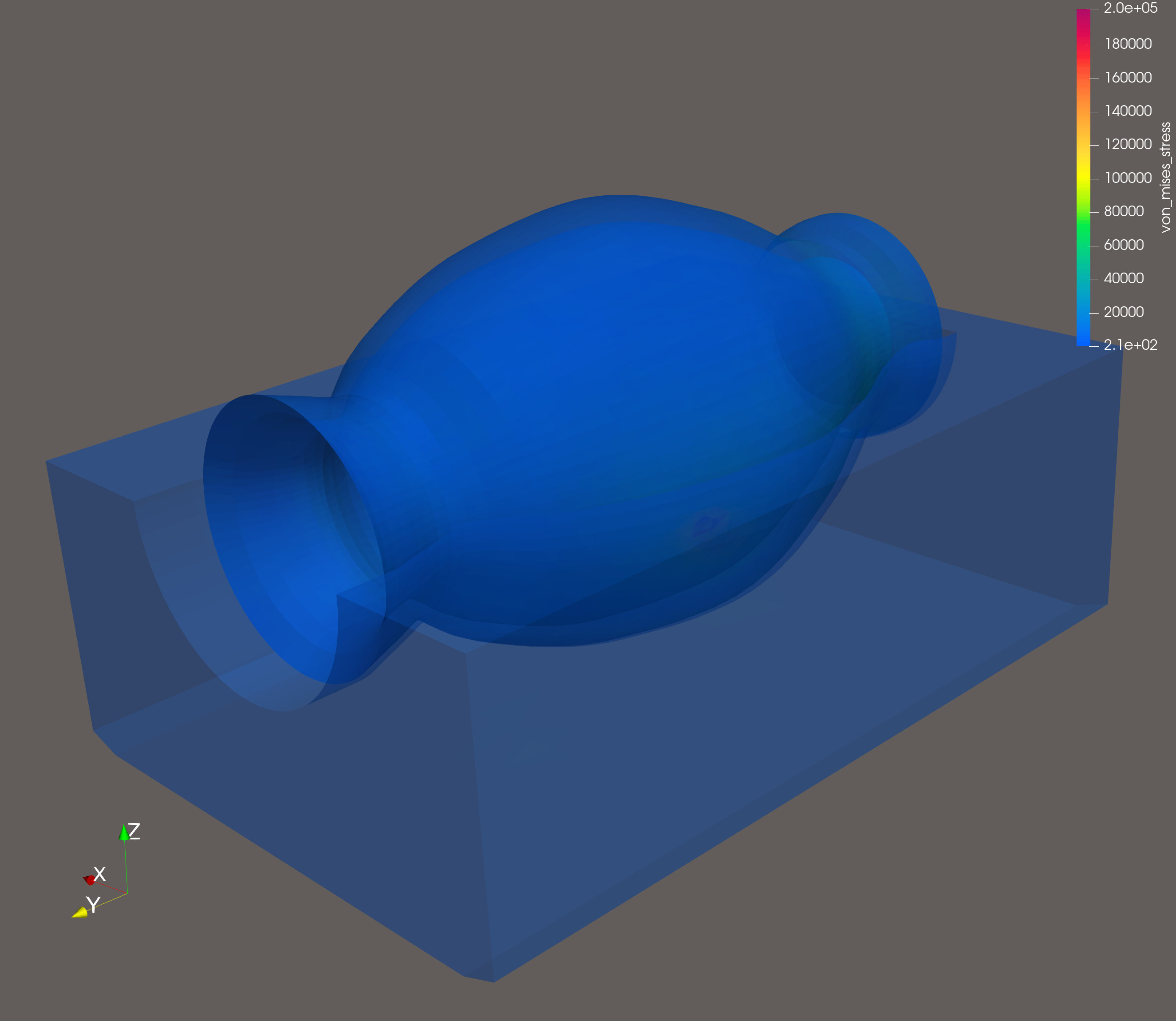}  &
    \maybeincludegraphics[width=0.315\linewidth]{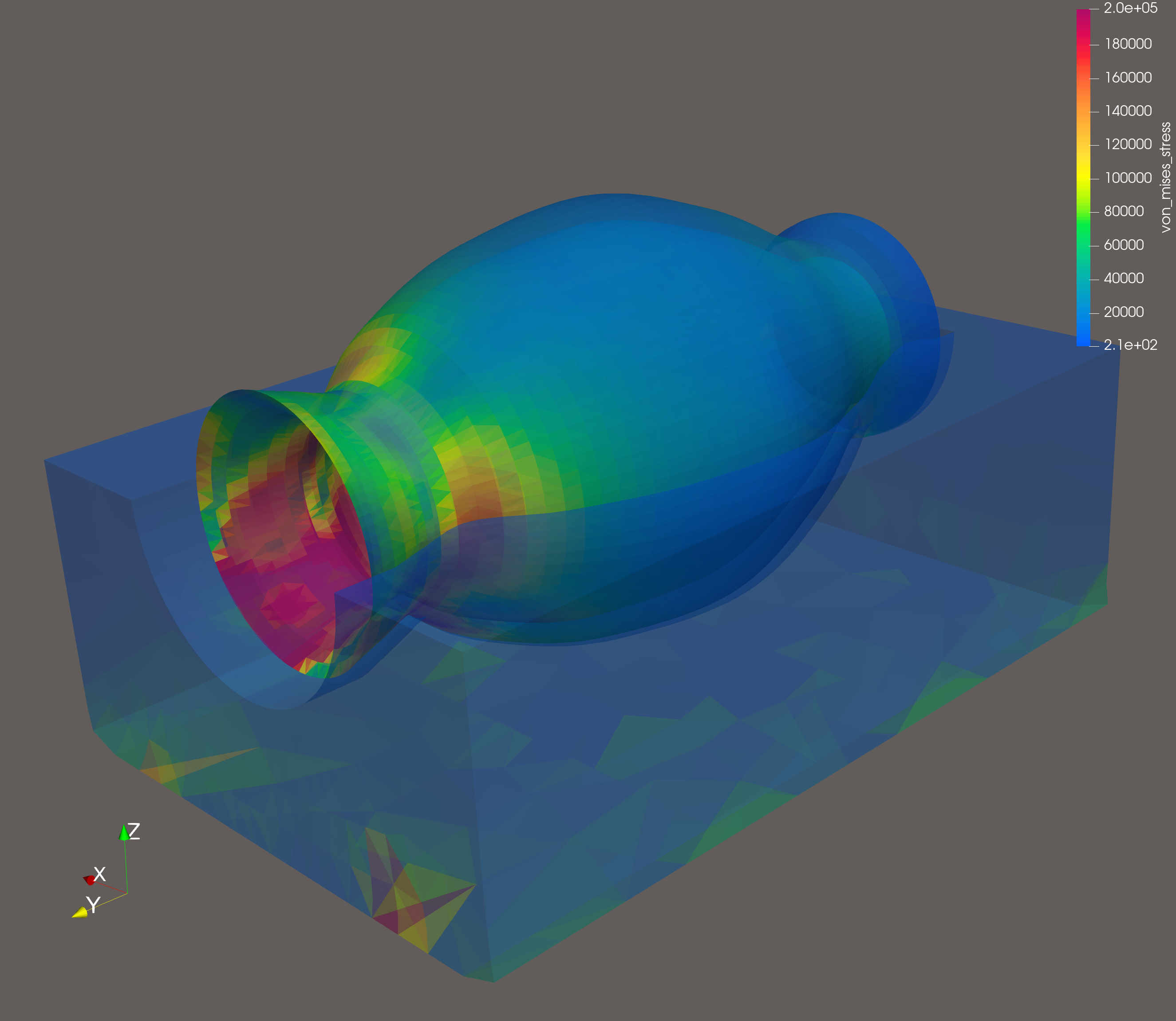} \\[1pt]
    \maybeincludegraphics[width=0.315\linewidth]{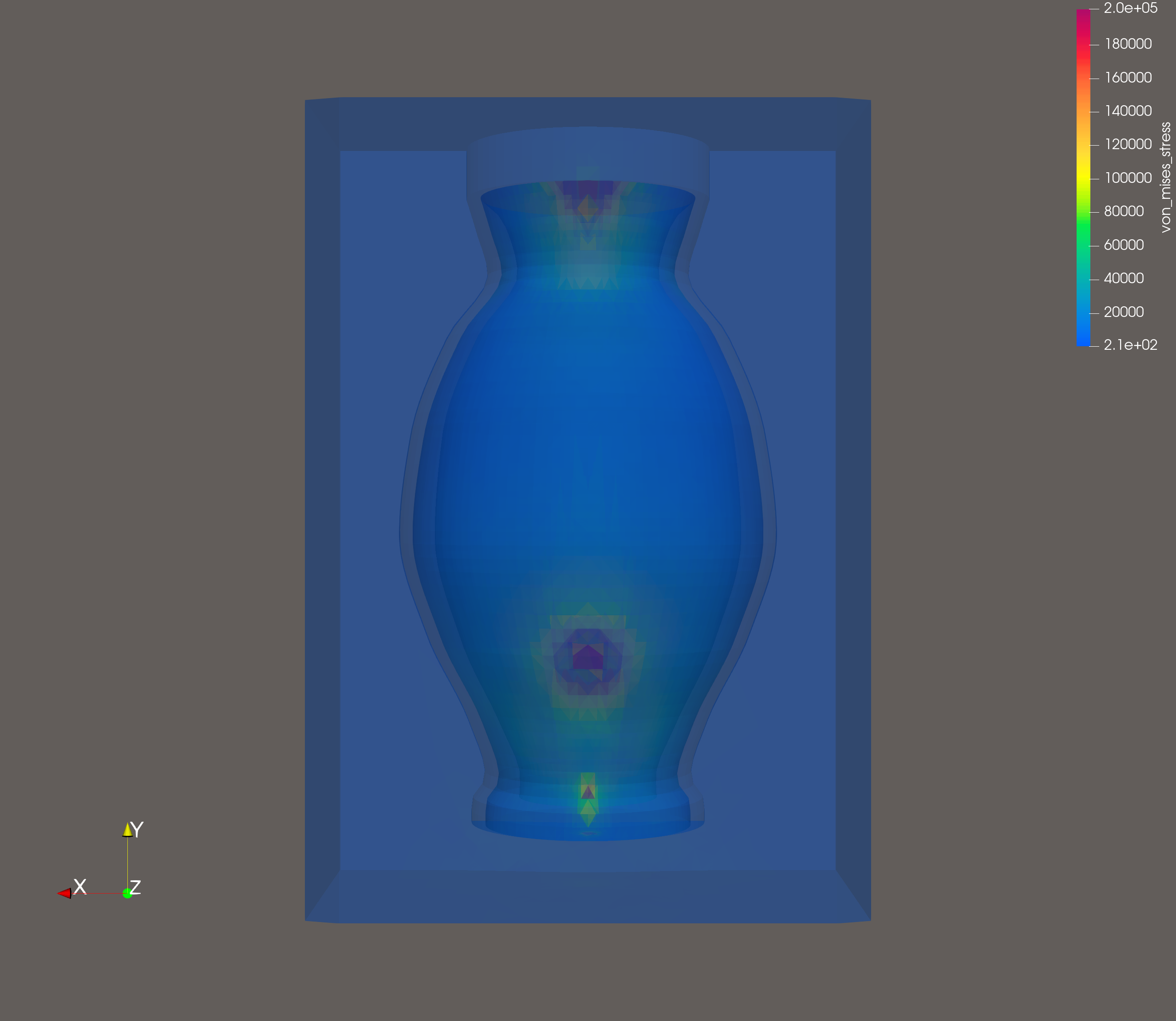} &
    \maybeincludegraphics[width=0.315\linewidth]{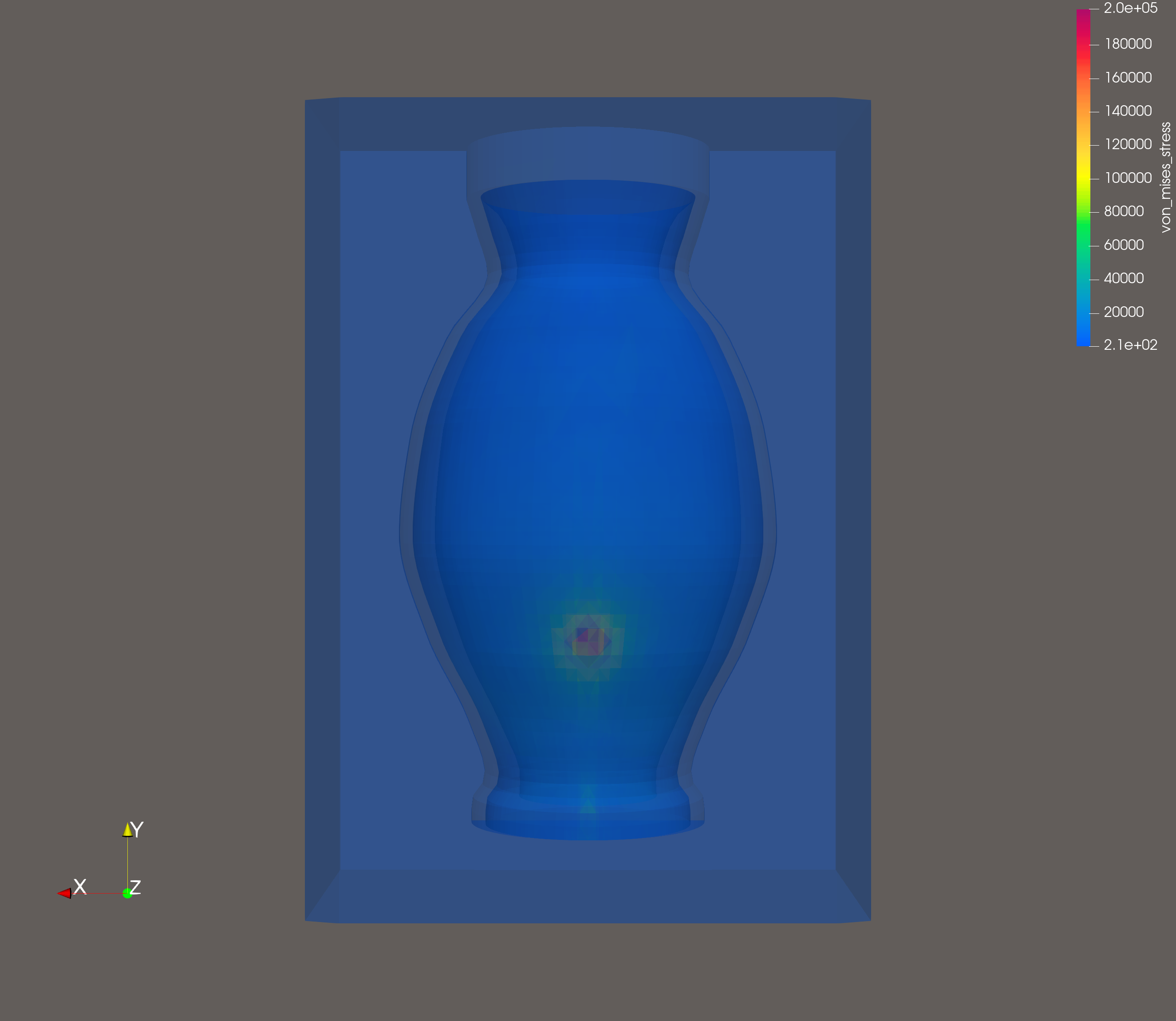}  &
    \maybeincludegraphics[width=0.315\linewidth]{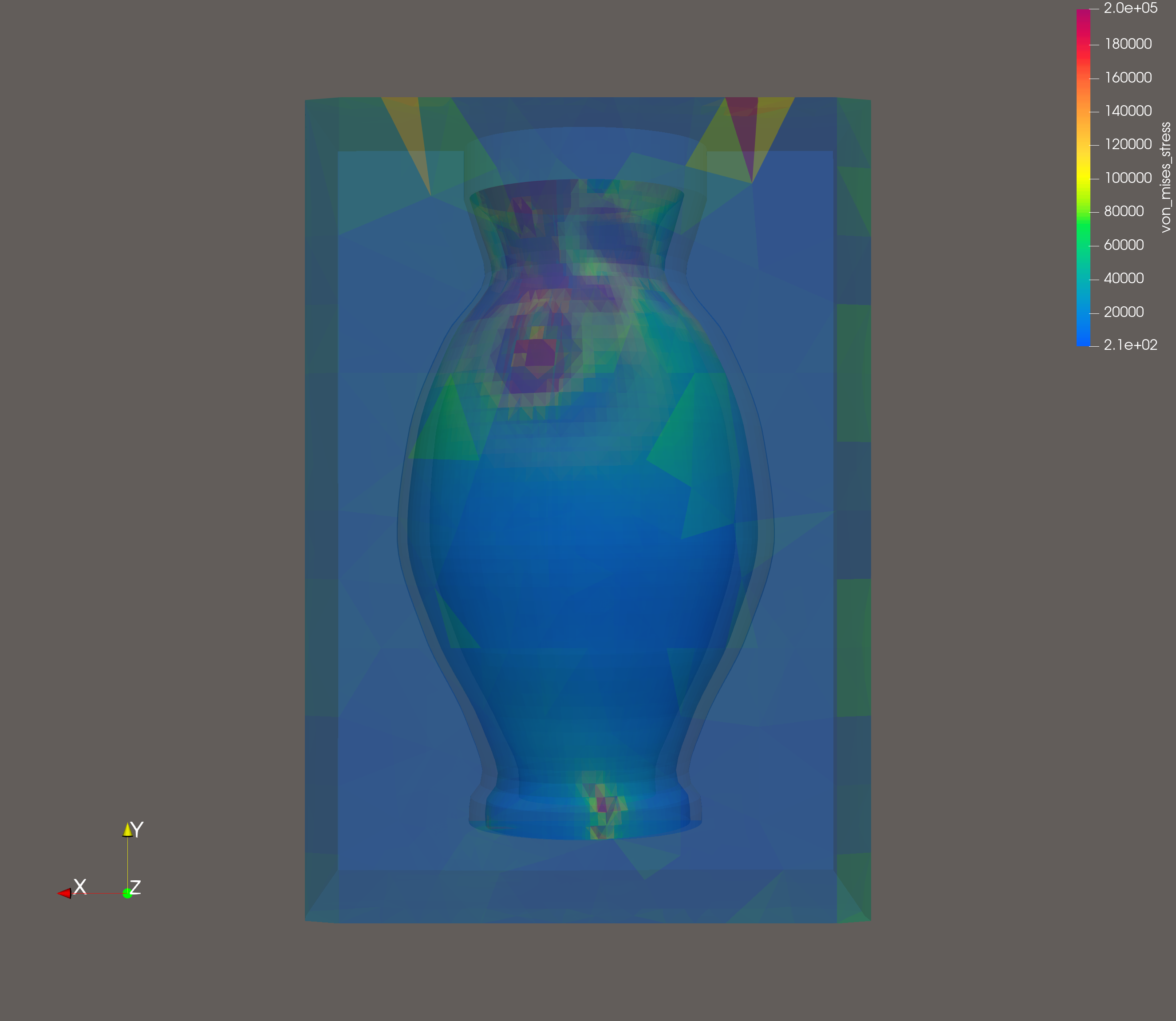}
  \end{tabular}
  \caption{Von Mises stress fields on the vase (EVA80 insert) at three time
    instants. Top row: three-quarter ($45^\circ$) view; bottom row: underside
    ($+z$) view. Columns correspond to post-drop pre-settling, settled before
    shaking, and mid-shaking contact redistribution}
  \label{fig:vase_eva80_snapshots}
\end{figure}

Representative von Mises stress fields for the EVA80 case are shown in
Fig.~\ref{fig:vase_eva80_snapshots} at three mechanically distinct instants:
post-impact before full settling ($t=0.08\,\mathrm{s}$), the supported
configuration just before shaking begins ($t=0.10\,\mathrm{s}$), and an
instant of active contact redistribution during the shaking phase
($t=0.25\,\mathrm{s}$).

\begin{figure}[t]
  \centering
  \maybeincludegraphics[width=\textwidth]{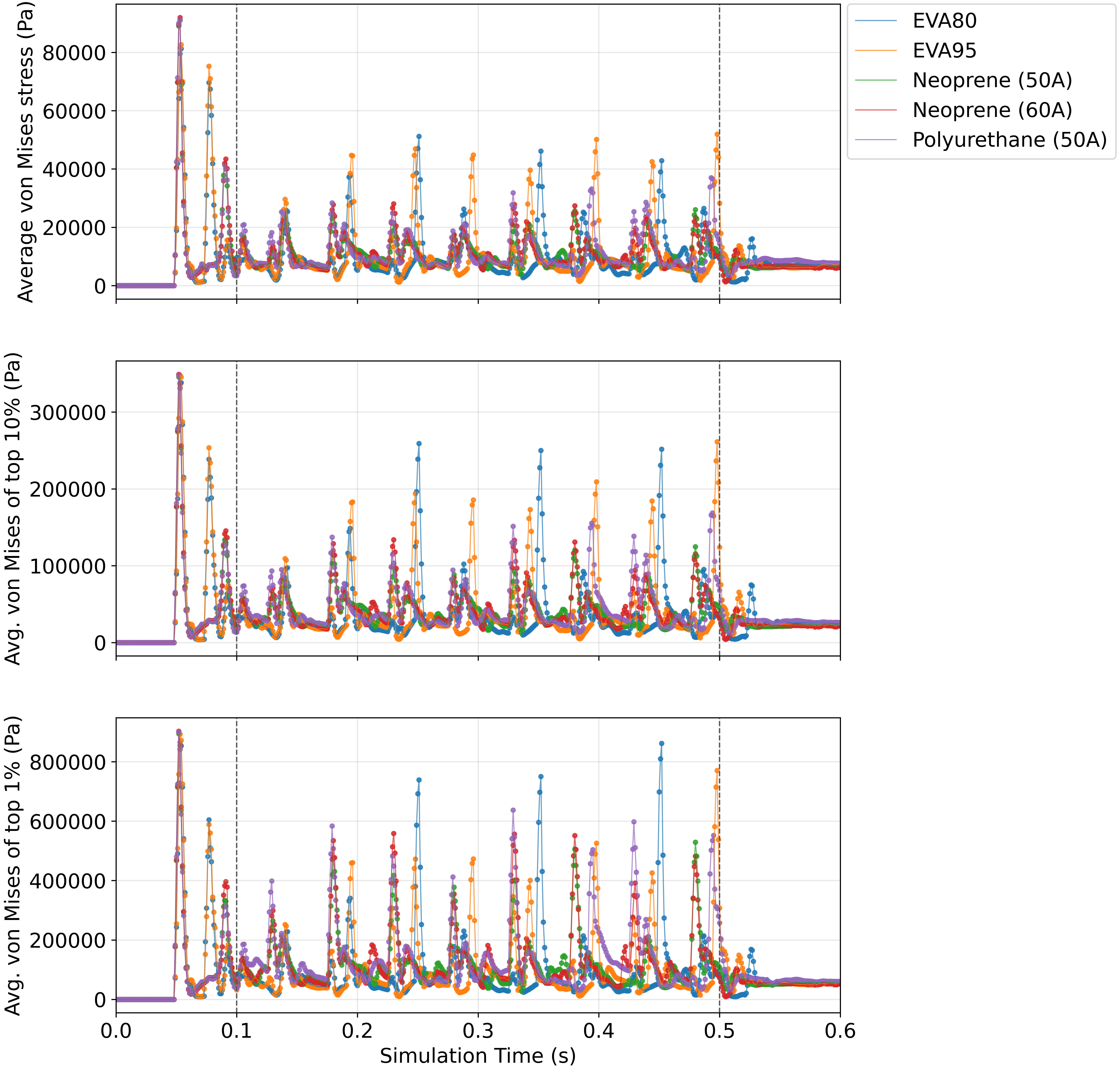}
  \caption{Vase von Mises stress histories for the five insert presets. The
    three panels show the whole-body average, the top-10\% hotspot average, and
    the top-1\% hotspot average. Phase~1 ($t\in[0,0.1]\,\mathrm{s}$): gravity
    settling; Phase~2 ($t\in[0.1,0.5]\,\mathrm{s}$): lateral shaking; Phase~3
    ($t\in[0.5,0.6]\,\mathrm{s}$): relaxation}
  \label{fig:vase_von_mises_comparison}
\end{figure}

To assess how insert material affects vase stress, four complementary metrics
are extracted over the shaking interval $t\in[0.1,0.5]\,\mathrm{s}$: (i) the
whole-body average von Mises stress; (ii) the spatial top-10\% and (iii)
top-1\% hotspot averages (cells ranked by stress at each step, then averaged
over time); and (iv) temporal tail averages from the smoothed top-1\% history,
retaining the highest 1\% and 2\% of timesteps. Global measures capture the
typical transmitted loading, while temporal tail metrics isolate rare but severe
re-contact events. The resulting summary is given in
Table~\ref{tab:vase_stress_summary}, and the full stress time histories are
shown in Fig.~\ref{fig:vase_von_mises_comparison}.

\begin{table}[t]
\centering
\small
\renewcommand{\arraystretch}{1.05}
\setlength{\tabcolsep}{5pt}
\caption{Vase von Mises stress metrics over the shaking interval
  $t\in[0.1,0.5]\,\mathrm{s}$ (all values in Pa). ``Top 10\%'' and ``Top 1\%''
  are time-averaged spatial hotspot measures; ``Tail 1\%'' and ``Tail 2\%'' are
  temporal upper-tail averages of the smoothed top-1\% hotspot history.}
\label{tab:vase_stress_summary}
\begin{tabular}{lrrrrr}
\toprule
Preset & Avg.\ VM & Top 10\% & Top 1\% & Tail 1\% & Tail 2\% \\
\midrule
EVA 80           &  9,698 &  36,446 &  93,545 & 771,887 & 714,611 \\
EVA 95           & 10,365 &  38,968 & 101,700 & 626,119 & 558,398 \\
Neoprene 50A     & 10,761 &  41,874 & 123,396 & 491,122 & 467,567 \\
Neoprene 60A     & 11,265 &  44,346 & 133,391 & 544,109 & 525,577 \\
Polyurethane 50A & 11,578 &  46,137 & 142,849 & 583,371 & 556,896 \\
\bottomrule
\end{tabular}
\end{table}

The interval-averaged metrics exhibit a clear compliance-ordered ranking: EVA80
transmits the lowest mean stress, followed by EVA95, the two neoprene variants
(60A slightly stiffer than 50A), and polyurethane 50A, which produces the
highest values. This ordering is consistent across all three spatial hotspot
measures and reflects the higher compressibility of the EVA surrogates relative
to the near-incompressible elastomeric presets. The temporal tail metrics,
however, reverse this trend: EVA80 produces the \emph{largest} tail values,
indicating that although softer inserts reduce the typical transmitted load,
they permit greater vase excursions that generate sharper re-contact events
during a small fraction of timesteps. EVA95 occupies an intermediate position,
offering moderate average stress reduction with a less pronounced tail
amplification than EVA80. Taken together, the results highlight a
compliance--tail-severity trade-off: optimizing insert material solely against
mean transmitted stress risks underestimating the intensity of rare, localized
impact events.

\subsection{Mixed Item Dropping}
\label{sec:tire_drop}

This test examines a cluttered, large-scale contact scenario in which nine
deformable tire bodies are dropped into a fixed open container and allowed
to interact through repeated impact, rolling, and re-contact with one another,
with the container walls, and with a compliant cantilevered ANCF3443 shell
beam positioned inside the box as a deformable support surface.
Unlike the controlled two-body scenario of Section~\ref{sec:vase_foam}, this
configuration is primarily a stress test of the full simulation pipeline,
exercising the multi-block Newton solver, the asynchronous GPU collision
detection pipeline of Section~\ref{sec:collision}, and the coupled ANCF3443
shell and T10 solid element formulations simultaneously under rapidly evolving
many-body contact with mixed element discretizations.

\begin{figure}[ht]
  \centering
  \small
  \begin{minipage}[t]{0.48\linewidth}
    \centering
    \maybeincludegraphics[width=\linewidth]{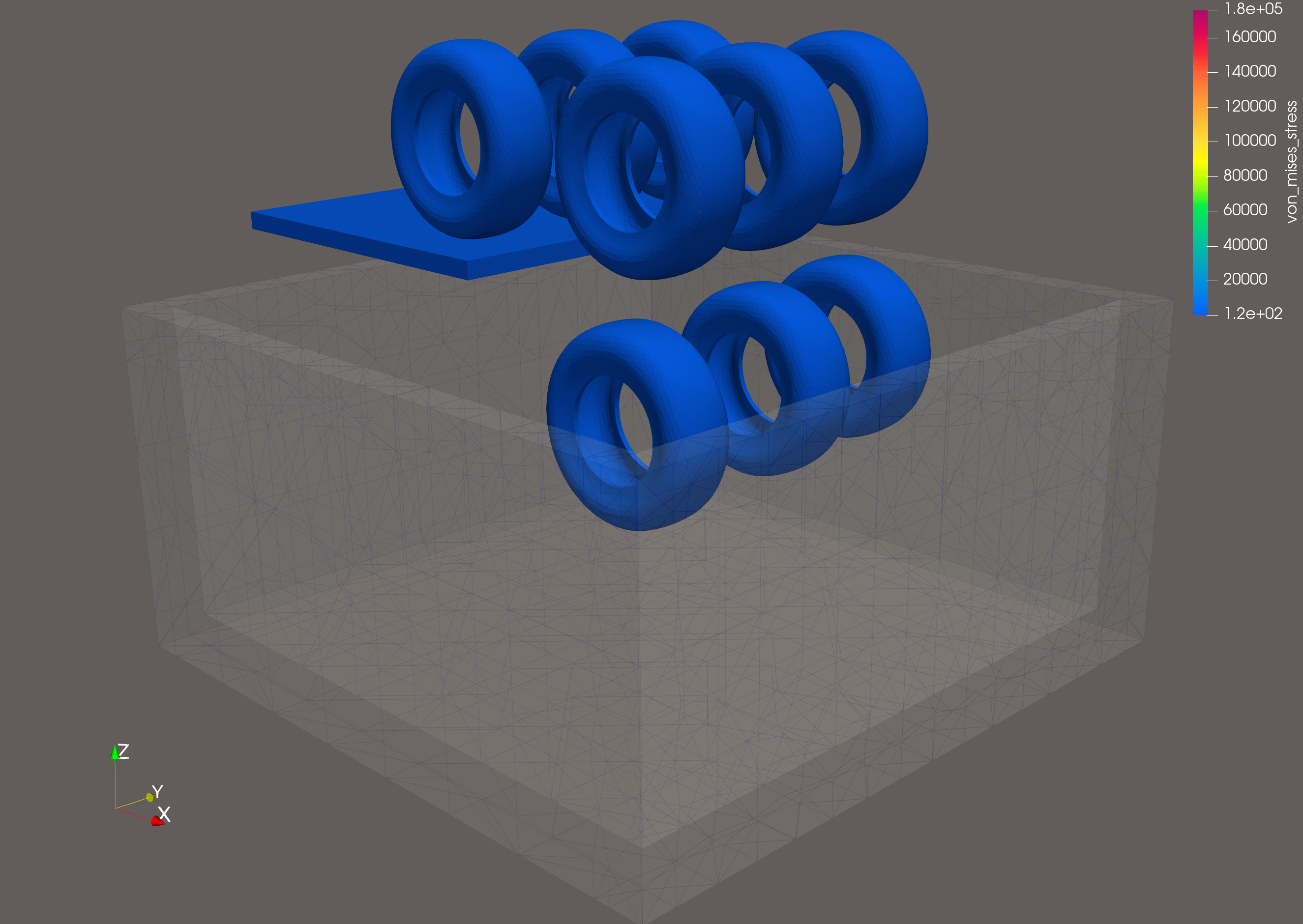}\\[3pt]
    $t=0\,\mathrm{s}$: nine tires in free fall; shell beam unloaded.
  \end{minipage}
  \hfill
  \begin{minipage}[t]{0.48\linewidth}
    \centering
    \maybeincludegraphics[width=\linewidth]{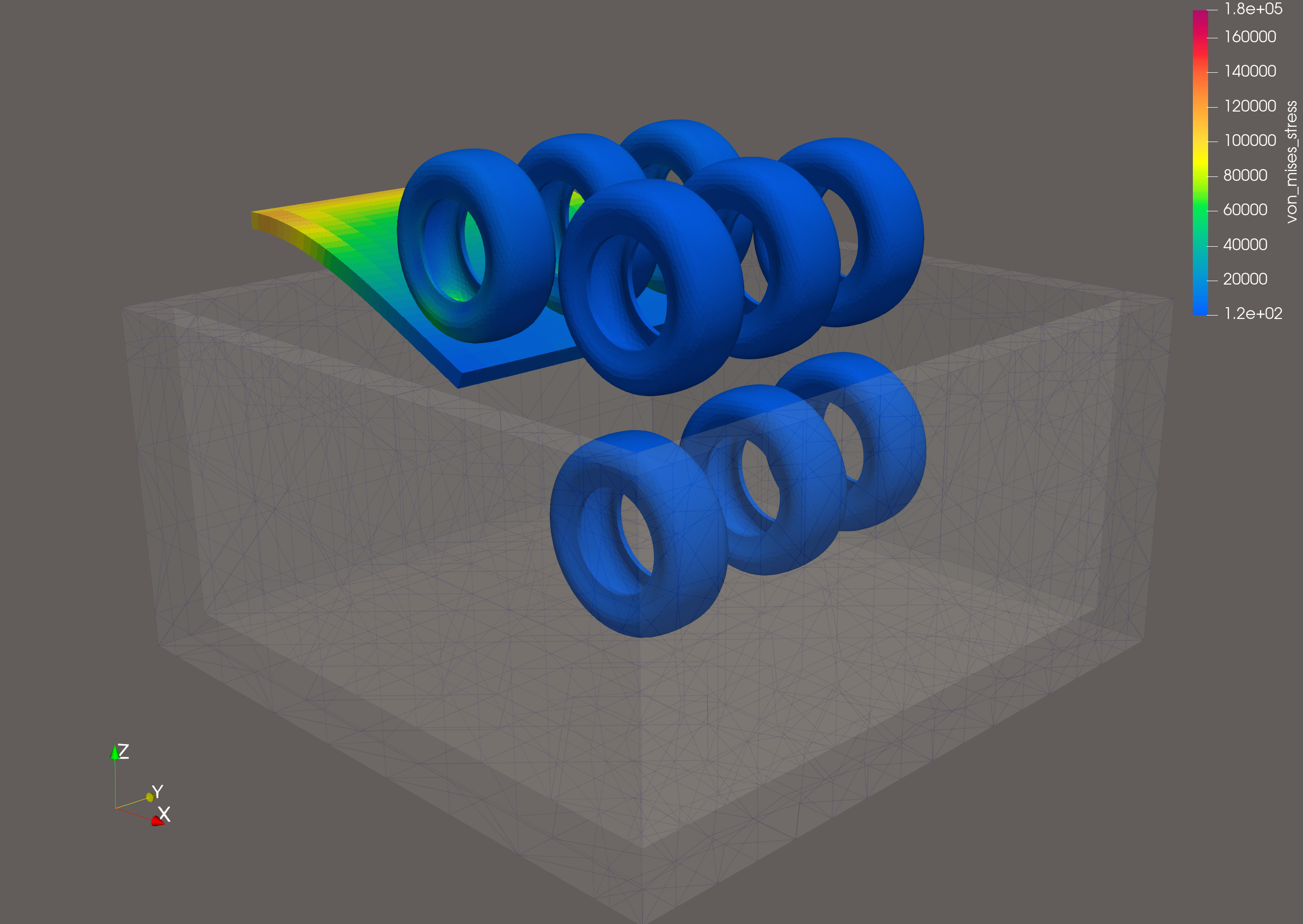}\\[3pt]
    $t=0.145\,\mathrm{s}$: first contact; localized bending stress at the contact patch.
  \end{minipage}\\[8pt]
  \begin{minipage}[t]{0.48\linewidth}
    \centering
    \maybeincludegraphics[width=\linewidth]{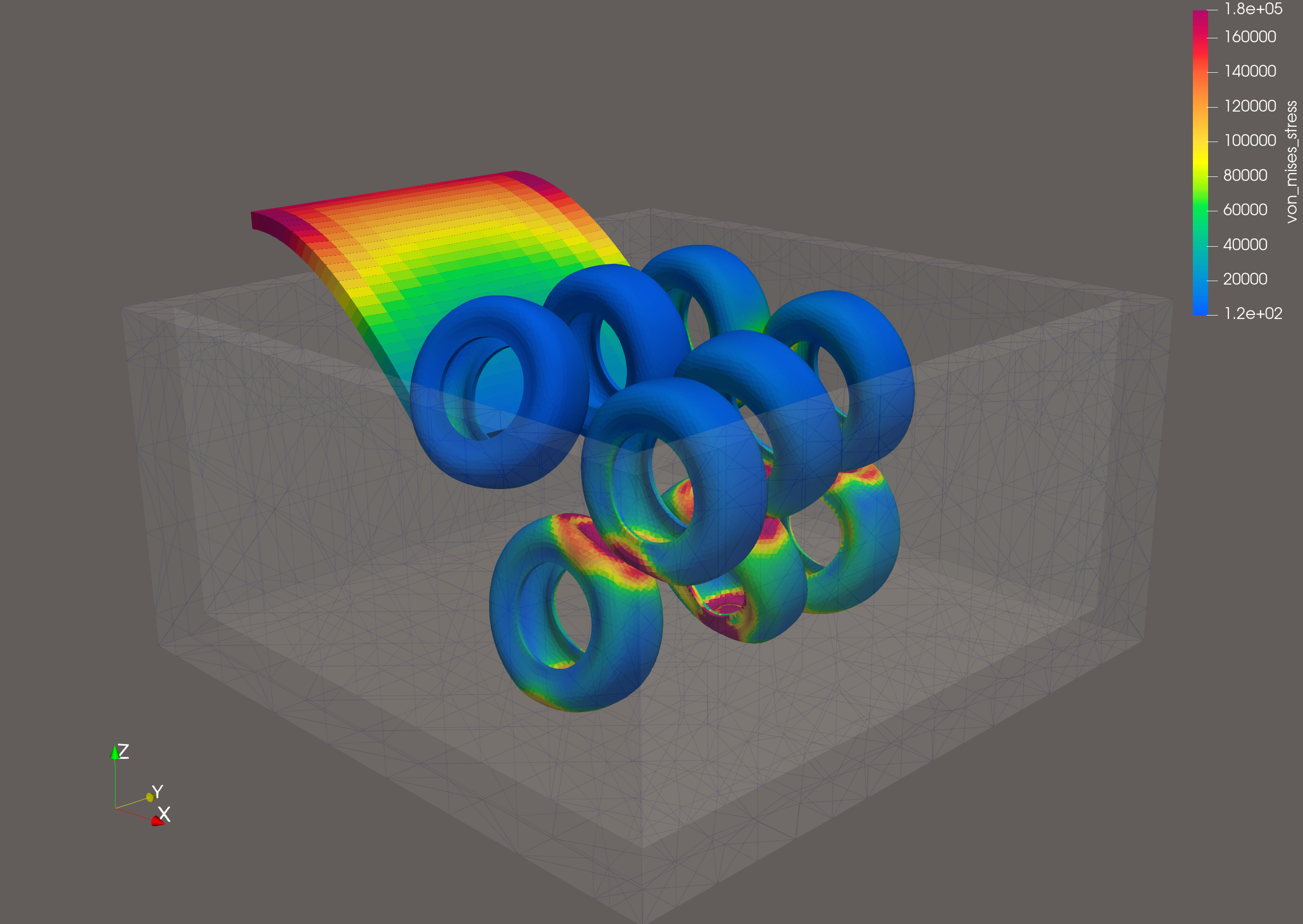}\\[3pt]
    $t=0.262\,\mathrm{s}$: dense contact; shell beam heavily deformed with stress concentrations at inter-tire contact points.
  \end{minipage}
  \caption{Von Mises stress field (Pa) at three representative instants of the mixed
    item-dropping simulation. The color scale is shared across all panels; container
    walls shown as transparent wireframe}
  \label{fig:tire-drop-snapshots}
\end{figure}

\paragraph{Problem setup}
The container is a rigid open box discretized with 4{,}474 nodes and
2{,}170 T10 tetrahedral elements; all its nodes are fixed throughout the
simulation.
A cantilevered ANCF3443 shell beam with planform dimensions
$0.40\times0.40$\,m and thickness 0.02\,m is meshed with $30\times12$
shell elements (360~elements, 403~nodes, 4{,}836~displacement DOFs), and
clamped along one longitudinal edge (156~constrained DOFs).
Nine deformable tire bodies, each discretized with the same T10 mesh
introduced in Section~\ref{sec:benchmark_complex} (34{,}414~nodes,
17{,}167~elements per tire), are arranged in two groups: six tires in an
upper cluster near the beam surface, and three additional tires in a lower
row so that the descending upper bodies also interact with an already
populated region of the container.
The combined system contains 314{,}603 nodes and 947{,}436 displacement DOFs,
of which 13{,}578 are constrained (container walls and beam clamp), leaving
933{,}858 free---a nearly one-million-DOF nonlinear transient contact problem
with two coupled element technologies operating simultaneously.
Mesh statistics, material parameters, and solver settings are collected in
Tables~\ref{tab:tire-drop-mesh}--\ref{tab:tire-drop-solver} in Appendix~\ref{sec:appendix_benchmark_params}.
Representative snapshots of the simulation are shown in
Fig.~\ref{fig:tire-drop-snapshots}.

\begin{figure}[t]
  \centering
  \maybeincludegraphics[width=0.82\linewidth]{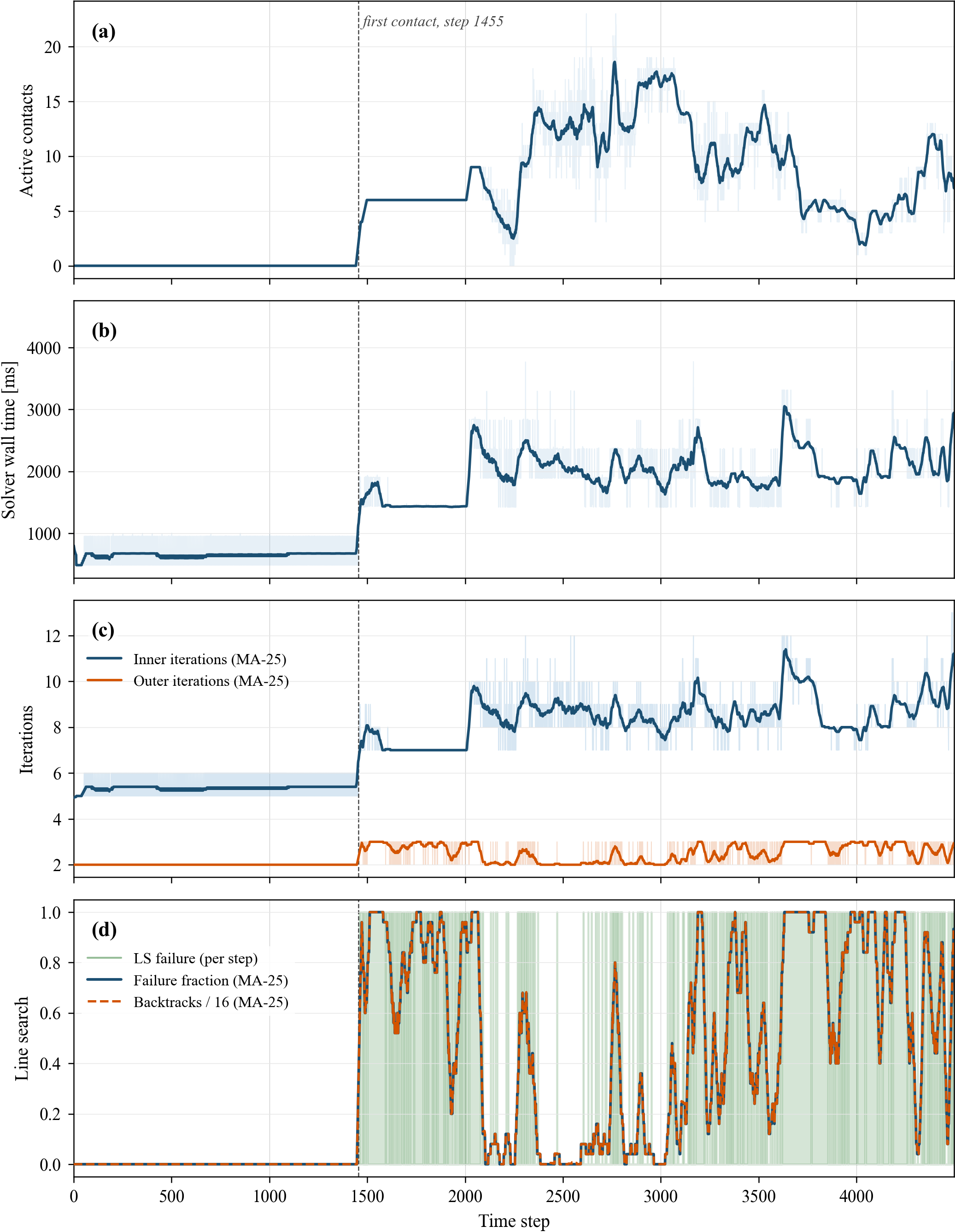}
  \caption{Overview diagnostics for the first 4{,}500 steps (0.45\,s) of the mixed
    item-dropping test.  Thin traces are per-step values; heavy lines are 25-step
    moving averages; the dashed vertical line marks first contact at step~1{,}455.
    (a)~Active contact count, peaking at~23 (step~2{,}620).
    (b)~Solver wall time: mean rises from 645\,ms to 1{,}969\,ms after contact onset;
    the costliest step (3{,}782\,ms, step~4{,}487) has only 10~contacts, confirming
    that cost tracks local geometry, not contact count.
    (c)~Iteration counts: inner grows from~5.3 to~8.9; outer stays between~2 and~3.
    (d)~Line-search: bimodal pattern---most post-contact steps either accept
    immediately or exhaust the backtracking budget}
  \label{fig:tire-drop-overview}
\end{figure}

\paragraph{Diagnostics over the first 4\,500 steps}
The first 4{,}500 time steps (0.45\,s of physical time) capture the
transition from a pre-contact free-fall phase to a sustained multi-contact
settling regime; the corresponding diagnostics are shown in
Fig.~\ref{fig:tire-drop-overview}.
No contacts are detected until step~1{,}455, after which the active contact
count rises rapidly and remains nonzero for approximately 67.6\% of the
analyzed interval.
The contact population reaches a maximum of 23~active contacts at
step~2{,}620, indicating that the most geometrically crowded configuration
occurs in the middle of the analyzed window rather than at first impact, i.e., a
behavior consistent with the intended role of the test as a cluttering
experiment: the dominant challenge is not first touchdown, but the subsequent
sequence of rearrangements as bodies compete for space inside the container.

Contact onset produces a marked increase in nonlinear-solve cost without
loss of numerical control.
Excluding the initialization step, the mean solver wall time rises from
approximately 645\,ms before first contact to approximately 1{,}969\,ms
afterward, i.e., a roughly threefold increase.
At $\Delta t = 10^{-4}$\,s, these correspond to a mean RTF of approximately
6{,}450 in the pre-contact phase and 19{,}690 in the post-contact phase.
The mean inner-iteration count follows the same trend, growing from 5.34 in
the pre-contact regime to 7.91 during impact buildup, 8.52 in the
dense-contact interval (steps 2{,}500--3{,}499), and 8.93 during late
rearrangement (steps 3{,}500--4{,}499).
Notably, the most expensive steps do not coincide with the maximum contact
count: the largest non-initialization solver wall time occurs at
step~4{,}487 (3{,}781.9\,ms) with only 10~active contacts, whereas the
contact-count peak at step~2{,}620 is comparatively less costly.
This indicates that solver difficulty is governed more by unfavorable local
contact geometry and repeated reorganization than by the raw number of active
contact pairs.

\begin{figure}[t]
  \centering
  \maybeincludegraphics[width=0.82\linewidth]{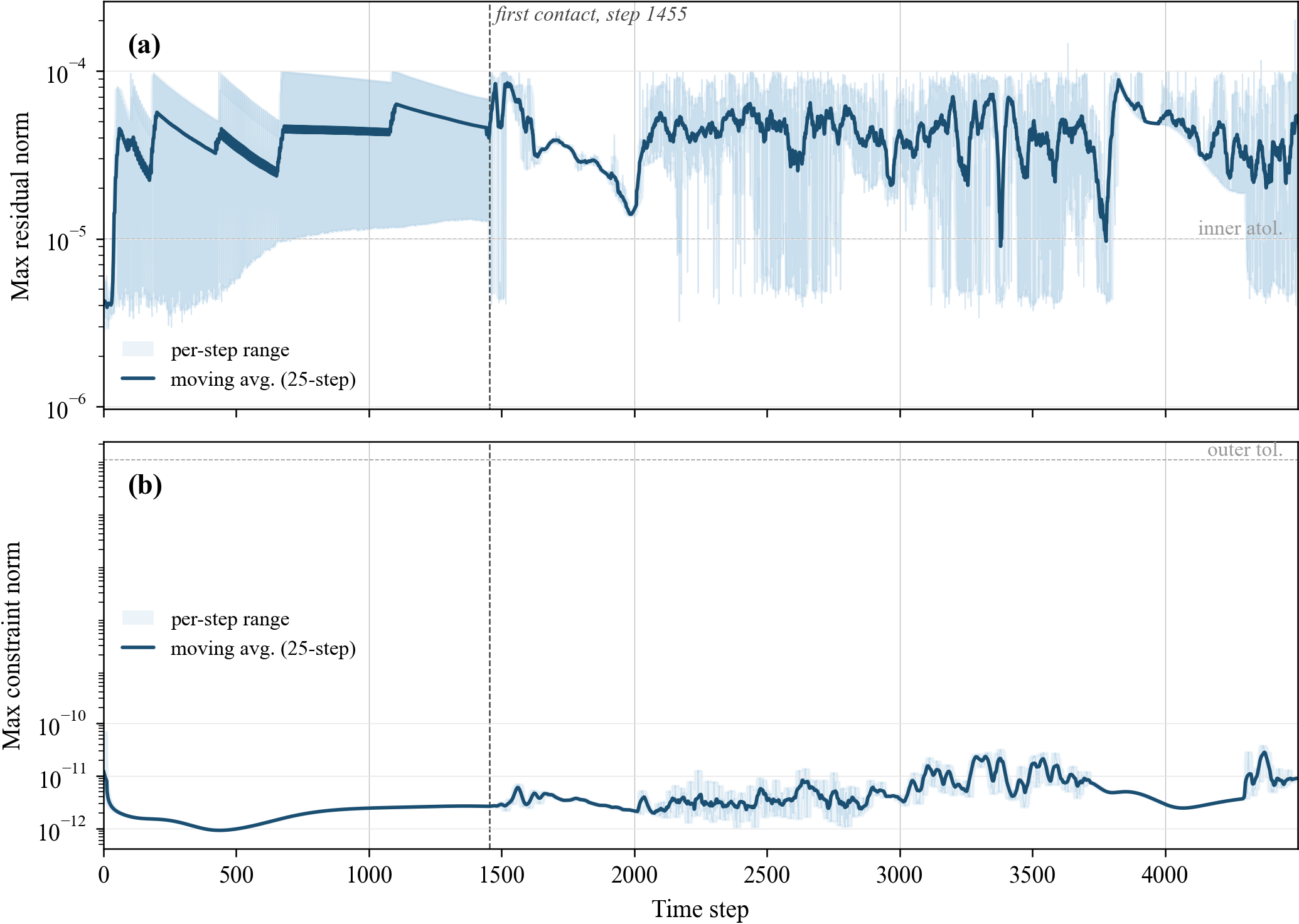}
  \caption{Convergence diagnostics for the first 4{,}500 steps (0.45\,s).
    Shaded bands are per-step ranges; solid lines are 25-step moving averages;
    dashed lines mark solver tolerances.
    (a)~Maximum residual norm: stays at or below $\varepsilon_{\mathrm{in}}=10^{-4}$
    except for five steps (0.1\%), largest $2.0\times10^{-4}$ at step~4{,}490.
    (b)~Maximum constraint norm: confined to $10^{-13}$--$10^{-11}$, far below the
    outer ALM tolerance $\varepsilon_{\mathrm{out}}=10^{-5}$, confirming negligible
    constraint drift throughout}
  \label{fig:tire-drop-convergence}
\end{figure}

Convergence indicators remain bounded throughout the analyzed interval
(Fig.~\ref{fig:tire-drop-convergence}).
The maximum constraint norm stays in the range $10^{-13}$--$10^{-11}$,
many orders of magnitude below the outer ALM tolerance, while the maximum
residual norm remains controlled; its largest excursion
(${\approx}\,2.0\times10^{-4}$ at step~4{,}490) appears near the end of
the window during late-stage rearrangement rather than at first impact.
The most prominent numerical signature is the line-search pattern: many steps
either accept the full Newton step immediately or reach the backtracking cap,
suggesting strongly bimodal globalization behavior under dense frictional
contact, which is qualitatively consistent with the step-acceptance observations
reported for the joint-constraint benchmarks in
Section~\ref{sec:joint_validation}.

Taken together, these results support two conclusions.
First, the framework stably advances a geometrically rich, many-body, mixed
shell--solid deformable-contact scene over 0.45\,s of physical time at close
to one million free DOFs, demonstrating end-to-end pipeline capability at
this problem scale.
Second, the test reveals that late rearrangement and re-contact---not peak
contact count alone---constitute the dominant source of nonlinear difficulty
in cluttered dropping scenarios, a finding that complements the more
controlled scaling benchmarks and unit tests reported in
Sections~\ref{sec:performance_benchmark} and~\ref{sec:experiments}.

\section{Limitations and Future Work}
\label{sec:future_work}

The benchmarks and large-scale demonstrations presented in the preceding sections establish the viability of the GPU-accelerated framework for implicit, second-order finite element simulation of flexible multibody systems with contact. At the same time, the current implementation has limitations that point toward concrete directions for future work.

\paragraph{Explicit contact coupling}
The frictional contact forces are treated as external loads within the augmented Lagrangian formulation. They are assembled into $\mathbf{f}_{\mathrm{ext}}$ and held fixed during the inner Newton solve rather than being incorporated into the Hessian $\mathbf{H}$. This explicit operator-splitting treatment of contact decouples the contact constitutive response from the structural Newton system, avoiding the complexity of computing and assembling a contact Jacobian, but at the cost of conditional stability: large contact stiffness values or rapidly evolving contact states can induce oscillations or divergence unless the time step is sufficiently small. 
The bimodal line-search behavior observed and reported in Fig.~\ref{fig:tire-drop-overview}, where most post-contact steps either accept the full Newton step immediately 
or exhaust the backtracking budget, is a direct consequence of this explicit 
treatment: contact forces assembled into $\mathbf{f}_{\mathrm{ext}}$ at the 
start of a step can be inconsistent with the configuration reached by the Newton 
solve, producing gradient directions that the line search cannot reconcile without 
a timestep reduction.
A fully implicit treatment, which linearizes the contact force model and contributes its Jacobian to $\mathbf{H}$, would remove this restriction and is a natural direction for future work.

\paragraph{AdamW solver hyperparameter sensitivity}
The AdamW solver was explored as a first-order alternative to Newton precisely because, in the Total Lagrangian FEA setting, it requires only gradient evaluations, without Hessian assembly or sparse factorization. As described in Section~\ref{sec:solvers_first_order}, each inner iteration reduces to an embarrassingly parallel per-DOF update: every thread independently evaluates its gradient entry and applies the bias-corrected AdamW rule without any inter-thread communication, making the solver a natural fit for GPU execution. However, as noted in Section~\ref{sec:performance_benchmark}, the first-order convergence rate causes the gradient norm to plateau within any practical iteration budget at large problem sizes, making tight tolerances unattainable without a prohibitive increase in iteration count. Beyond convergence rate, the solver is sensitive to the choice of hyperparameters, in particular the learning rate and weight decay coefficient, whose optimal values depend on element type, material model, mesh resolution, and time step size. The benchmarks in Section~\ref{sec:performance_benchmark} demonstrate competitive performance for the structured beam-sagging cases, but these parameters were tuned specifically for those configurations and do not transfer automatically to new problem setups. This limits the out-of-the-box usability of the AdamW solver relative to the Newton solver, the latter requiring no such problem-specific tuning. Developing adaptive or self-tuning strategies for the AdamW hyperparameters is a natural direction for future work.

\paragraph{Shared memory staging for GPU assembly kernels}
During implementation, several strategies employing GPU shared memory as an intermediate buffer within the internal force and Hessian assembly kernels were explored, motivated by the expectation that staging intermediate results in on-chip shared memory would reduce global memory traffic and improve kernel throughput. In practice, however, these approaches yielded negative or negligible performance gains in benchmarks. A key observation was that each element carries a substantial amount of element-specific data, e.g., precomputed reference-configuration quantities, quadrature-point stress tensors, and shape function gradient arrays, and loading all of this into shared memory requires a large shared memory allocation per thread block. On the RTX~5090, this high per-block shared memory footprint directly reduces the number of thread blocks that can reside concurrently on a streaming multiprocessor (SM), lowering SM occupancy and offsetting any latency benefit from on-chip staging. This suggests that the assembly kernels operate in a regime where shared memory tiling, as applied here, is not effective due to the volume of per-element state. A more targeted investigation, which identifies which specific intermediate quantities benefit from shared memory staging without significantly increasing the per-block footprint, is left as future work.

\section{Conclusions and Contributions}
\label{sec:contributions}

This paper presents the numerical methods and GPU-accelerated implementation of the Total Lagrangian finite element framework formulated in the companion paper~\cite{json-ganesh-danTLFEA-1-2026}. The central thrust is a fully GPU-resident implicit solver for large-deformation flexible multibody dynamics with bilateral constraints and frictional contact, built around a fixed-sparsity Newton method within an augmented Lagrangian outer loop and paired with a BVH-free asynchronous collision detection pipeline. Systematic scaling benchmarks and quantitative validation tests, described in Sections~\ref{sec:performance_benchmark}--\ref{sec:large_scale}, support the following specific contributions.

\paragraph{GPU-resident precomputation and fixed-sparsity assembly} All reference configuration quantities, i.e.,  shape-function values and gradients at quadrature points, element Jacobians, and the consistent mass matrix, are precomputed once and cached in device memory. The nonzero sparsity pattern of every global operator (mass matrix, constraint Jacobian, Newton Hessian) is constructed once at initialization and held fixed for the duration of the simulation. This fixed-sparsity strategy eliminates repeated symbolic analysis, allows cuDSS to reuse its fill-reducing permutation and factor buffers, and reduces each subsequent Newton iteration to a cheaper numerical refactorization.

\paragraph{Two-stage GPU parallelization of internal force and tangent stiffness} Stage~1 assigns one GPU thread per element--quadrature-point pair to evaluate and cache the first Piola--Kirchhoff stress; threads write to disjoint memory locations, so the stage is fully data-parallel with no synchronization. Stage~2 assigns one thread per element--node pair and accumulates force contributions into the global vector via atomic scatter. Tangent stiffness assembly reuses the Stage~1 stress cache, eliminating redundant constitutive evaluations. This decomposition scales to arbitrary mesh sizes and element types without restructuring.

\paragraph{GPU Newton solver within an augmented Lagrangian outer loop} Bilateral constraints commonly encountered in multibody dynamics problems are enforced by an ALM outer loop that alternates an inner velocity solve with a dual-ascent multiplier update, decoupling penalty-parameter sensitivity from the inner solve. The inner solver assembles the sparse global Hessian on the GPU and delegates the linear solve to cuDSS via Cholesky factorization, exploiting the symmetric positive-definite structure of the Newton system. Systematic benchmarks across three element types (T10, ANCF3243, ANCF3443) and six mesh resolutions show that the Newton solver achieves approximately one order of magnitude reduction in real-time factor relative to CPU baselines at the largest resolutions tested. An exploratory first-order AdamW solver is also implemented and benchmarked; it is competitive at small scale but is superseded by Newton at large scale due to first-order convergence saturation and hyperparameter sensitivity (Section~\ref{sec:solvers_first_order}).

\paragraph{BVH-free GPU collision detection with asynchronous pipelining} A GPU-native broad-phase algorithm decomposes triangle meshes into soups and identifies contact candidates through a bin-based spatial partitioning that maps directly to GPU thread blocks, avoiding the per-step tree updates required by bounding-volume hierarchies. Narrow-phase resolution uses projection-based triangle--triangle overlap queries, and contact normals are smoothed by patch-level area-weighted reduction. The algorithm is structured as two asynchronous CPU-managed threads --- a kinematics thread for broad-phase detection and a dynamics thread for narrow-phase resolution --- each controlling a dedicated GPU stream and launching kernels, that communicate through shared buffers, so that collision detection cost overlaps with physics integration.

\paragraph{Quantitative validation of contact and bilateral constraints} The frictional contact model is validated against closed-form rigid-body predictions in two unit tests: a quasi-static brick-on-slope experiment spanning four Coulomb regimes (including the stick--slip threshold), and a dynamic oblique-impact sweep comparing the tangential coefficient of restitution and post-impact angular velocity against analytical theory from $50^\circ$ to $87^\circ$. The bilateral constraint and joint-force recovery machinery is validated through deformable double-pendulum simulations with revolute and spherical joints, comparing trajectory and constraint residuals against Project Chrono rigid-body references and verifying ALM-recovered joint forces against analytical quasi-static equilibrium predictions.

\backmatter

\section*{Statements and Declarations}

\begin{itemize}
\item \textbf{Funding}\enspace This work was supported in part through a gift from Artfit Technologies founder Bob Crozier.
\item \textbf{Competing Interests}\enspace The authors declare no competing
interests.
\item \textbf{Code availability}\enspace The GPU-accelerated Total Lagrangian
finite element implementation described in this work is openly available as
an open-source research code~\cite{TotalLagrangianFEA2025}.
\item \textbf{Data availability}\enspace Not applicable.
\end{itemize}

\begin{appendices}

\section{Performance Benchmark Parameters}
\label{sec:appendix_benchmark_params}
This appendix collects the material and time-integration parameters for every benchmark in Sections~\ref{sec:performance_benchmark}--\ref{sec:large_scale}. RTF tables and mesh-statistics tables remain in the main text.

\subsection*{Scaling Benchmarks (T10, ANCF3243, ANCF3443)}
\label{sec:appendix_scaling_params}
\label{sec:appendix_t10_params}
\label{sec:appendix_ancf3243_params}
\label{sec:appendix_ancf3443_params}

All three element-type scaling benchmarks use the SVK constitutive model with identical material parameters: $E = 7.0\times10^{8}$\,Pa, $\nu = 0.33$, $\rho_0 = 2700$\,kg/m$^3$.
The ANCF3243 beam additionally requires a cross-section specification: rectangular, $W = 0.1$\,m, $H = 0.1$\,m, element length $L = 0.2$\,m.
Time-integration parameters are summarized in Table~\ref{tab:scaling_sim_params}.
The T10 benchmark additionally evaluates the compressible Mooney--Rivlin model; its parameters are given in Table~\ref{tab:t10_mr_params}.

\begin{table}[htb]
  \centering
  \small
  \renewcommand{\arraystretch}{1.1}
  \setlength{\tabcolsep}{6pt}
  \begin{tabular}{@{}lrrr@{}}
    \toprule
    Benchmark & $\Delta t$ (s) & Steps & Duration (s) \\
    \midrule
    T10      & $10^{-3}$        & 50  & 0.05 \\
    ANCF3243 & $10^{-3}$        & --- & ---  \\
    ANCF3443 & $5\times10^{-4}$ & 200 & 0.10 \\
    \bottomrule
  \end{tabular}
  \caption{Time-integration parameters for the three element-type scaling benchmarks.
    Each resolution level is run independently; ``---'' indicates that the step count is
    resolution-dependent and is not fixed in the appendix.}
  \label{tab:scaling_sim_params}
\end{table}

\begin{table}[htb]
  \centering
  \small
  \renewcommand{\arraystretch}{1.1}
  \setlength{\tabcolsep}{6pt}
  \begin{tabular}{@{}ll@{}}
    \toprule
    Parameter & Value \\
    \midrule
    $C_{10}$ ($\mu_{10}$) & $7.89\times10^{7}$\,Pa \\
    $C_{01}$ ($\mu_{01}$) & $5.26\times10^{7}$\,Pa \\
    $\kappa$              & $1.03\times10^{9}$\,Pa \\
    $\rho_0$              & $2700$\,kg/m$^3$ \\
    \bottomrule
  \end{tabular}
  \caption{Mooney--Rivlin material parameters for the T10 beam-sagging benchmark.
    These constants correspond to an equivalent small-strain pair of $E=7.0\times10^8$\,Pa
    and $\nu=0.33$, matching the SVK case above.}
  \label{tab:t10_mr_params}
\end{table}

\subsection*{Geometrically Complex Mesh Benchmarks}
\label{sec:appendix_complex_params}

All three cases use the SVK constitutive model with the Newton solver.
Time-integration parameters are given in Table~\ref{tab:complex_sim_params};
material parameters are given in Table~\ref{tab:complex_materials}.

\begin{table}[htb]
  \centering
  \small
  \renewcommand{\arraystretch}{1.1}
  \setlength{\tabcolsep}{6pt}
  \begin{tabular}{@{}lrrr@{}}
    \toprule
    Case & $\Delta t$ (s) & Steps & Duration (s) \\
    \midrule
    Utah Teapot     & $5\times10^{-4}$ & 1{,}000 & 0.5 \\
    Stanford Bunny  & $1\times10^{-3}$ & 8{,}000 & 8.0 \\
    Deformable Tire & $5\times10^{-4}$ & 1{,}200 & 0.6 \\
    \bottomrule
  \end{tabular}
  \caption{Time-integration parameters for the three geometrically complex mesh benchmarks.}
  \label{tab:complex_sim_params}
\end{table}

\begin{table}[htb]
  \centering
  \small
  \renewcommand{\arraystretch}{1.1}
  \setlength{\tabcolsep}{5pt}
  \begin{tabular}{@{}lrrr@{}}
    \toprule
    Parameter & Utah Teapot & Stanford Bunny & Deformable Tire \\
    \midrule
    $E$ (Pa)      & $2.0\times10^{6}$ & $3.0\times10^{8}$ & $1.0\times10^{7}$ \\
    $\nu$         & $0.35$            & $0.40$            & $0.35$            \\
    $\rho_0$ (kg/m$^3$) & $1000$      & $920$             & $1000$            \\
    $\eta_\mathrm{damp}$ (Pa$\cdot$s) & $5.0\times10^{4}$ & $0$ & $5.0\times10^{3}$ \\
    $\lambda_\mathrm{damp}$ (Pa$\cdot$s) & $5.0\times10^{4}$ & $0$ & $5.0\times10^{3}$ \\
    \bottomrule
  \end{tabular}
  \caption{Material parameters for the three geometrically complex mesh benchmarks (SVK model with Kelvin--Voigt viscous damping).}
  \label{tab:complex_materials}
\end{table}

\subsection*{Mixed Item Dropping Benchmark}
\label{app:tire-drop-params}

Nine deformable tires are dropped into a fixed rigid container alongside a
cantilevered ANCF3443 shell beam. The system is advanced at
$\Delta t = 1\times10^{-4}$\,s for 50{,}000 steps (5.0\,s total). Both
subsystems use the SVK constitutive model with Kelvin--Voigt viscous damping.
Mesh statistics, material parameters, and solver settings are given in
Tables~\ref{tab:tire-drop-mesh}--\ref{tab:tire-drop-solver}.

\begin{table}[htb]
  \centering
  \small
  \renewcommand{\arraystretch}{1.1}
  \setlength{\tabcolsep}{6pt}
  \begin{tabular}{@{}lrrrr@{}}
    \toprule
    Component & Nodes & Elements & DOFs & Constrained DOFs \\
    \midrule
    ANCF3443 shell beam   & 403       & 360       & 4{,}836    & 156     \\
    T10 tire ($\times$9)  & 309{,}726 & 154{,}503 & 929{,}178  & ---     \\
    T10 open box          & 4{,}474   & 2{,}170   & 13{,}422   & 13{,}422\\
    \midrule
    \textbf{Total}        & 314{,}603 & 157{,}033 & 947{,}436  & 13{,}578\\
    \bottomrule
  \end{tabular}
  \caption{Mesh statistics for the mixed item-dropping benchmark.}
  \label{tab:tire-drop-mesh}
\end{table}

\begin{table}[htb]
  \centering
  \small
  \renewcommand{\arraystretch}{1.1}
  \setlength{\tabcolsep}{5pt}
  \begin{tabular}{@{}lrr@{}}
    \toprule
    Parameter & Tire bodies ($\times$9) & Shell beam \\
    \midrule
    $E$ (Pa)      & $5.0\times10^{6}$ & $8.0\times10^{6}$ \\
    $\nu$         & $0.40$            & $0.33$            \\
    $\rho_0$ (kg/m$^3$) & $900$       & $1200$            \\
    $\eta_\mathrm{damp} = \lambda_\mathrm{damp}$ (Pa\,s) & $2.0\times10^{4}$ & $2.0\times10^{4}$ \\
    \bottomrule
  \end{tabular}
  \caption{Material parameters for the mixed item-dropping benchmark (SVK model).}
  \label{tab:tire-drop-material}
\end{table}

\begin{table}[htb]
  \centering
  \small
  \renewcommand{\arraystretch}{1.1}
  \setlength{\tabcolsep}{6pt}
  \begin{tabular}{@{}llr@{}}
    \toprule
    Category & Parameter & Value \\
    \midrule
    Time integration & $\Delta t$                 & $1\times10^{-4}$\,s \\
                     & Steps (total)              & 50{,}000 (5.0\,s)   \\
                     & Gravity                    & $-9.81$\,m/s$^{2}$  \\
    \midrule
    Nonlinear solve  & Inner absolute tolerance   & $1\times10^{-4}$    \\
                     & Inner relative tolerance   & $1\times10^{-6}$    \\
                     & Outer tolerance            & $1\times10^{-5}$    \\
                     & Line search                & enabled             \\
                     & ALM penalty $\rho$         & $1\times10^{14}$    \\
                     & Max outer iterations       & 5                   \\
                     & Max inner iterations       & 20                  \\
                     & Solver blocks              & 2                   \\
    \midrule
    Contact          & $\mu_s$                    & 0.6                 \\
                     & $\mu_k$                    & 0.5                 \\
                     & Contact stiffness          & $1\times10^{7}$\,Pa/m \\
                     & Restitution coefficient    & 0.5                 \\
                     & Self-collision             & disabled            \\
    \midrule
    DEM coupling     & Force clamp                & 50{,}000\,N         \\
                     & Distribution parameter $K$ & 8                   \\
    \bottomrule
  \end{tabular}
  \caption{Solver, time-integration, and contact parameters for the mixed item-dropping benchmark.}
  \label{tab:tire-drop-solver}
\end{table}

\subsection*{Joint Constraint Validation Benchmarks}
\label{sec:appendix_joint_params}

All joint pendulum tests use two identical T10 beams
($0.5\,\mathrm{m}\times0.04\,\mathrm{m}\times0.04\,\mathrm{m}$,
$m_b = 0.96$\,kg per link). Motion tests are advanced for 5{,}000 steps
at $\Delta t = 5\times10^{-4}$\,s (2.5\,s total); pull tests for 1{,}000
steps at the same $\Delta t$ (0.5\,s total). All cases use the Newton solver
with ALM $\rho = 1\times10^{10}$, outer tolerance $1\times10^{-8}$, inner
tolerance $1\times10^{-6}$, at most 8 outer iterations, and at most 10 inner
iterations. Material parameters are given in Table~\ref{tab:joint_material_params}.

\begin{table}[htb]
  \centering
  \small
  \renewcommand{\arraystretch}{1.1}
  \setlength{\tabcolsep}{5pt}
  \begin{tabular}{@{}lrr@{}}
    \toprule
    Parameter & Motion Tests & Pull Tests \\
    \midrule
    Model  & SVK & SVK \\
    $E$ (Pa)      & $2\times10^{6}$  & $1\times10^{7}$  \\
    $\nu$         & $0.30$           & $0.30$           \\
    $\rho_0$ (kg/m$^3$) & $1200$     & $1200$           \\
    $\eta_\mathrm{damp} = \lambda_\mathrm{damp}$ (Pa$\cdot$s)
                  & $\{0,\,10^2,\,10^3,\,10^4\}$ & $1\times10^{4}$ \\
    \bottomrule
  \end{tabular}
  \caption{Material parameters for the joint constraint validation benchmarks
    (revolute and spherical, motion and pull tests).}
  \label{tab:joint_material_params}
\end{table}

\end{appendices}

\bibliography{BibFiles/refsGraphics,BibFiles/refsSensors,BibFiles/refsAutonomousVehicles,BibFiles/refsChronoSpecific,BibFiles/refsDEM,BibFiles/refsFSI,BibFiles/refsMBS,BibFiles/refsRobotics,BibFiles/refsSBELspecific,BibFiles/refsTerramech,BibFiles/refsCompSci,BibFiles/refsNumericalIntegr,BibFiles/refsMLPhysics,BibFiles/refsSurfaceTension,BibFiles/refsStatsML,BibFiles/refsOddsEnds,BibFiles/refsML-AI}

\end{document}